\documentclass[conference]{IEEEtran}
\usepackage{eso-pic} 
\usepackage{algorithm}
\usepackage{algpseudocode}
\usepackage{amsmath}
\usepackage{amssymb}
\usepackage{url}
\usepackage{epsfig}
\usepackage{graphicx}
\usepackage{caption}
\usepackage{subcaption}

\makeatletter
\renewcommand{\ALG@beginalgorithmic}{\small}
\makeatother

\algnotext{EndIf}
\algnotext{EndFor}
\algnotext{EndWhile}

\ifCLASSINFOpdf
\else
\fi
\begin{document}
\sloppy
%
% paper title
% can use linebreaks \\ within to get better formatting as desired
\title{Private Link Exchange over Social Graphs}

% author names and affiliations
% use a multiple column layout for up to three different
% affiliations
\author{\IEEEauthorblockN{Hiep H. Nguyen, Abdessamad Imine, and Micha\"{e}l Rusinowitch}
\IEEEauthorblockA{LORIA/INRIA Nancy-Grand Est, France}
Email: \{huu-hiep.nguyen,michael.rusinowitch\}@inria.fr, abdessamad.imine@loria.fr
}

% conference papers do not typically use \thanks and this command
% is locked out in conference mode. If really needed, such as for
% the acknowledgment of grants, issue a \IEEEoverridecommandlockouts
% after \documentclass

% for over three affiliations, or if they all won't fit within the width
% of the page, use this alternative format:
% 
%\author{\IEEEauthorblockN{Michael Shell\IEEEauthorrefmark{1},
%Homer Simpson\IEEEauthorrefmark{2},
%James Kirk\IEEEauthorrefmark{3}, 
%Montgomery Scott\IEEEauthorrefmark{3} and
%Eldon Tyrell\IEEEauthorrefmark{4}}
%\IEEEauthorblockA{\IEEEauthorrefmark{1}School of Electrical and Computer Engineering\\
%Georgia Institute of Technology,
%Atlanta, Georgia 30332--0250\\ Email: see http://www.michaelshell.org/contact.html}
%\IEEEauthorblockA{\IEEEauthorrefmark{2}Twentieth Century Fox, Springfield, USA\\
%Email: homer@thesimpsons.com}
%\IEEEauthorblockA{\IEEEauthorrefmark{3}Starfleet Academy, San Francisco, California 96678-2391\\
%Telephone: (800) 555--1212, Fax: (888) 555--1212}
%\IEEEauthorblockA{\IEEEauthorrefmark{4}Tyrell Inc., 123 Replicant Street, Los Angeles, California 90210--4321}}

% use for special paper notices
%\IEEEspecialpapernotice{(Invited Paper)}

\maketitle
\begin{abstract}
Currently, most of the online social networks (OSN) keep their data secret and in centralized manner. Researchers are allowed to crawl the underlying social graphs (and data) but with limited rates, leading to only partial views of the true social graphs. To overcome this constraint, we may start from user perspective, the contributors of the OSNs. More precisely, if users cautiously collaborate with one another, they can use the very infrastructure of the OSNs to exchange noisy friend lists with their neighbors in several rounds. In the end, they can build local subgraphs, also called local views of the true social graph. In this paper, we propose such protocols for the problem of \textit{private link exchange} over social graphs.

The problem is unique in the sense that the disseminated data over the links are the links themselves. However, there exist fundamental questions about the feasibility of this model. The first question is how to define simple and effective privacy concepts for the link exchange processes. The second question comes from the high volume of link lists in exchange which may increase exponentially round after round. While storage and computation complexity may be affordable for desktop PCs, communication costs are non-trivial. We address both questions by a simple $(\alpha,\beta)$-exchange using Bloom filters. 
\end{abstract}

% IEEEtran.cls defaults to using nonbold math in the Abstract.
% This preserves the distinction between vectors and scalars. However,
% if the conference you are submitting to favors bold math in the abstract,
% then you can use LaTeX's standard command \boldmath at the very start
% of the abstract to achieve this. Many IEEE journals/conferences frown on
% math in the abstract anyway.

% no keywords

% For peer review papers, you can put extra information on the cover
% page as needed:
% \ifCLASSOPTIONpeerreview
% \begin{center} \bfseries EDICS Category: 3-BBND \end{center}
% \fi
%
% For peerreview papers, this IEEEtran command inserts a page break and
% creates the second title. It will be ignored for other modes.
\IEEEpeerreviewmaketitle
\section{Introduction}

\label{sec:introduction}
Online social networks (OSN) have grown significantly over the last ten years with billions of active users using a variety of social network services. OSNs have revolutionized the way people interact. People join social networking sites to connect and communicate with their friends in real-time. They share interests and activities across political, economic, and geographic borders. %OSNs like Facebook also become an important channel for e-commerce.
As social network sites continue to develop both in number and size, the service providers accumulate unprecedented amount of information about OSN users. As a result, social networks are a valuable data source for research on information societies. In particular, underlying social graphs play a key role in understanding how people form communities, how the OSNs suggest friendship to two users who do not know each other but have many common friends, etc. However, social graphs are not published in clear form due to serious privacy concerns. Instead, they are anonymized in various forms and published to third party consumers such as sociologists, epidemiologists, advertisers and criminologists. Alternatively, social networking sites provide APIs \footnote{https://developers.facebook.com/docs/graph-api} for data crawlers at limited rates and within privacy constraints (e.g. only public friend lists are available). Using this method, the data crawlers can collect friendship information and build a partial (local) view of the target social graph.

To overcome the constraints set by the service providers, we can start from user perspective, i.e. the contributors of OSNs. More precisely, if users cautiously collaborate with one another, they can exchange \textit{noisy} friend lists (containing fake friendships) with their neighbors in several rounds to get better views of the true social graph. Our ideas are based on the fact that user IDs are public (e.g. Facebook profiles are searchable \cite{fb-dir}) but the friendships are not so, except when a user leaves his friend list in public mode. Using public user IDs, any user can claim fake links from himself to the users not in his friend list.

The aggregation problem in this paper is unique in the sense that the disseminated data over the links are the links themselves. However, there exist fundamental questions about the feasibility of this model. The first question is how to define simple and effective privacy concepts for the link exchange processes. The second question comes from the high volume of link lists in exchange which may increase exponentially round after round. While storage and computation complexity may not be big problems, communication costs are non-trivial. We address both questions by a simple $(\alpha,\beta)$-exchange protocol with or without Bloom filters. To protect true links from inference attacks, we add fake links which are $beta$-fraction of true links. Furthermore, we realize the attenuated propagation of links via the parameter $\alpha \leq 1$.

Basically, we assume that users are \textit{honest-but-curious} (HbC), i.e. they follow the protocol but try to figure out true friendships among noisy friend lists. To preserve link privacy, each node obfuscates its friend list by adding fake links originating from itself to a number of nodes not in its friend list. Then in exchange stage, nodes share only with their friends a fraction of noisy links they possess.

Our contributions are summarized as follows:
\begin{itemize}
\item We introduce a novel private link exchange problem as an alternative to social graph crawling and centralized anonymization of data. The problem is distributed and provides a privacy/utility trade-off for all nodes.
\item We present two schemes for $(\alpha,\beta)$-exchange protocol: Baseline and Bloom filter based. We protect the true links by adding fake links and requiring the propagation probability of links to be attenuated by distance. We analyze the advantages and disadvantages of each scheme.
\item We evaluate our proposed schemes on various synthetic graph models and draw a number of critical findings.
\end{itemize}

The paper is organized as follows. We review the related work for information dissemination in social graphs, distributed anonymization, social trust models and Bloom filter in Section \ref{sec:related}. Section \ref{sec:pre} briefly introduces key concepts of Bloom filter and our link exchange model. In Section \ref{sec:link-exchange}, we present Baseline $(\alpha,\beta)$-exchange that realizes the exchange model by sending noisy link lists in clear form. Section \ref{sec:bf} describes Bloom filter version of $(\alpha,\beta)$-exchange with constant complexities and better privacy. We validate the proposed schemes in Section \ref{sec:eval}. Finally, we present our remarks and suggest future work in Section \ref{sec:conclusion}.

Table \ref{tab:notation} summarizes notations used in this paper.
 
\begin{table}[htb]
\small
\centering
\caption{List of notations} \label{tab:notation}
\begin{tabular}{|c|l|}
\hline
\textbf{Symbol} &\textbf{Definition} \\
\hline
$G=(V,E)$ & social graph with $N=|V|$ and $M=|E_{G}|$\\
\hline
$A(G)$ & adjacency matrix of $G$\\
\hline
$D$ & degree sequence of $G$ (column vector)\\
\hline
$Diam(G)$ & diameter of $G$\\
\hline
$N(u)$ & neighbors of node $u$ in $G$, $d_u = |N(u)|$ \\
\hline
$T$ & number of exchange rounds \\
\hline
$(v,w)$ & true link between $v$ and $w$ \\
\hline
$(v\rightarrow w)$ & fake link generated by $v$ \\
\hline
$L_u(t)$ & set of links possessed by $u$ at round $t$\\
\hline
$L_{uv}(t)$ & set of links $u$ sends to $v$ at time $t$\\
\hline
$\propto$ & uniformly at random sampling without replacement\\
\hline
$\alpha$ & fraction of links shared between a pair of nodes\\
\hline
$\beta$ & fraction of fake links generated at $t = 0$ \\
\hline
$m$ & number of bits in Bloom filter \\
\hline
$k$ & number of hash functions used in Bloom filter \\
\hline
$n$ & number of elements in Bloom filter \\
\hline
$Bf_u(t)$ & Bloom filter possessed by $u$ at round $t$\\
\hline
$Bf_{uv}(t)$ & Bloom filter $u$ sends to $v$ at time $t$\\
\hline

\end{tabular}
\end{table}

% %
\section{Related Work}
\label{sec:related}
%\subsection{Decentralized Social Networks}
%\subsection{Peer-to-Peer}
%\subsection{Gossip Based Computation}
%\subsection{Distributed Anonymization}
%\subsection{Social Trust}
%\subsection{Bloom Filter}
%\subsection{Network Resilience}

Epidemic spreading \cite{pastor2001epidemic,moreno2002epidemic} is the most related to our work. In \cite{pastor2001epidemic}, Pastor-Satorras et al. study the spreading of infections on scale-free (power-law) networks via the susceptible-infected-susceptible (SIS) model \cite{bailey1975mathematical}. They find the absence of an epidemic threshold ($\lambda_c = 0$) and its associated critical behavior when the number of nodes goes to infinity using mean-field approximation. Moreno et al. \cite{moreno2002epidemic} provide a detailed analytical and numerical study of susceptible-infected-removed (SIR) on Watts-Strogatz (WS) small-world model and Barab\'{a}si-Albert (BA) scale-free model. WS graphs with exponentially distributed degrees can be considered as a \textit{homogeneous} model in which each node hash the same number of links. WS graphs have finite epidemic thresholds. On the contrary, BA graphs with power-law distributed degrees are \textit{heterogeneous} and they expose the weaker resistance to epidemics starting on highly connected nodes. 

Giakkoupis et al. \cite{giakkoupis2015privacy} introduce a distributed algorithm RIPOSTE for disseminating information in a social network that preserves privacy of nodes. Whenever the information reaches a node, the node decides to either forward the information to his neighbors or drop it. RIPOSTE uses two global parameters $\delta$ and $\lambda$ and satisfies differential privacy by applying Randomized Response Technique (RRT) \cite{dwork2014algorithmic}. Our work is also a form of information dissemination over graphs but it spreads a large number of links, not a single item.

Gossip-based protocols \cite{ganesh2003peer,voulgaris2005cyclon} aim at providing alternatives to network-level multicast with good scalability and reliability properties. In these protocols, message redundancy for high reliability is ensured by the fact each member forwards each message to a set of other, randomly chosen, group members. Ganesh et al. \cite{ganesh2003peer} propose a fully decentralized and self-configuring protocol SCAMP that provides each member with a partial view of group membership. As the number of participating nodes changes, the size of partial views automatically adapts to the value required to support a gossip algorithm reliably. CYCLON \cite{voulgaris2005cyclon} is a protocol for construction of reliable overlay networks. It is targeted to overlays that have low diameter, low clustering, highly symmetric node degrees and highly resilient to massive node failures. These properties belong to random graphs. CYCLON employs enhanced shuffling operation in which nodes select neighbors for cache exchange based on their age.

By exchanging noisy link lists, our schemes are related to distributed graph anonymization \cite{campan2008clustering,tassa2013anonymization}. However, rather than producing a single global anonymized graph as in \cite{tassa2013anonymization}, link exchange protocols result in multiple local outputs. In addition, link exchange operates at finest-grained level (node-level) whereas previous works consider a small number of data holders who manage disjoint sets of nodes. 

The idea of adding fake links to hide true links appears in a number of earlier studies, e.g. \cite{shokri2009preserving,nguyen2015anonymizing}. Shokri et al. \cite{shokri2009preserving} propose a method for privacy preservation in collaborative filtering recommendation systems. They develop a model where each user stores locally an offline profile on his own side, hidden from the server, and an online profile on the server from which the server generates the recommendations. Each user arbitrarily contacts other users over time, and modifies his own offline profile through aggregating ratings from other users. The more ratings a user aggregates, the higher privacy he is but lower accuracy in recommendations. Nguyen et al. \cite{nguyen2015anonymizing} present a centralized graph anonymization scheme based on edge uncertainty semantics. Fake links are added to probabilistically hide true links. They consider distance-2 fake links to keep higher utility.

% %
\section{Preliminaries}
\label{sec:pre}
In this section, we present the exchange model and attack model. Then we review key concepts about Bloom filter.

\subsection{Exchange Model}
\label{subsec:exchange-model}
%Regarding the current model of Facebook, we make the following assumptions

We consider a distributed exchange model in which each node possesses his friend list and all nodes participate in the exchange protocol. We work on the following assumptions
\begin{itemize}
\item \textbf{Assumption 1} The space of node IDs is public. A node can generate fake links to any node. All friend lists (true links) are private, i.e. the existence of true link $(u,v)$ is surely known to $u$ and $v$ only.
\item \textbf{Assumption 2} A node exchanges messages with its neighbors only. Interacting with neighbors is based on an intuition of trusted relationships: we trust our friends more than any stranger. 
\item \textbf{Assumption 3} A synchronous model is guaranteed by \textit{round-tagged} messages. It means a node prepares the message for round $t+1$ if and only if it has received all $t$-th round messages from his friends.
\item \textbf{Assumption 4} All nodes are honest-but-curious. They follow the protocol but try to infer true links among noisy links.
\end{itemize}

\begin{figure}
\centering
\includegraphics[height=2.3in]{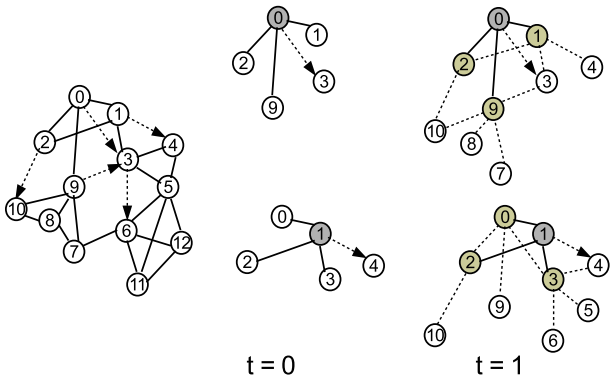}
\setlength{\abovecaptionskip}{-10pt}
\caption{Link exchange with $\alpha = 1$, $\beta = 1/3$}
\vspace{-1.0em}
\label{fig:link-exchange}
\end{figure}

Fig. \ref{fig:link-exchange} illustrates the exchange model. At round $t = 0$ (initial round), each node $u$ prepare a noisy friend list by adding some fake links $(u,v)$ (i.e. links from $u$ to some people not in his friend list). This is feasible because all user IDs are public (e.g. \cite{fb-dir}). For example, node 0 adds a fake link (0,3) and his noisy friend list \{(0,1), (0,2), (0,3)\} is ready to be exchanged. Similarly, the other nodes prepare their noisy friend lists as in Fig \ref{fig:link-exchange}. At round $t = 1$, all nodes send and receive noisy friend lists from their neighbors. The local views of nodes 0 and 1 at $t = 1$ are shown in Fig. \ref{fig:link-exchange} where the solid lines (resp. the dashed arrows) are the true links (resp. fake links) known by the node and the dashed lines represent noisy links received at the node.

\subsection{Attack Model}
\label{subsec:attack-model}
We consider honest-but-curious users (nodes) who follow the protocol but try to infer true links among noisy links. We propose a simple inference attack based on frequencies of links arriving to a node. Given a link $(v,w)$ (a true link or a fake link) arriving to node $u$, if $(v,w)$ does not exist in $u$'s local view, it will be added. Otherwise, its frequency is increased by 1. At the end of the protocol, each node sorts all links in its local view by frequency and selects top $K$ links as true links. How to select the value of $K$ will be discussed later.

By splitting noisy links into two sets of links as above, the inference capability of each node is evaluated by common measures \cite{fawcett2006introduction}: True Positives (TP), True Negatives (TN), False Positives (FP), False Negatives (FN). As we will see in Section \ref{sec:link-exchange}, the parameter $\alpha$ introduces an \textit{attenuation effect} for link propagation when $\alpha < 1$. Given a link $e$, nodes farther from $e$ have lower chance of getting this link. This effect adds another dimension to our privacy model.

\subsection{Bloom Filter}
The Bloom filter is a space-efficient probabilistic data structure that supports set membership queries. It was first conceived by Burton Howard Bloom in 1970 \cite{bloom1970space}. It is used to test whether an element is a member of a set and can result in false positives (claiming an element to belong to the set when it was not inserted), but never in false negatives (reporting an inserted element not in the set).

\begin{figure}
\centering
\includegraphics[height=1.2in]{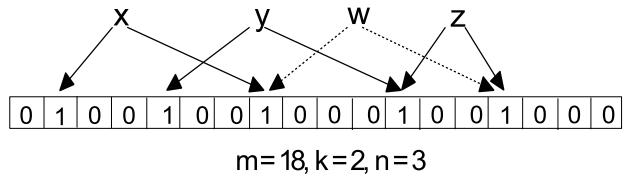}
\setlength{\abovecaptionskip}{-20pt}
\caption{Bloom filter}
\vspace{-1.0em}
\label{fig:bloom-filter}
\end{figure}

An empty Bloom filter is an array of $m$ bits, all set to zero. There must also be $k$ different hash functions defined, each of which maps or hashes some set element $x$ to one of the $m$ array positions with a \textit{uniform} random distribution. The number of elements inserted into the Bloom filter is $n$. Fig. \ref{fig:bloom-filter} gives an example of Bloom filter with $m=18$, $k=2$ and $n=3$. The MD5 hash algorithm is a popular choice for the hash functions. When an element not in the set $w$ is looked up, it will be hashed by the $k$ hash functions into bit positions. If one of the positions is zero, we conclude that $w$ is not in the set. It may happen that all the bit positions of an element have been set. When this occurs, the Bloom filter will erroneously report that the element is a member of the set, also known as false positives. Fig. \ref{fig:bloom-filter} shows $w$ as a false positive.

Given the three parameters $m$, $k$ and $n$, the false positive probability is (see \cite{broder2004network}).
\begin{equation}
p = \left( 1 - (1-\frac{1}{m})^{kn}\right)^{k} \approx (1 - e^{-kn/m})^{k}
\end{equation}

The false positive probability decreases as $m$ increases or $n$ decreases. Given $m$ and $n$, the probability of false positives $(1 - e^{-kn/m})^{k}$ is minimized at $k = k_{opt} = \frac{m}{n} \ln 2$ (see \cite{broder2004network}). In this case, the false positive rate $p = (1/2)^k$ or equivalently
\begin{equation}
k = -\log_2{p} \label{eqn:k}
\end{equation}

\section{Baseline $(\alpha,\beta)$-exchange}
\label{sec:link-exchange}
In this section, we present the main ideas of our proposed $(\alpha,\beta)$-exchange and the improvements using Bloom filters.

\subsection{Overview}
As shown in Section \ref{subsec:exchange-model}, the link exchange protocol is straightforward. At the beginning of the protocol, all the nodes agree on the number of rounds $T$ and the two parameters $\alpha \in [0,1]$, $\beta \geq 0$. Then, each node $u$ prepares his own noisy friend list $L_u(0)$ by setting $L_u(0) = \{(u,v)| v \in N(u)\}$ and adding $\beta N(u)$ fake links in the form $(u \rightarrow w)$ where $w \notin N(u)$. At $t = 1$, each node $u$ makes a noisy list $L_{uv}(1)$ for every neighbor $v$ so that $L_{uv}(1)$ contains $\alpha |L_u(0)|$ links sampled from $L_u(0)$. Similarly, node $v$ prepares a noisy list $L_{vu}(1)$ for $u$. All the nodes sends and receives noisy link lists. Next, each node aggregates noisy link lists by removing duplicate links (if any) and obtains his local view of graph by $L_u(1)$. The round $t = 1$ finishes. 

At $t = 2$, the process repeats: all nodes $u$ makes a noisy list $L_{uv}(2)$ for every neighbor $v$ that contains $\alpha |L_u(1)|$ links sampled from $L_u(1)$. They exchange noisy link lists and after receiving all $L_{vu}(2)$ from his friends, node $u$ updates his local view and gets $L_u(2)$. When $t = T$, the protocol terminates.

\subsection{Baseline Scheme}
The idea in the previous section is called Baseline $(\alpha,\beta)$-exchange as all noisy link lists are in clear form. Algorithm \ref{algo-baseline} shows steps for Baseline $(\alpha,\beta)$-exchange.

\begin{algorithm}
\caption{Baseline $(\alpha,\beta)$-exchange}
\label{algo-baseline}
\begin{algorithmic}[1]
	\Require undirected graph $G=(V,E)$, parameters $\alpha \in [0,1]$, $\beta \geq 0$, number of rounds $T$
	\Ensure noisy local views of graph $L_u(T), u \in V$ 

	\State // initialization stage
	\For {$u \in V$}
		\State $Fa(u) = \{(u \rightarrow w) | w \notin N(u) \}$ s.t. $|Fa(u)| = \beta |N(u)|$
		\State $L_u(0) = \{(u,v)| v \in N(u)\} \cup Fa(u)$
	\EndFor
	\State // exchange stage
	\For {$t = 1..T$}
		\For {$(u,v) \in E$}
			\State $u$ : $L_{uv}(t) \propto L_u(t-1)$ s.t. $|L_{uv}(t)| = \alpha |L_u(t-1)|$
			\State $v$ : $L_{vu}(t) \propto L_v(t-1)$ s.t. $|L_{vu}(t)| = \alpha |L_v(t-1)|$	
			\State $u$ sends $L_{uv}(t)$ to $v$
			\State $v$ sends $L_{vu}(t)$ to $u$	
		\EndFor
		\For {$u \in V$}
			\State $L_u(t) = L_u(t-1) \cup \bigcup\limits_{v \in N(u)} L_{vu}(t)$
		\EndFor
	\EndFor
	\Return $L_u(T), u \in V$ 
\end{algorithmic}
\end{algorithm}

Given the graph structure $G=(V,E)$, two parameters $\alpha \in [0,1]$, $\beta \geq 0$ and the number of rounds $T$. The protocol takes place in two stages. In initialization stage, each node $u$ prepares his own noisy friend list $L_u(0)$ by adding $\beta N(u)$ fake links in the form $(u,w)$ where $w \notin N(u)$ (Lines 3 and 4). In exchange stage (Lines 6-13), at round $t$, each node $u$ makes a noisy list $L_{uv}(t)$ for every neighbor $v$ that contains $\alpha |L_u(t)|$ links sampled from $L_u(t)$. The exchange happens on every relationship (true link). Each node takes the union of all noisy links he receives before starting the next round.

\subsubsection{Faster Simulation in A Single PC}
For simulation in a single PC, storing all link lists for all nodes in clear form is a costly solution. Moreover, the union operation on lists is time-consuming. We present here a technique to reduce the memory footprint and processing time using bit sets. 

\begin{figure}
\centering
\includegraphics[height=2.0in]{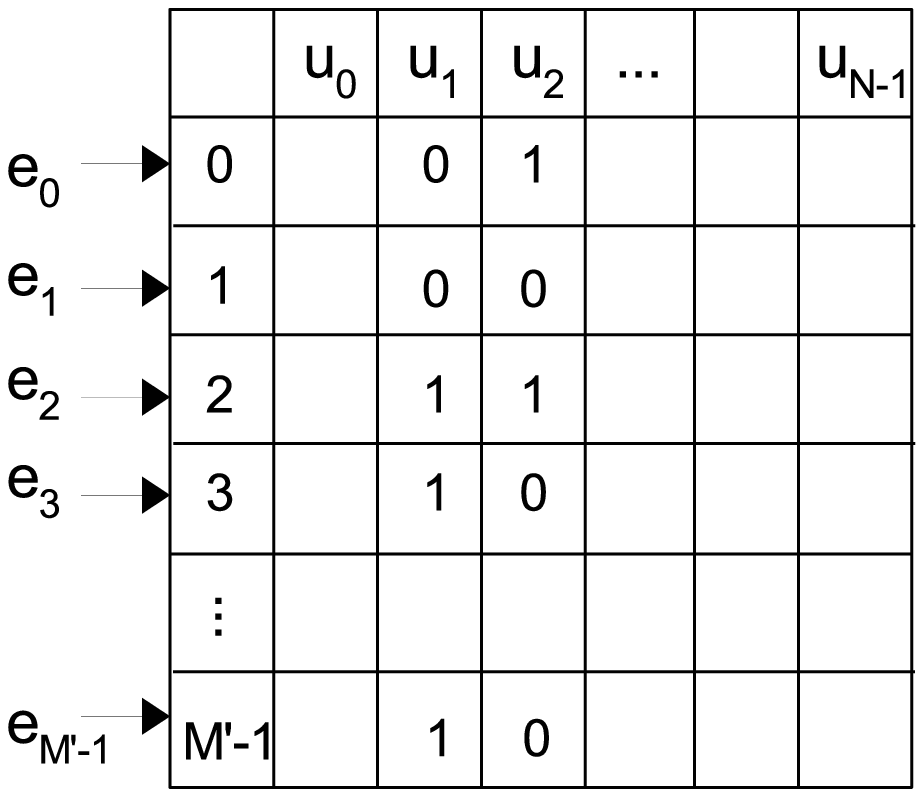}
\setlength{\abovecaptionskip}{-10pt}
\caption{Fast simulation using bit sets (column vectors)}
\vspace{-1.0em}
\label{fig:bit-set}
\end{figure}

Fig. \ref{fig:bit-set} outlines the idea. We have $M' = (1+2\beta)|E_G|$ distinct links consisting of $|E_G|$ true links and $2\beta|E_G|$ fake links. By indexing $M'$ links from 0 to $M'-1$, the noisy link list at each node is stored in a bit set of size $M'$. The union of link lists (Line 13 Algorithm \ref{algo-baseline}) is equivalent to an OR operation between bit sets. To prepare $L_{uv}(t)$ for link exchange in round $t$, node $u$ must recover link IDs in its bit set.

We emphasize that indexing links and storing link IDs in bit sets are only for simulation. In reality, the number of links are unknown to all nodes, so they must run Baseline or Bloom filter (Section \ref{sec:bf}) protocol.

\begin{figure}
\centering
\includegraphics[height=1.4in]{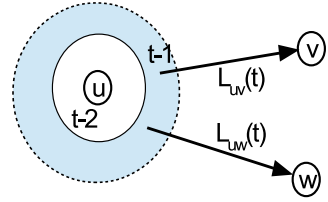}
\setlength{\abovecaptionskip}{-10pt}
\caption{Incremental volume for $\alpha = 1$}
\vspace{-1.0em}
\label{fig:incremental}
\end{figure}

For the case $\alpha = 1$, the exchange volume is reduced further if each node $u$ sends only ``new'' links, i.e. the links that do not exist in $u$'s list in the previous round. Fig. \ref{fig:incremental} visualizes this idea in which ``new'' links are in shaded region and old links are in white region. Note that the incremental volume is valid only for $\alpha = 1$. When $\alpha < 1$, the phenomenon of multipath propagation (Fig. \ref{fig:edge-propagation}) requires both new and old links to be sampled with probability $\alpha$.

%\subsubsection{Early Termination}
%In reality, the diameter of the graph is unknown to all nodes. Given the number of rounds $T$, each node may decide to 

\subsubsection{Utility-Oriented Initialization}
\label{subsec:util-oriented}
Baseline scheme in the previous section lets a node $u$ generate fake links by connecting $u$ to a certain number of nodes not in $u$'s friend list. This initialization may make local sub graphs at the final round have distorted path distributions due to many fake links connecting faraway nodes. Distorted path distributions reduce the ``utility'' perceived at each node.
Based on the observation of using fake links connecting nearby nodes \cite{nguyen2015anonymizing}, we suggest a utility-oriented improvement by two-round initialization. We call a fake link $(u\rightarrow v)$ \textit{distance-2 link} if $d(u,v) = 2$. For example, $(0 \rightarrow 3)$ is a distance-2 fake link while $(2\rightarrow 10)$ is not. Correspondingly, $v$ is called a \textit{distance-2 node} w.r.t $u$.

We introduce a new parameter $\gamma \in [0,1]$ which stipulates that each node $u$ create $\gamma\beta d_u$ fake links at $t = 0$ and exchange $\alpha(1+\gamma\beta)d_u$ randomly chosen links to each of its neighbors. Node $u$ collects node IDs and save them in the set $ID_u$. At $t=1$, node $u$ uses node IDs in $ID_u$ to create $(1-\gamma)\beta d_u$ fake links. Algorithm \ref{algo-init-util} implements this idea. 

The number of distance-2 nodes that $u$ collects in Line 7 of Algorithm \ref{algo-init-util} is $\alpha (\sum_{v\in N(u)} d_v - d_u - 2 Tri(u))$ where $Tri(u)$ is the number of triangles with $u$ as a vertex. Assuming that the set $Fa_0(u)$ contains no distance-2 links (Line 3 Algorithm \ref{algo-init-util}). The number of non-distance-2 nodes that $u$ collects is $\sum_{v\in N(u)} \alpha\gamma\beta d_v$. The expected number of distance-2 links that $u$ can create is
\begin{equation}
L2(u) = \frac{(1-\gamma)(\sum_{v\in N(u)} d_v - d_u - 2 Tri(u))}{[\sum_{v\in N(u)} d_v - d_u - 2 Tri(u)] + \sum_{v\in N(u)} \gamma\beta d_v} \nonumber
\end{equation}

$L2(u)$ is a decreasing function of $\gamma$. All nodes have the highest (resp. lowest) number of distance-2 fake links at $\gamma = 0$ (resp. $\gamma = 1$). The case of $\gamma = 1$ reduces to standard initialization (Lines 2-4 Algorithm \ref{algo-baseline}).

\begin{algorithm}
\caption{Two-round Initialization}
\label{algo-init-util}
\begin{algorithmic}[1]
	\Require undirected graph $G=(V,E)$, parameters $\alpha,\gamma \in [0,1]$, $\beta \geq 0$
	\Ensure each node $u$ issues $\beta d_u$ fake links
	\State // t = 0
	\For {$u \in V$}
		\State $Fa_0(u) = \{(u \rightarrow w) | w \notin N(u) \}$ s.t. $|Fa_0(u)| = \gamma\beta |N(u)|$
		\State $L_u(0) = \{(u,v)| v \in N(u)\} \cup Fa_0(u)$
	\EndFor
	\State // t = 1
	\For {$(u,v) \in E$}
		\State $u$ and $v$ exchange $\alpha$-fraction of their links
	\EndFor
	\For {$u \in V$}
		\State $u$ aggregates all links it knows into $L_u(1)$
		\State $ID_u = \{w| w=v_1 \wedge w=v_2, (v_1,v_2) \in L_u(1) \} \setminus \{u, N(u)\}$
		\State $Fa_1(u) = \{(u \rightarrow w) | w \in ID_u \}$ 
		\State $\;\;\;\;$ s.t. $|Fa_1(u)| = (1-\gamma)\beta |N(u)|$
		\State $L_u(1) = L_u(1) \cup Fa_1(u)$
	\EndFor
\end{algorithmic}
\end{algorithm}

\subsection{Complexity Analysis}
Let $A$ be the adjacency matrix of $G$ and $D$ be the column vector of degree sequence of nodes, the number of links at all nodes is upper bounded by the following vector, where $I_N$ is the identity matrix of size $N$.
\begin{equation}
LU(t) = (I_N + \alpha A)^t (1+\beta) D  \label{eqn:upperbound}
\end{equation}

We say $LU(t)$ is an ``upper-bound'' because $LU(t)$ accepts duplicate links. More precisely, let $LU_u(t)$ and $LU_{uv}(t)$ be the noisy link lists at node $u$ and for exchange without removing duplicate links as in Line 13 Algorithm \ref{algo-baseline}. We have $LU_u(t) = LU_u(t-1) + \sum\limits_{v \in N(u)} LU_{vu}(t)$, where ``+'' denotes \textit{multiset} semantics. Clearly, $L_u(t) < LU_u(t)$.

\begin{figure}
\centering
\includegraphics[height=1.4in]{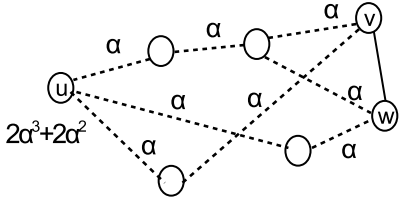}
\setlength{\abovecaptionskip}{-10pt}
\caption{Multipath link propagation}
\vspace{-1.0em}
\label{fig:edge-propagation}
\end{figure}

Note that the number of rounds $T$ can be small because of the following analysis. We have four simple facts (see Fig. \ref{fig:edge-propagation})
\begin{enumerate}
\item a true link $(v,w)$ is propagated to node $u$ at round $t$ if and only if $\min \{d(u,v), d(u,w)\} = t$ for $\alpha = 1$.
\item a fake link $(v\rightarrow w)$ is propagated to node $u$ at round $t$ if and only if $d(u,v) = t$ for $\alpha = 1$.
\item a true link $(v,w)$ is propagated to node $u$ at round $t$ with probability $\sum_{p_l \in P(u,v) \cup P(u,w)}  \alpha^l$ for $\alpha < 1$. Here $p_l$ is a path of length $l$ from $u$ to $v$ or $w$.
\item a fake link $(v\rightarrow w)$ is propagated to node $u$ at round $t$ with probability $\sum_{p_l \in P(u,v)}  \alpha^l$ for $\alpha < 1$.
\end{enumerate}

We consider three cases. 

\textbf{Case 1: $\alpha = 1, \beta = 0$} In this case, there is no fake links. Using Fact 1, we have $|L_u(Diam(G)-1)| = m$, i.e. every node $u$ receives all true links in $G$ after $(Diam(G)-1$ rounds.

\textbf{Case 2: $\alpha = 1, \beta > 0$} In this case, there are $2\beta m$ fake links. Using Facts 1 and 2, we have $|L_u(Diam(G))| = (1+2\beta)m$, i.e. every node $u$ receives all true links and fake links in $G$ after $Diam(G)$ rounds.

\textbf{Case 3: $\alpha < 1, \beta \geq 0$} In this case, there are $2\beta m$ fake links. Using Facts 3 and 4, every node $u$ receives all true links $(v,w)$ in $G$ after $T$ rounds if 
\begin{equation}
\sum_{t=1}^T [(\alpha A)^t]_{vu} + [(\alpha A)^t]_{wu}  \geq 1
\end{equation}
and all fake links $(v\rightarrow w)$ if 
\begin{equation}
\sum_{t=1}^T [(\alpha A)^t]_{vu}  \geq 1
\end{equation}

The protocol's complexity is measured in storage, computation and communication. Because all links are stored in clear form, all complexities increase round by round (except the trivial case $\alpha = 0$). They are also upper bounded by the total links in graph, which is $(1+2\beta)|E_G|$. Intuitively, low-degree nodes will incur lower complexities than high-degree nodes. However, as $t$ increases, the gap gets narrower. In Section \ref{sec:bf}, we will achieve constant complexities by using Bloom filters.

\subsection{Privacy Analysis}
\label{subsec:baseline-priv}
In this section, we discuss the link inference attacks that can be mounted by nodes. Each node has knowledge about the true links connecting itself to its neighbors and the fake links it creates before the first round as well as the fake links pointing to it. The remaining links (denoted as $B_u$) stored at node $u$ are subject to an inference attack by $u$. As discussed in Section \ref{subsec:attack-model}, $u$ may mount an inference attack by sorting links in $B_u$ by weight and picks top-$K$ links as true links.

\begin{figure}
\centering
\includegraphics[height=1.4in]{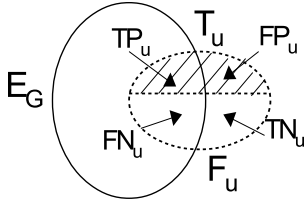}
\setlength{\abovecaptionskip}{-10pt}
\caption{Inference attack measures}
\vspace{-1.0em}
\label{fig:attack-measure}
\end{figure}

In Baseline $(\alpha,\beta)$-exchange, the ratio of true links over fake links is $\frac{1}{\beta}$. Each user, therefore, can set $K = \frac{|B_u|}{1+\beta}$ and divide $B_u$ into two sets $T_u$ (predicted true links) and $F_u$ (predicted fake links). The numbers of true positives, true negatives, false positives and false negatives are (see Fig. \ref{fig:attack-measure} for an illustration)

\begin{align}
TP_u &= |E_G \cap T_u| \;,	FP_u = |T_u \setminus E_G|	\\
FN_u &= |E_G \cap F_u| \;,	TN_u = |F_u \setminus E_G|
\end{align}
The precision, recall and F1 score are defined as $Prec=TP_u/(TP_u + FP_u)$, $Recall=TP_u/(TP_u + FN_u)$ and $F1 = 2*Prec*Recall/(Prec + Recall)$.

% %
\section{Bloom Filter Based Scheme}
\label{sec:bf}
\subsection{Motivation}
\label{subsec:bf-motivation}
Baseline $(\alpha,\beta)$-exchange has several drawbacks that motivate a better approach. First, all link lists are in clear form, allowing nodes to store link frequencies for inference attack (Section \ref{subsec:baseline-priv}). If we obfuscate link lists, this kind of attack may be mitigated. Hashing could be a solution. Second, sending link lists in clear form may incur high communication cost, especially at high degree nodes. Assuming that all node IDs are in range $\{0...2^{32}-1\}$, i.e. each ID needs 4 bytes, each link is encoded in 8 bytes. Given a link list, a better way to encode it is to store all links $(u,v_i)$ incident to $u$ by $\{u| \{v_i\}\}$. In this way, the message length for a link list can be reduced up to 50\%. In average, each link costs between 32-bit to 64-bit. Using Bloom filters, the number of bits per link may be reduced. For example, with $k = 4$, the number of bits per link is $k/\ln 2 \approx 5.8$. 

This section introduces a Bloom filter based approach. Compared to Baseline approach, it has several advantages and limitations. Bloom filters, by encoding links in compact forms, reduce the storage and communication costs. The computation at each node is also much simpler thanks to logical OR operation compared to set unions in Baseline. %Computation over Bloom filters is similar to homomorphic encryption \cite{homomorphic-enc}. However, Bloom filter approach needs an extra recovery step.

\subsection{Bloom Filter Based Scheme}
Algorithm \ref{algo-bf} describes steps of Bloom filter version of $(\alpha,\beta)$-exchange. As for inputs, we add a global false positive probability $p$ and the number of links $|E_G|$. As analyzed in \cite{broder2004network}, the number of hash functions $k$ is set to $\lceil -\log_2{p}\rceil$ (Line 2). The number of bits per link is $c = k/\ln 2$ (Line 3). The length of every Bloom filter is $m = c.|E_G|$ (Line 4). Then, each node $u$ initializes its Bloom filter $Bf_u(0)$ by hashing all links in $L_u(0)$ using $k$ hash functions. At the same time, all nodes send their noisy links $L_u(0)$ to the coordinator who will gather all links into the list $L$. This list will be used in the recovery stage.

In the exchange stage, each pair of nodes $(u,v)$ prepare and exchange noisy link lists in encoded form $Bf_{uv}(t)$ and $Bf_{vu}(t)$ (Lines 14-18). Before the next round, each node aggregates all Bloom filters sent to it by taking the OR operation. (Lines 19 and 20). Finally, the recovery stage helps each node to obtain its noisy local view $L_u(T)$. In this stage, the coordinator sends to $L$ to all nodes. If we omit the role of the coordinator (Lines 5,11 and 23), each node $u$ has to try hash $\frac{N(N-1)}{2}$ possible links against its final Bloom filter $Bf_u(T)$.

\begin{algorithm}
\caption{Bloom filter $(\alpha,\beta)$-exchange}
\label{algo-bf}
\begin{algorithmic}[1]
	\Require undirected graph $G=(V,E)$, parameters $\alpha \in [0,1]$, $\beta \geq 0$, number of rounds $T$, false positive probability $p$
	\Ensure noisy local views of graph $L_u(T), u \in V$ 
	\State // initialization stage
	\State $k = \lceil -\log_2{p}\rceil$		(see equation (\ref{eqn:k}))
	\State $c = k/\ln 2$	 
	\State $m = c.|E_G|$
	\State $L = \emptyset$
	\For {$u \in V$}
		\State $Bf_u(0) = \text{BloomFilter(k,m,c)}$
		\State $Fa(u) = \{(u \rightarrow w) | w \notin N(u) \}$ s.t. $|Fa(u)| = \beta |N(u)|$
		\State $L_u(0) = \{(u,v)| v \in N(u)\} \cup Fa(u)$
		\State Hash all $e \in L_u(0)$ into $Bf_u(0)$
		\State $L = L \cup L_u(0)$
	\EndFor
	\State // exchange stage
	\For {$t = 1..T$}
		\For {$(u,v) \in E$}
			\State $u$ prepares $Bf_{uv}(t) = \text{BitErasure}(Bf_u(t-1), \alpha)$
			\State $v$ prepares $Bf_{vu}(t) = \text{BitErasure}(Bf_v(t-1), \alpha)$	
			\State $u$ sends $Bf_{uv}(t)$ to $v$
			\State $v$ sends $Bf_{vu}(t)$ to $u$	
		\EndFor
		\For {$u \in V$}
			\State $Bf_u(t) = Bf_u(t-1) \vee \bigvee\limits_{v \in N(u)} Bf_{vu}(t)$
		\EndFor
	\EndFor
	\State // link recovery stage
	\For {$u \in V$}
		\State $L_u(T) = \text{Hash}(L, Bf_u(T))$
	\EndFor
	\Return $L_u(T), u \in V$ 
\end{algorithmic}
\end{algorithm}

\subsubsection{Bit Erasure}
Because Bloom filters store links information in encoded form, we have to simulate the $\alpha$-sampling steps (Lines 8 and 9, Algorithm \ref{algo-baseline}). 

$\alpha$-sampling is equivalent to ``deletion'' of $(1-\alpha)|Bf_u(t-1)|$ elements from $Bf_u(t-1)$. We can perform this operation by recovering elements in $Bf_u(t-1)$ then explicitly keeping $\alpha$-fraction of elements and hashing these elements to an empty Bloom filter. This approach, however, is costly because the node must try $\frac{N(N-1)}{2}$ possible links. As a result, an implicit removal of $(1-\alpha)$-fraction of elements is needed.

Resetting one bit causes one or several misses (false negatives) and possibly reduces false positives. For example, resetting the second bit in Bloom filter (Fig. \ref{fig:bloom-filter}) makes $x$ a false negative whereas resetting the 12th-bit makes both $y$ and $z$ disappear. Moreover, if the 8-th bit is reset, $x$ becomes a false negative and $w$ is no longer a false positive. 

Let $m_1$ be the number of 1-bits in Bloom filter $Bf_u(t-1)$ and $s$ be the number of randomly reset bits ($s < m_1$), the probability of a true positive remaining in Bloom filter is
\begin{equation}
(1-\frac{s}{m_1})^k		\label{eqn:remove-bit}
\end{equation}
If omitting the effect of false positives (which is reduced as illustrated above), the formula (\ref{eqn:remove-bit}) is exactly the sampling fraction $\alpha$. In other words,
\begin{equation}
\alpha = (1-\frac{s}{m_1})^k  \Rightarrow s = m_1 (1-\alpha^{1/k})
\end{equation}

\begin{figure}
\centering
\includegraphics[height=1.5in]{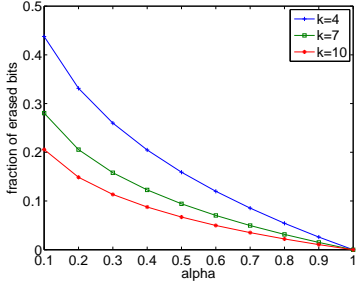}
\caption{Fraction of erased bits as a function of $\alpha$ and $k$}
\vspace{-1.0em}
\label{fig:bit-erasure}
\end{figure}

We can see that $s$ is a decreasing function of $\alpha$ and $k$. An illustration of this fact is shown in Fig. \ref{fig:bit-erasure}.

Algorithm \ref{algo-bit-erasure} realizes $\alpha$-sampling implicitly via bit erasure.

\begin{algorithm}
\caption{Bit Erasure}
\label{algo-bit-erasure}
\begin{algorithmic}[1]
	\Require Bloom filter $B$, parameter $\alpha \in [0,1]$, number of hashes $k$
	\Ensure Bloom filter $B'$ that contains approximately $\alpha$ fraction of elements in $B$
	\State $B' = B$
	\State $M_1 = \{i | B(i) = 1\}$
	\State $m_1 = |M_1|$
	\State $s = \lfloor m_1 (1-\alpha^{1/k} \rfloor$
	\State randomly reset $s$ bits in $m_1$ positions of $B'$ \\
	\Return $B'$ 
\end{algorithmic}
\end{algorithm}

\subsubsection{Bloom Filter Compression} 
\label{subsec:bf-compress}
In Algorithm \ref{algo-bf}, all Bloom filters stored at nodes and transmitted between nodes are of length $m$ bits where $m = |E_G|k/\ln 2$. For $p = 0.1$, we have $k = 4$ and $m \approx 5.8 |E_G|$. For $p = 0.01$, we have $k = 7$ and $m \approx 10.1|E_G|$. For million-scale graphs with hundreds of millions of links, the length of Bloom filters would be hundreds of megabytes. This is undesirable for message transmission although storage and computation are not big problems. However, we observe that as in Baseline $(\alpha,\beta)$-exchange, not all messages have the length of $\Theta(E_G)$. Thus, lossless data compression is a useful tool for Bloom filter exchange.

Arithmetic coding \cite{moffat1998arithmetic} is such a lossless compression scheme. Arithmetic coding differs from other forms of entropy encoding, such as Huffman coding \cite{huffman1952method}. Huffman coding separates the input into component symbols with symbol probabilities approximated by negative powers of two and replaces each with a code. Arithmetic coding encodes the entire message into a single number, a fraction $f$ where $0.0 \leq f < 1.0$.

\subsection{Complexity and Privacy Analysis}
Thanks to constant sizes of bit arrays and constant time for OR operations, the total communication cost of Bloom Filter scheme is constant and the aggregation of noisy link lists is constant too. However, Bloom Filter scheme incurs an extra recovery step at all nodes. Each node needs to download the full noisy link set $L$ from the coordinator. As we confirm in Section \ref{subsec:eval-bf}, the exchange time of Bloom Filter scheme is much lower than that of Baseline, but the recovery step costs higher time complexity.

As mentioned in Section \ref{subsec:bf-motivation}, all link lists are obfuscated in Bloom filters, frequency-based inference attacks may be mitigated if the set of all links $L$ is revealed to all nodes only after the final round. The ratio of true links over fake links in Bloom Filter scheme is almost identical to that of Baseline. The reason lies in the independence of all links in exchange protocols. All links have the same probability to be sampled and sent to neighbors of nodes. Interestingly, Bloom Filter helps reduce the true/fake link ratio faster than Baseline for small $\alpha$ (Section \ref{subsec:eval-vol-inference}) thanks to its inherent false positives as well as false negatives caused by bit erasure.

\section{Evaluation}
\label{sec:eval}
In this section, we empirically evaluate the performance of our proposed schemes on synthetic graphs. All algorithms are implemented in Java and run on a desktop PC with $Intel^{\circledR}$ Core i7-4770@ 3.4Ghz, 16GB memory. 

Two kinds of synthetic graphs are generated: Barab\'{a}si-Albert power-law (PL) graphs and Erd\"{o}s-R\'{e}nyi (ER) random graphs \cite{newman2003structure}. Table \ref{tab:dataset} lists six synthetic graphs used in our experiments. Each test case is run 10 times. We abbreviate the two schemes Baseline (BS) and BloomFilter-based (BF).

We choose $\alpha \in \{0.25, 0.5, 0.75, 1.0\}$ and $\beta \in \{0.5, 1.0\}$. The default number of hash functions $k$ is 4. 

\begin{table}[htb]
\small
\centering
\caption{Synthetic graphs} \label{tab:dataset}
\begin{tabular}{|c|r|r|r|}
\hline
\textbf{Graph} &\textbf{\#Nodes} & \textbf{\#Links} & \textbf{Diameter}\\
\hline
 PL1	&	10,000	& 	29,990 & 7	\\
\hline
 \textbf{PL2}	&	\textbf{10,000}	& 	\textbf{49,970} & \textbf{6}	\\
\hline
 PL3	&	10,000	& 	99,872 & 5	\\
\hline
 ER1	&	10,000	& 	30,076 & 10	\\
\hline
 \textbf{ER2}	&	\textbf{10,000}	& 	\textbf{50,424} & \textbf{7}	\\
\hline
 ER3	&	10,000	& 	99,615 & 5	\\     
\hline
\end{tabular}
\end{table}

\subsection{Message Volume and Inference Attacks}
\label{subsec:eval-vol-inference}
\begin{figure*}
 	\centering
         \begin{subfigure}[b]{0.22\textwidth}
                 \centering
                 \epsfig{file=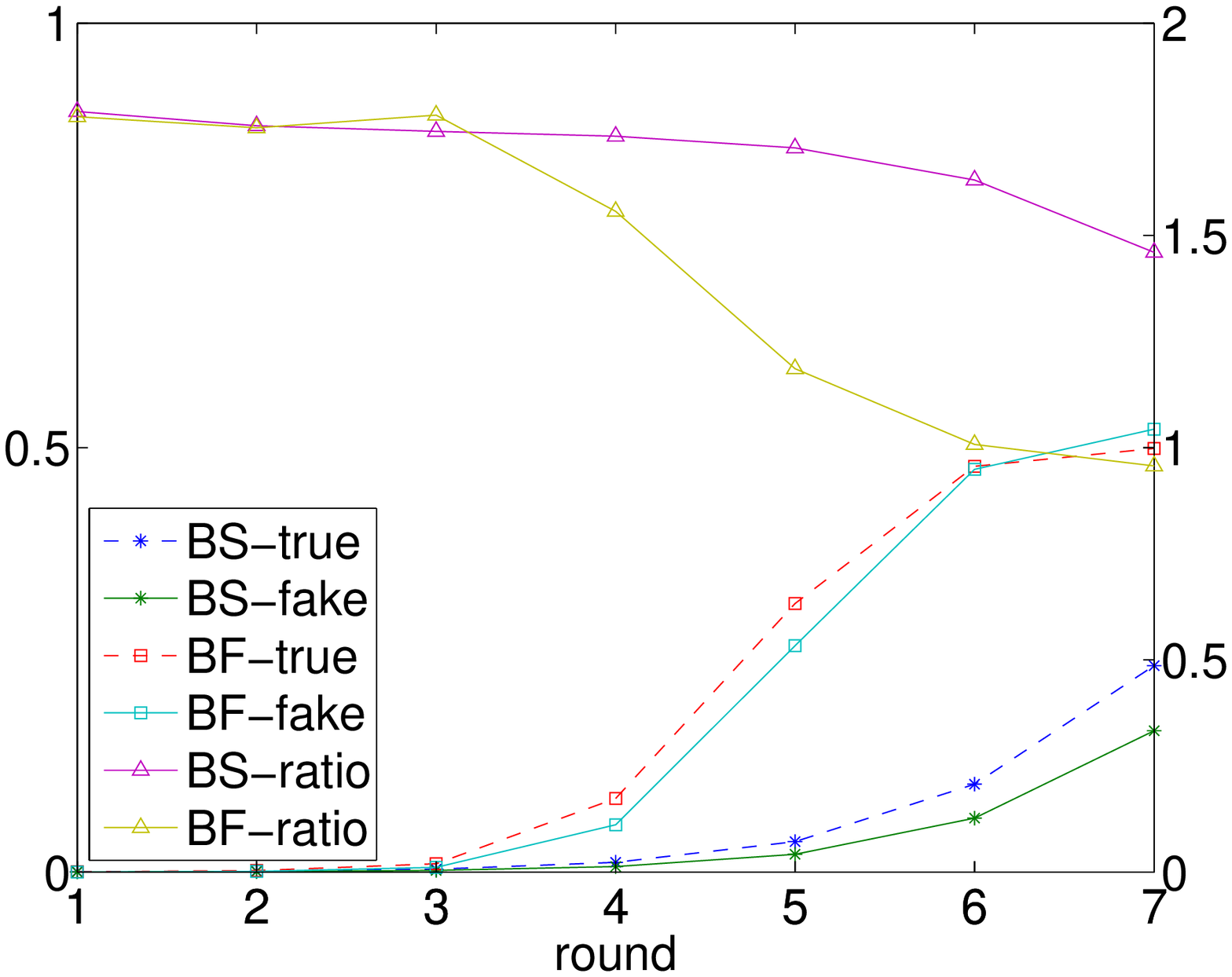, height=1.2in}
                 \setlength{\abovecaptionskip}{0pt}
                 \caption{$\alpha=0.25, \beta=0.5$}	
                 \label{fig:er2-1-1}
         \end{subfigure}
         \hfill
         \begin{subfigure}[b]{0.22\textwidth}
                 \centering
                 \epsfig{file=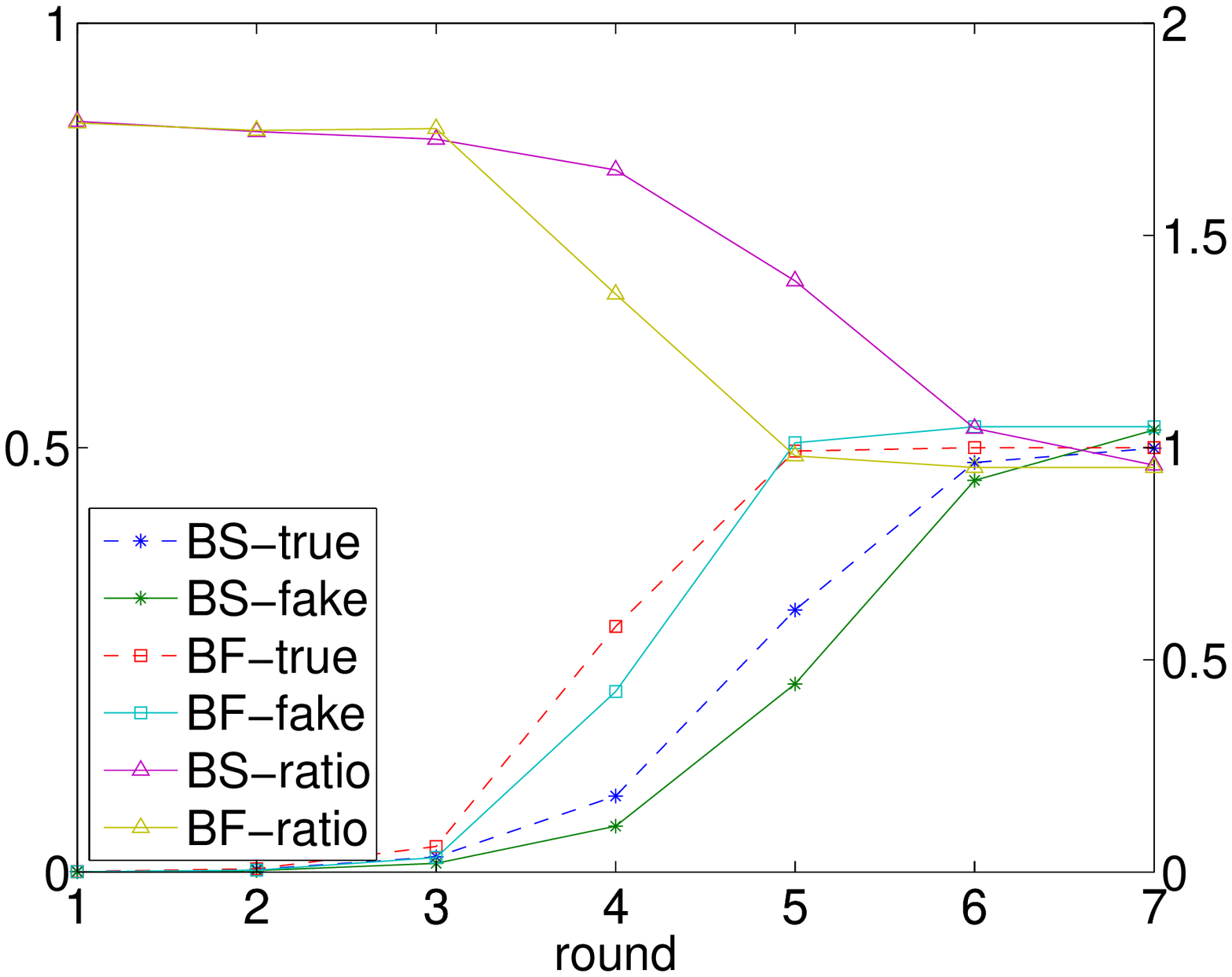, height=1.2in}
                 \setlength{\abovecaptionskip}{0pt}
                 \caption{$\alpha=0.5, \beta=0.5$}
                 \label{fig:er2-2-1}
         \end{subfigure}
         \hfill         
         \begin{subfigure}[b]{0.22\textwidth}
                 \centering
                 \epsfig{file=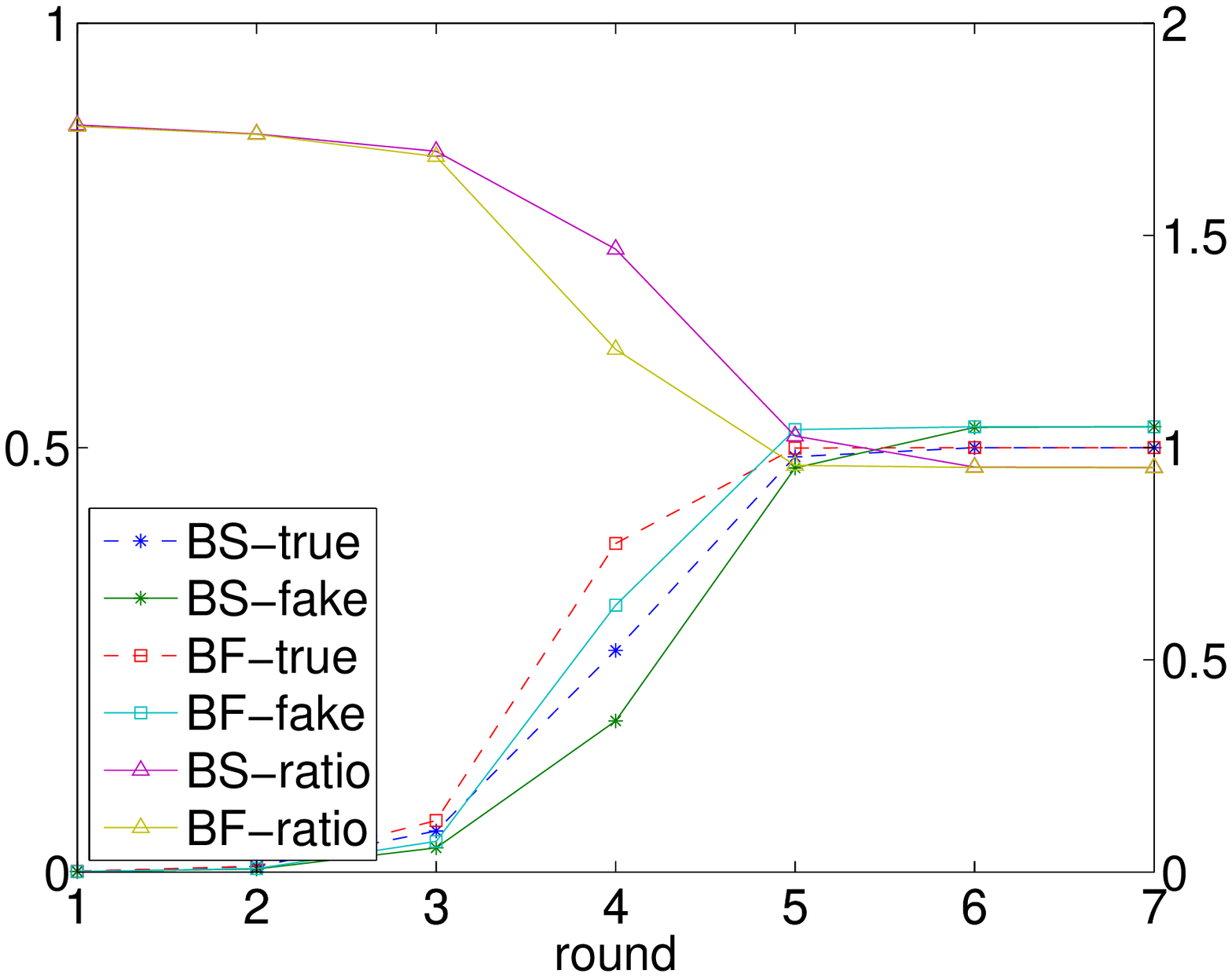, height=1.2in}
                 \setlength{\abovecaptionskip}{0pt}
                 \caption{$\alpha=0.75, \beta=0.5$}	
                 \label{fig:er2-3-1}
         \end{subfigure}
         \hfill         
         \begin{subfigure}[b]{0.22\textwidth}
                  \centering
                 \epsfig{file=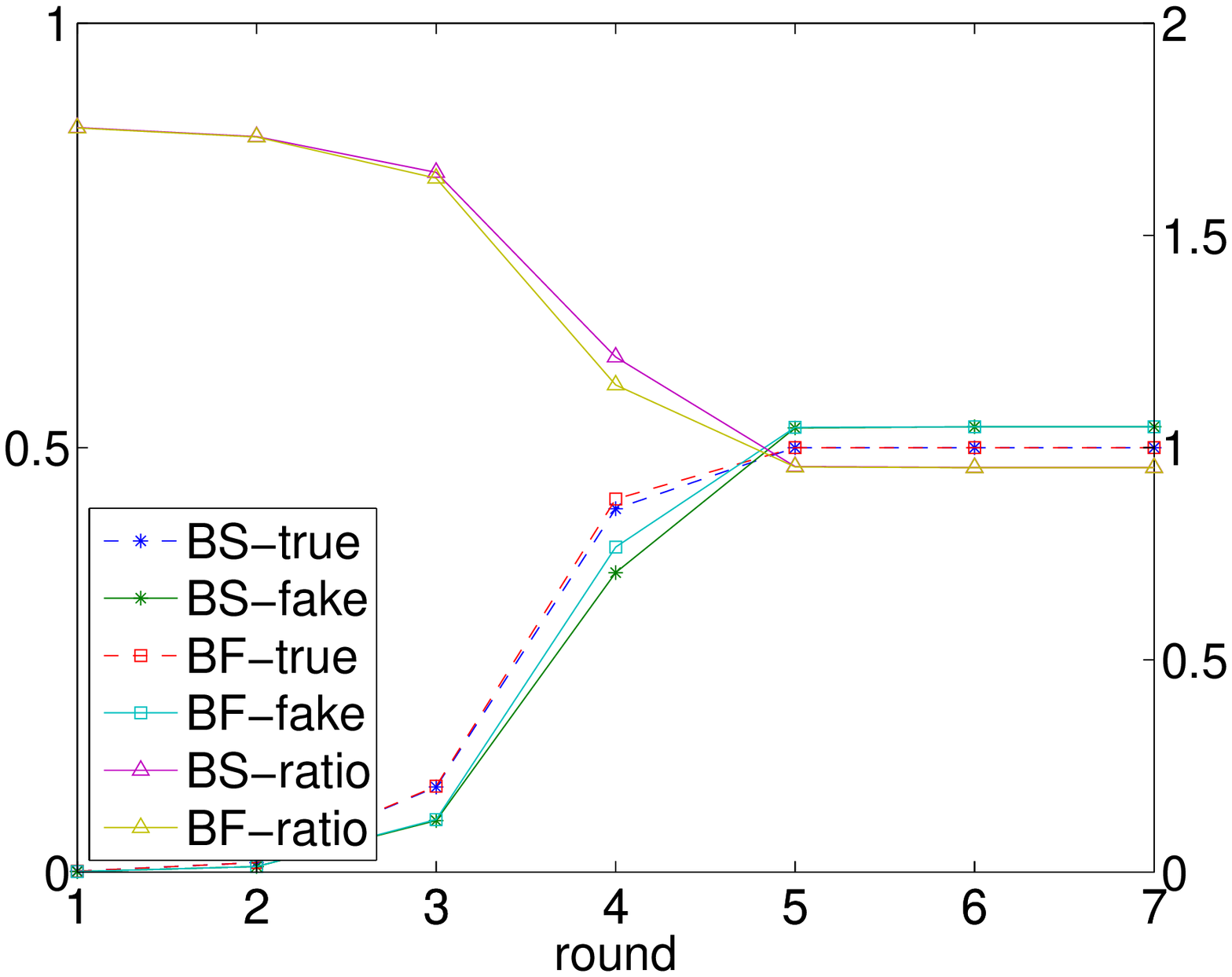, height=1.2in}
                 \setlength{\abovecaptionskip}{0pt}
                 \caption{$\alpha=1.0, \beta=0.5$}
                 \label{fig:er2-4-1}
         \end{subfigure}
         
         \begin{subfigure}[b]{0.22\textwidth}
                 \centering         
                 \epsfig{file=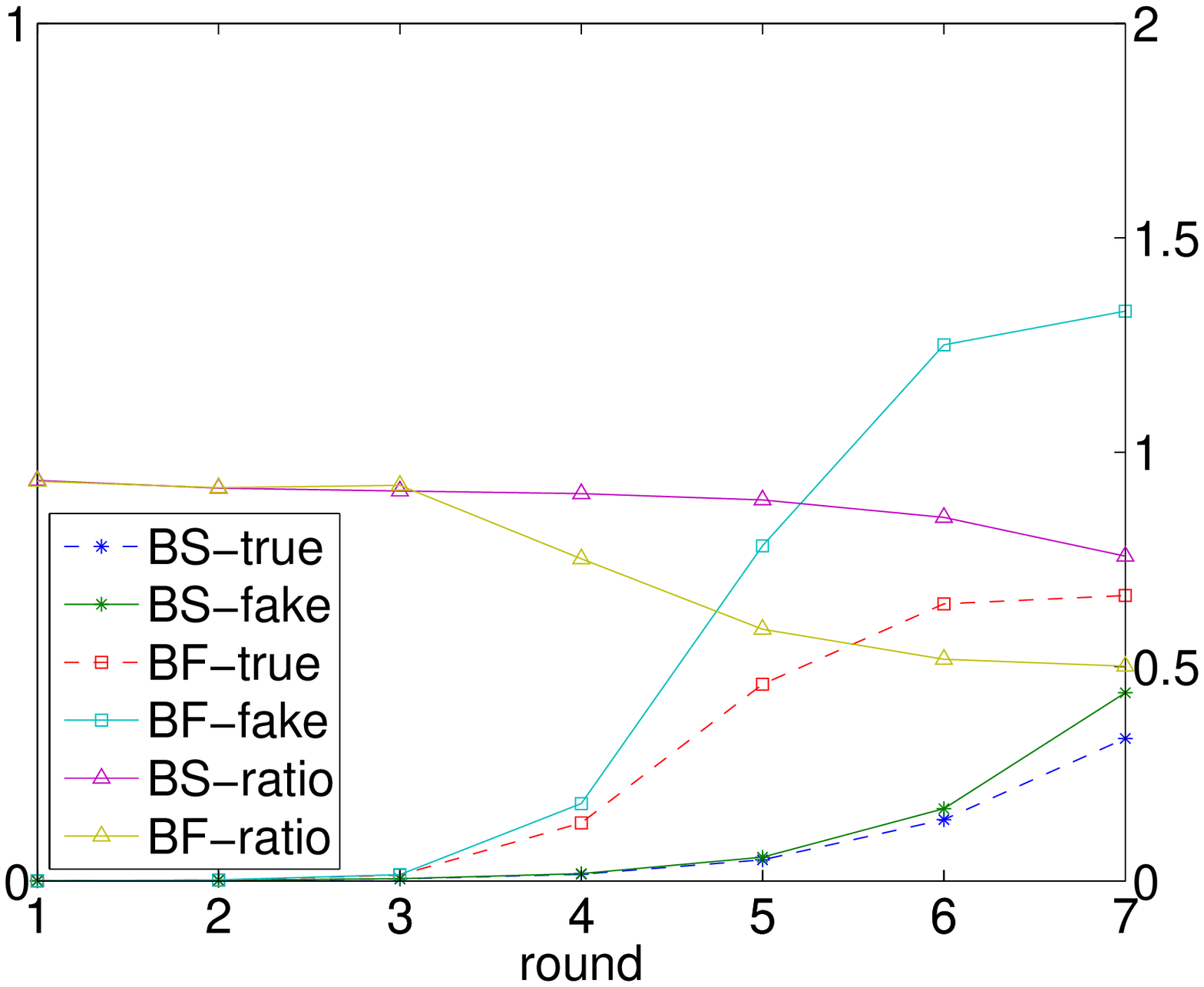, height=1.2in}
                 \setlength{\abovecaptionskip}{0pt}
                 \caption{$\alpha=0.25, \beta=1.0$}
                 \label{fig:er2-1-2}
         \end{subfigure}
         \hfill
         \begin{subfigure}[b]{0.22\textwidth}
                 \centering
                 \epsfig{file=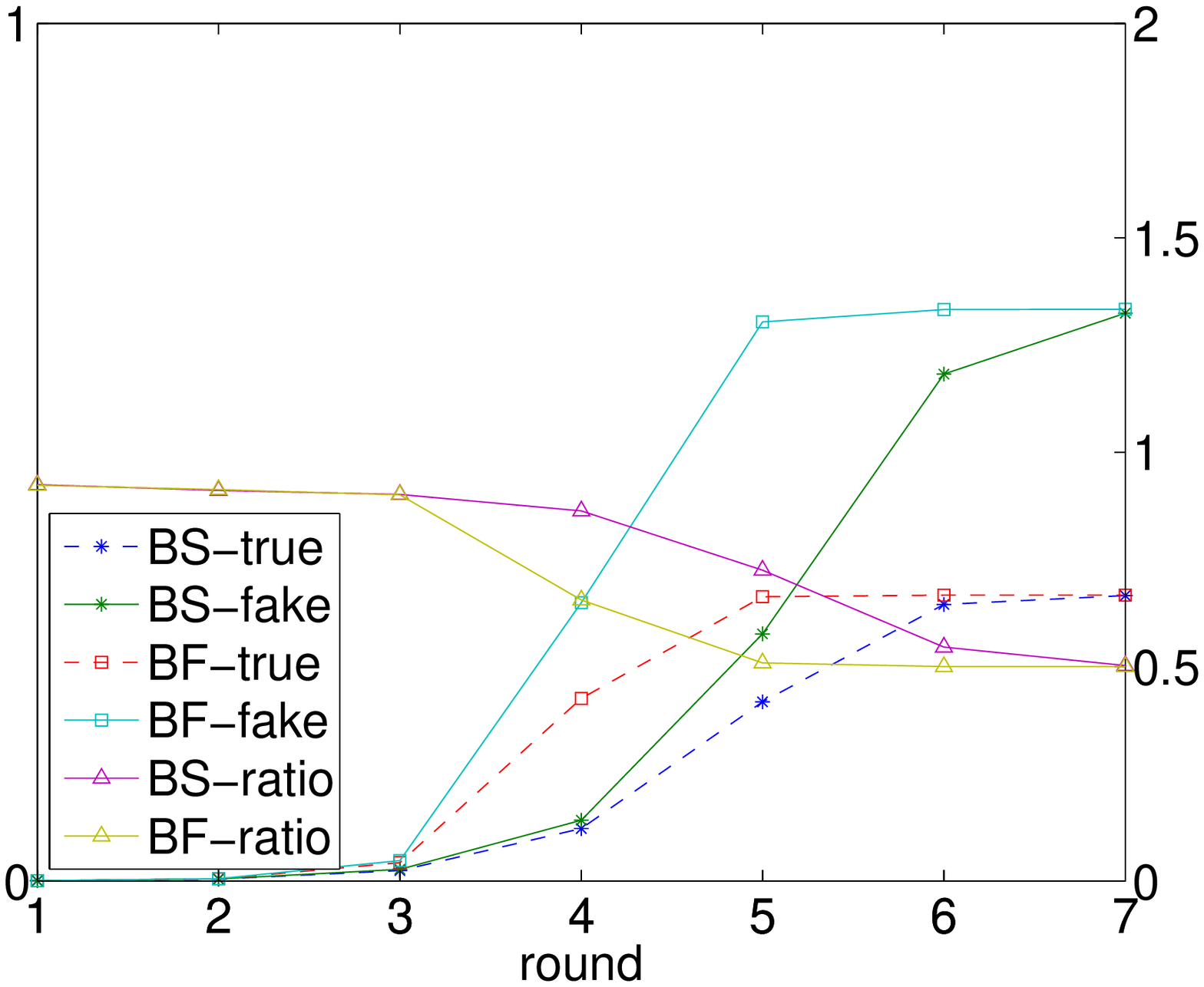, height=1.2in}
				 \setlength{\abovecaptionskip}{0pt}
				 \caption{$\alpha=0.5, \beta=1.0$}
				 \label{fig:er2-2-2}
         \end{subfigure}
         \hfill         
         \begin{subfigure}[b]{0.22\textwidth}
                 \centering                  
                 \epsfig{file=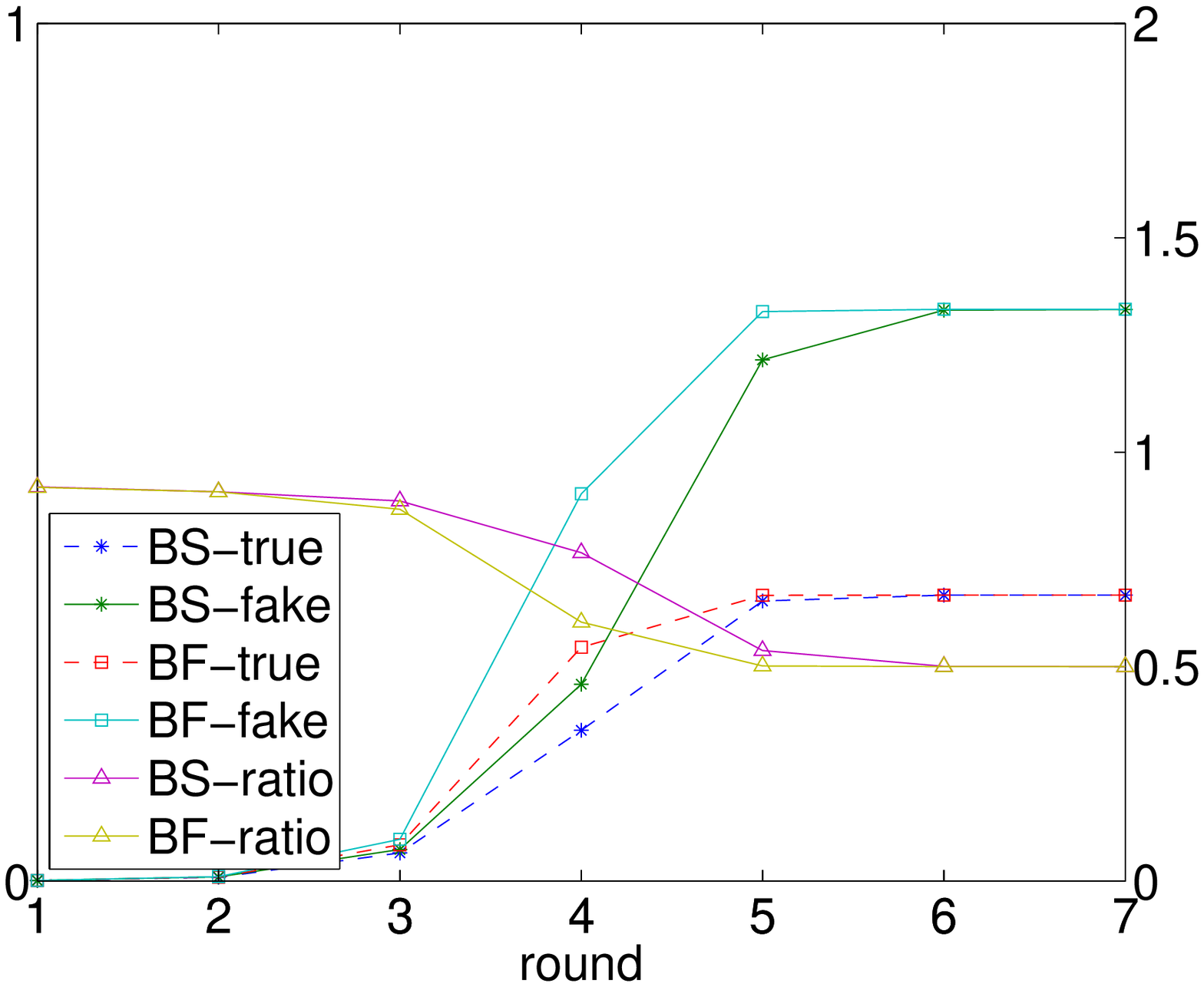, height=1.2in}
                 \setlength{\abovecaptionskip}{0pt}
                 \caption{$\alpha=0.75, \beta=1.0$}
                 \label{fig:er2-3-2}
         \end{subfigure}
         \hfill         
         \begin{subfigure}[b]{0.22\textwidth}
                  \centering
                  \epsfig{file=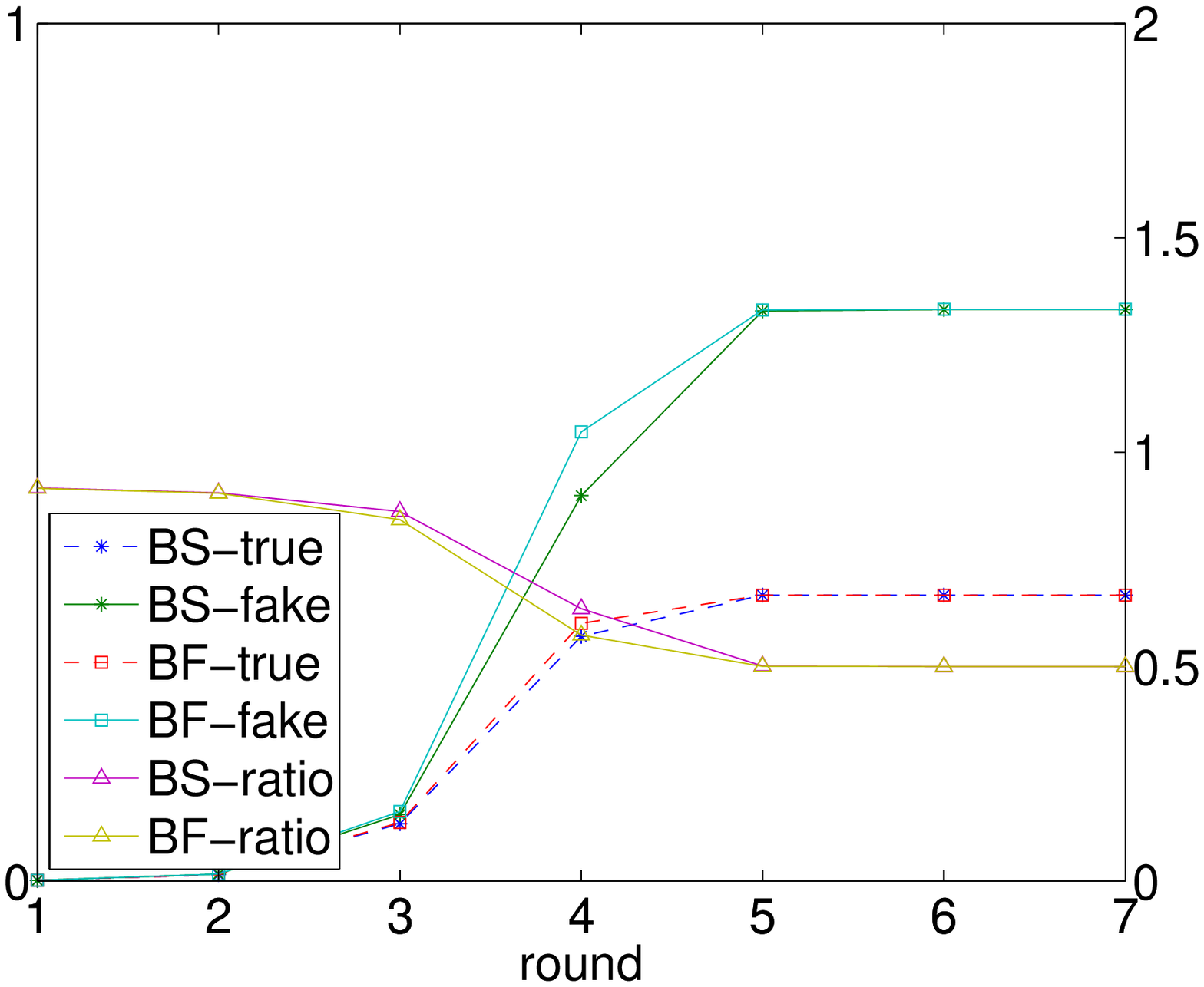, height=1.2in}
                  \setlength{\abovecaptionskip}{0pt}
                 \caption{$\alpha=1.0, \beta=1.0$}
                  \label{fig:er2-4-2}
         \end{subfigure}        	
     \caption{Normalized number of true/fake links and link ratios on ER2}
     \label{fig:er2}
 \end{figure*}

\begin{figure*}
 	\centering
         \begin{subfigure}[b]{0.22\textwidth}
                 \centering
                 \epsfig{file=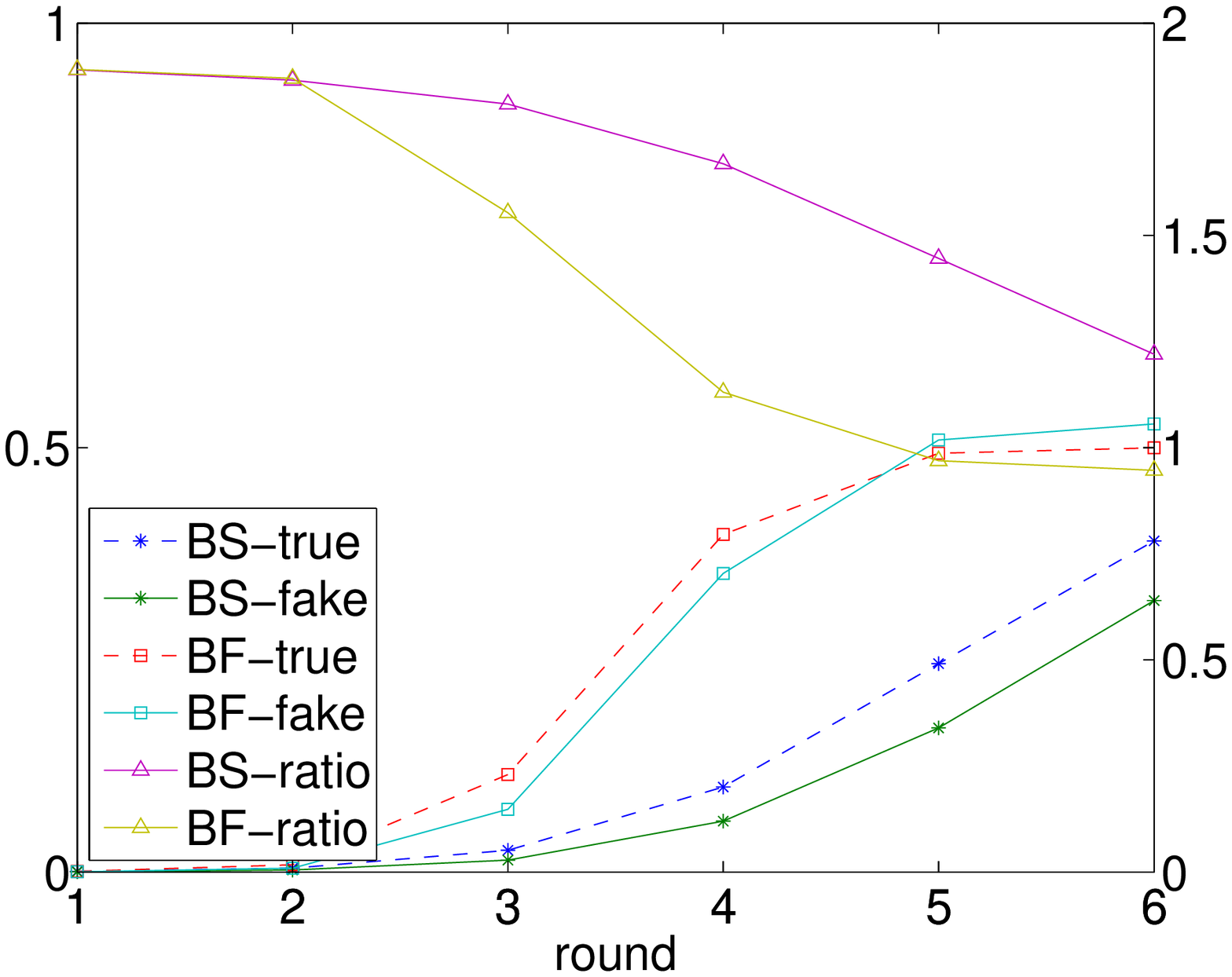, height=1.2in}
                 \setlength{\abovecaptionskip}{0pt}
                 \caption{$\alpha=0.25, \beta=0.5$}	
                 \label{fig:pl2-1-1}
         \end{subfigure}
         \hfill
         \begin{subfigure}[b]{0.22\textwidth}
                 \centering
                 \epsfig{file=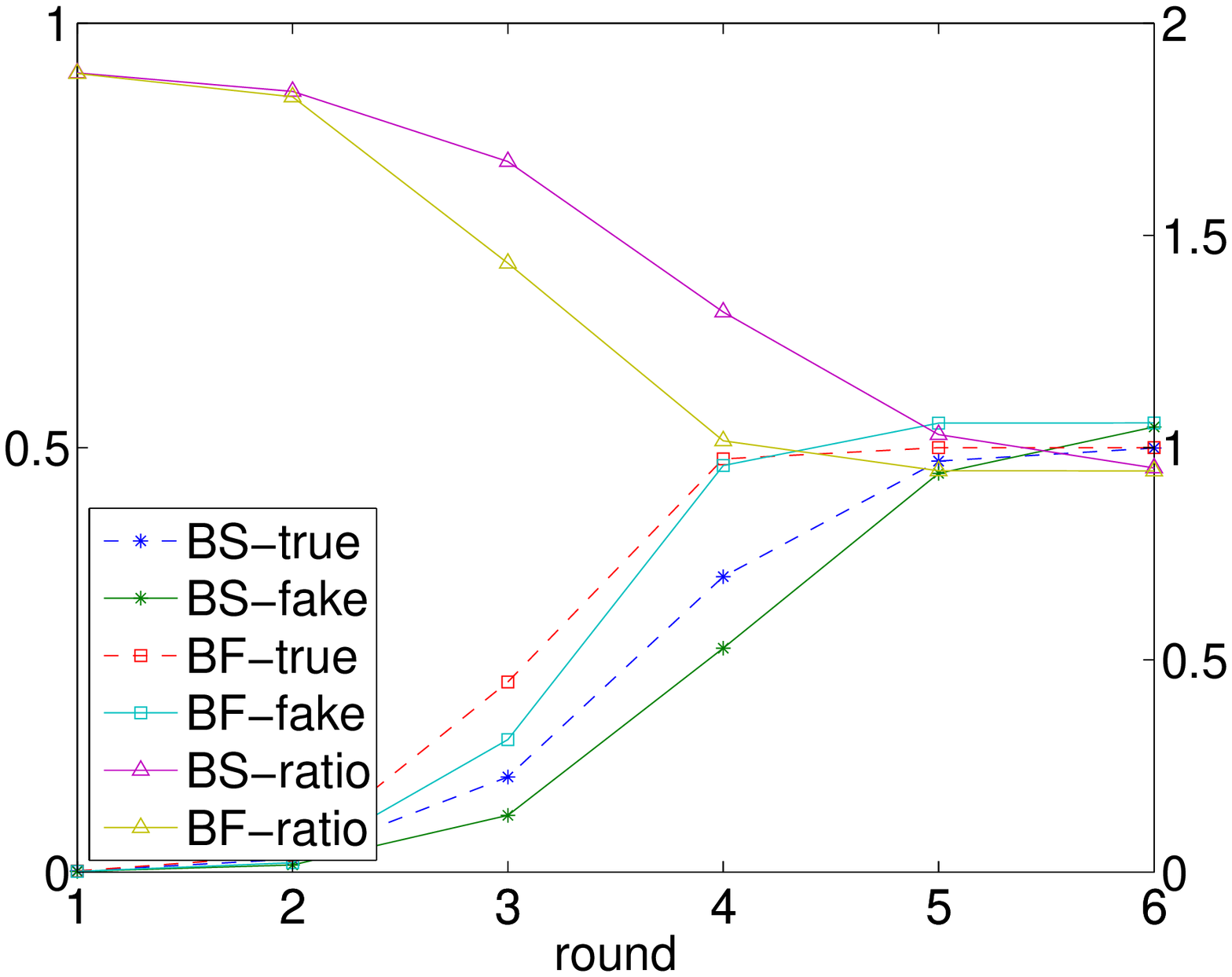, height=1.2in}
                 \setlength{\abovecaptionskip}{0pt}
                 \caption{$\alpha=0.5, \beta=0.5$}
                 \label{fig:pl2-2-1}
         \end{subfigure}
         \hfill         
         \begin{subfigure}[b]{0.22\textwidth}
                 \centering
                 \epsfig{file=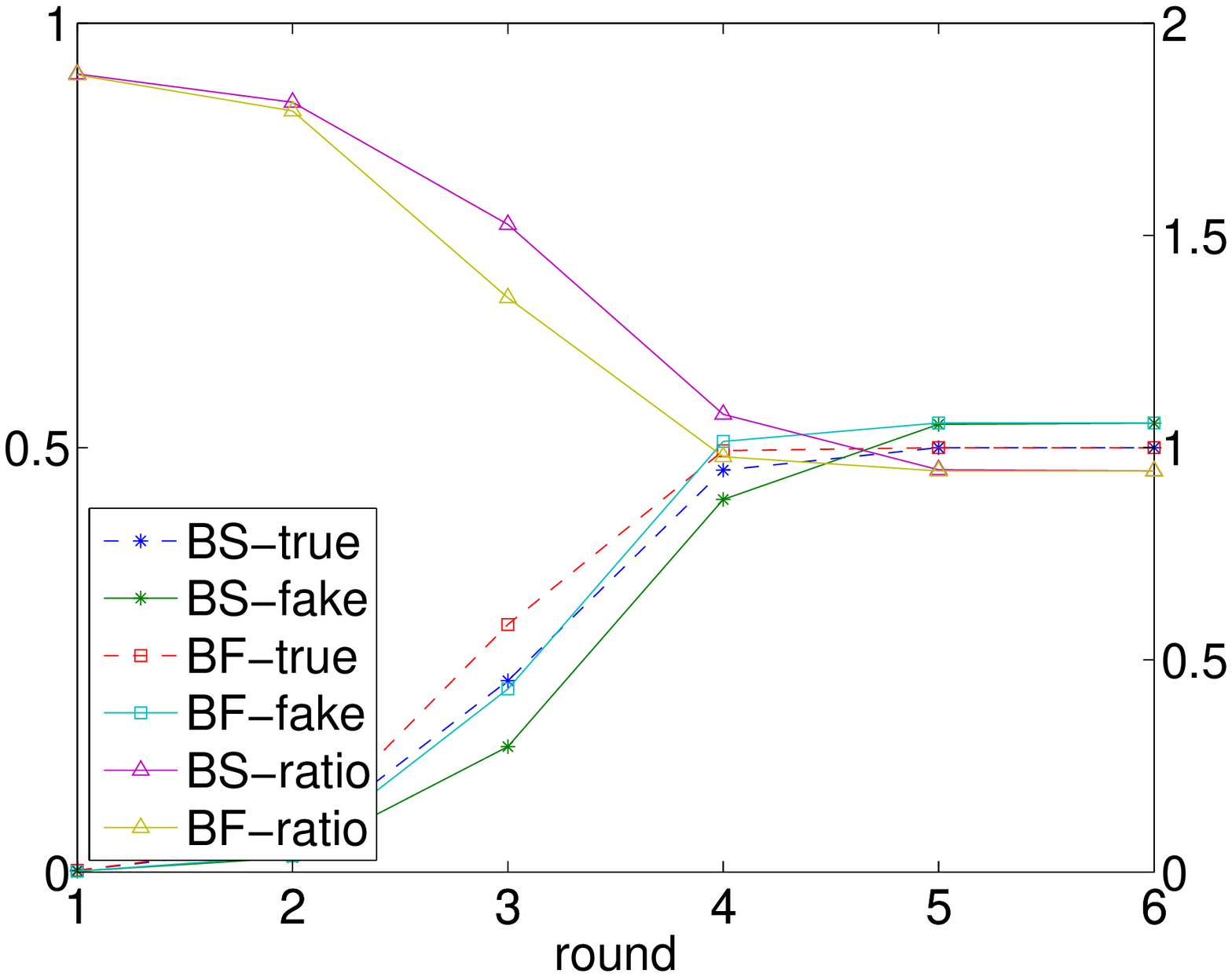, height=1.2in}
                 \setlength{\abovecaptionskip}{0pt}
                 \caption{$\alpha=0.75, \beta=0.5$}	
                 \label{fig:pl2-3-1}
         \end{subfigure}
         \hfill         
         \begin{subfigure}[b]{0.22\textwidth}
                  \centering
                 \epsfig{file=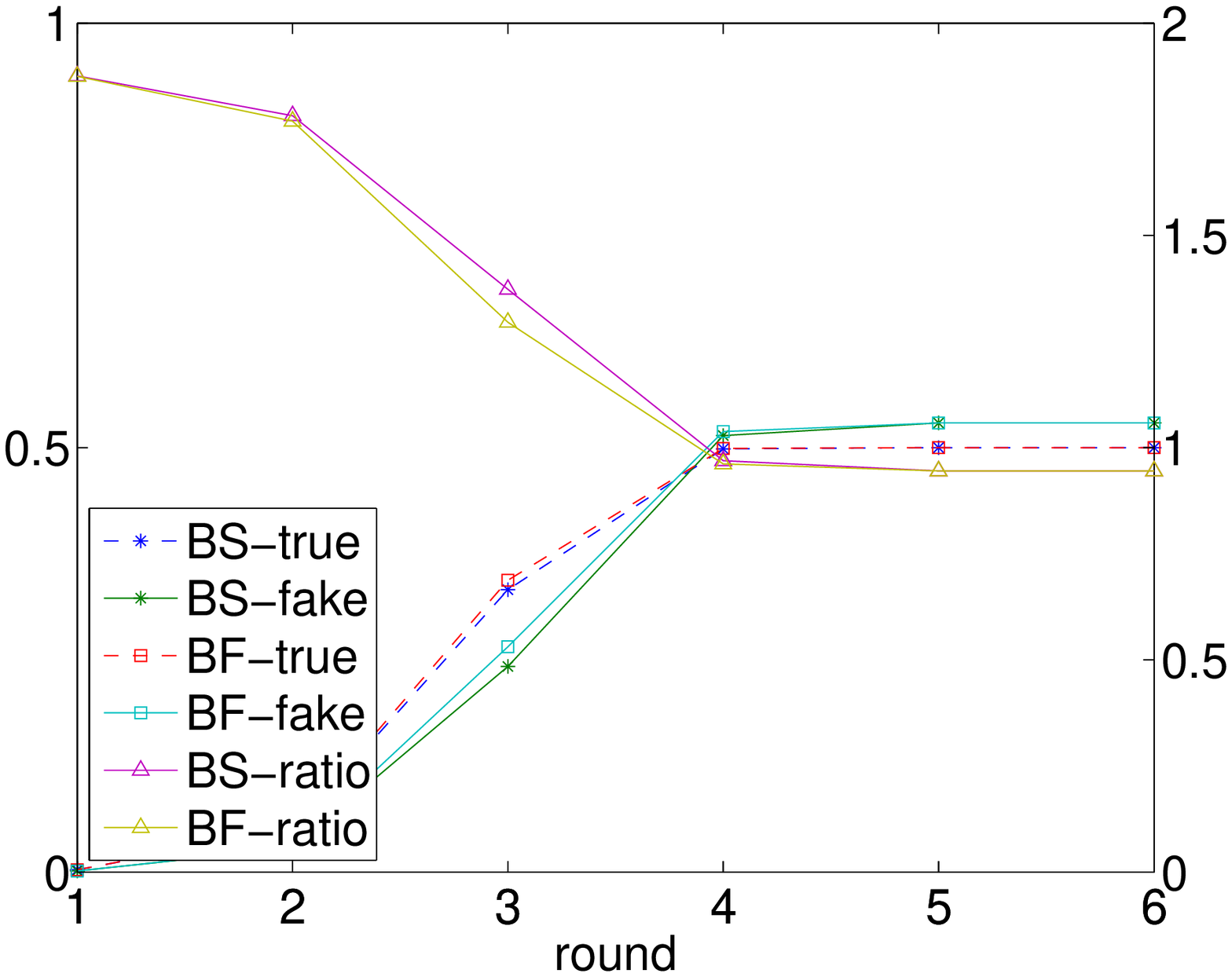, height=1.2in}
                 \setlength{\abovecaptionskip}{0pt}
                 \caption{$\alpha=1.0, \beta=0.5$}
                 \label{fig:pl2-4-1}
         \end{subfigure}
         
         \begin{subfigure}[b]{0.22\textwidth}
                 \centering
                 \epsfig{file=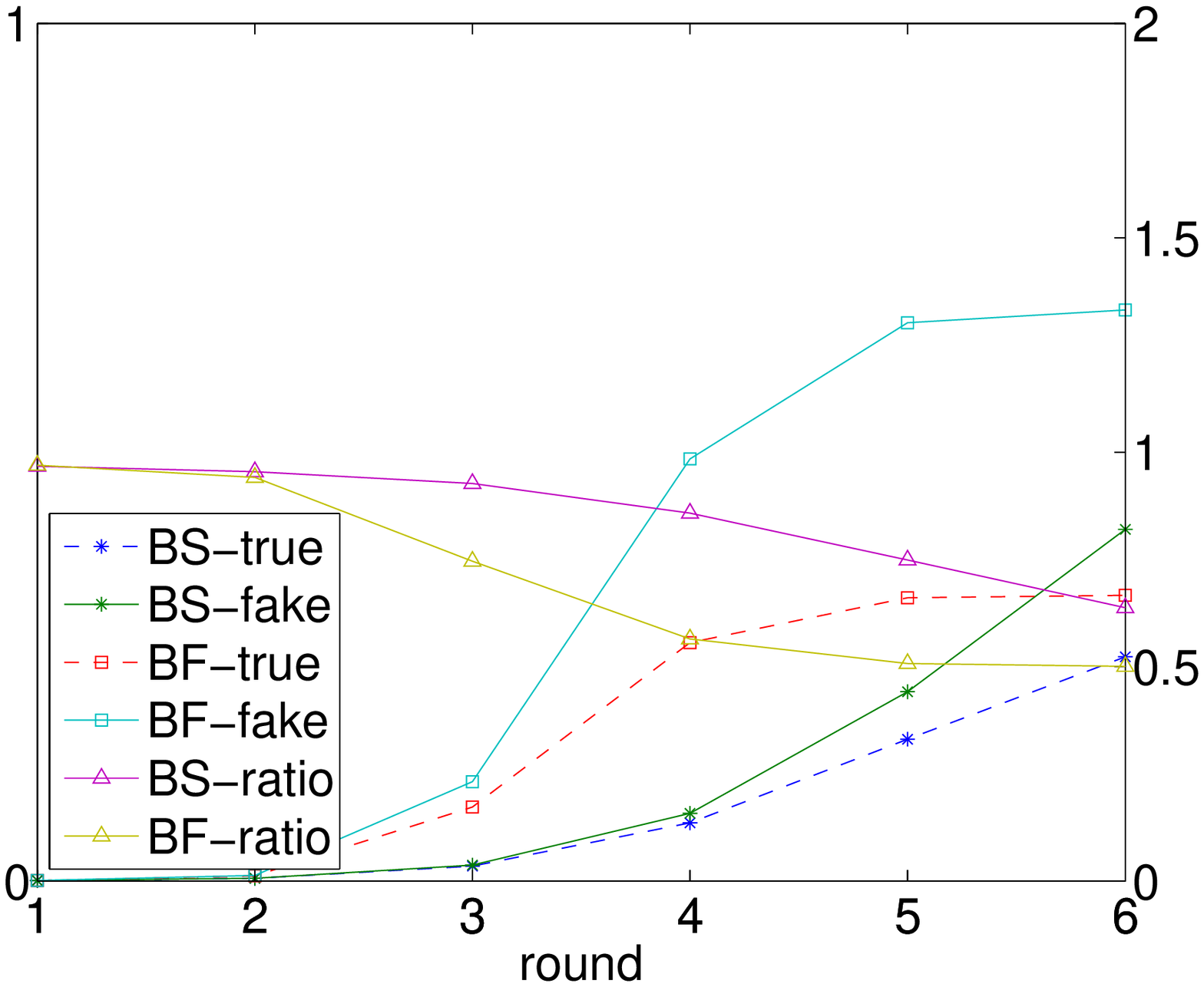, height=1.2in}
                 \setlength{\abovecaptionskip}{0pt}
                 \caption{$\alpha=0.25, \beta=1.0$}
                 \label{fig:pl2-1-2}
         \end{subfigure}
         \hfill
         \begin{subfigure}[b]{0.22\textwidth}
                 \centering
                  \epsfig{file=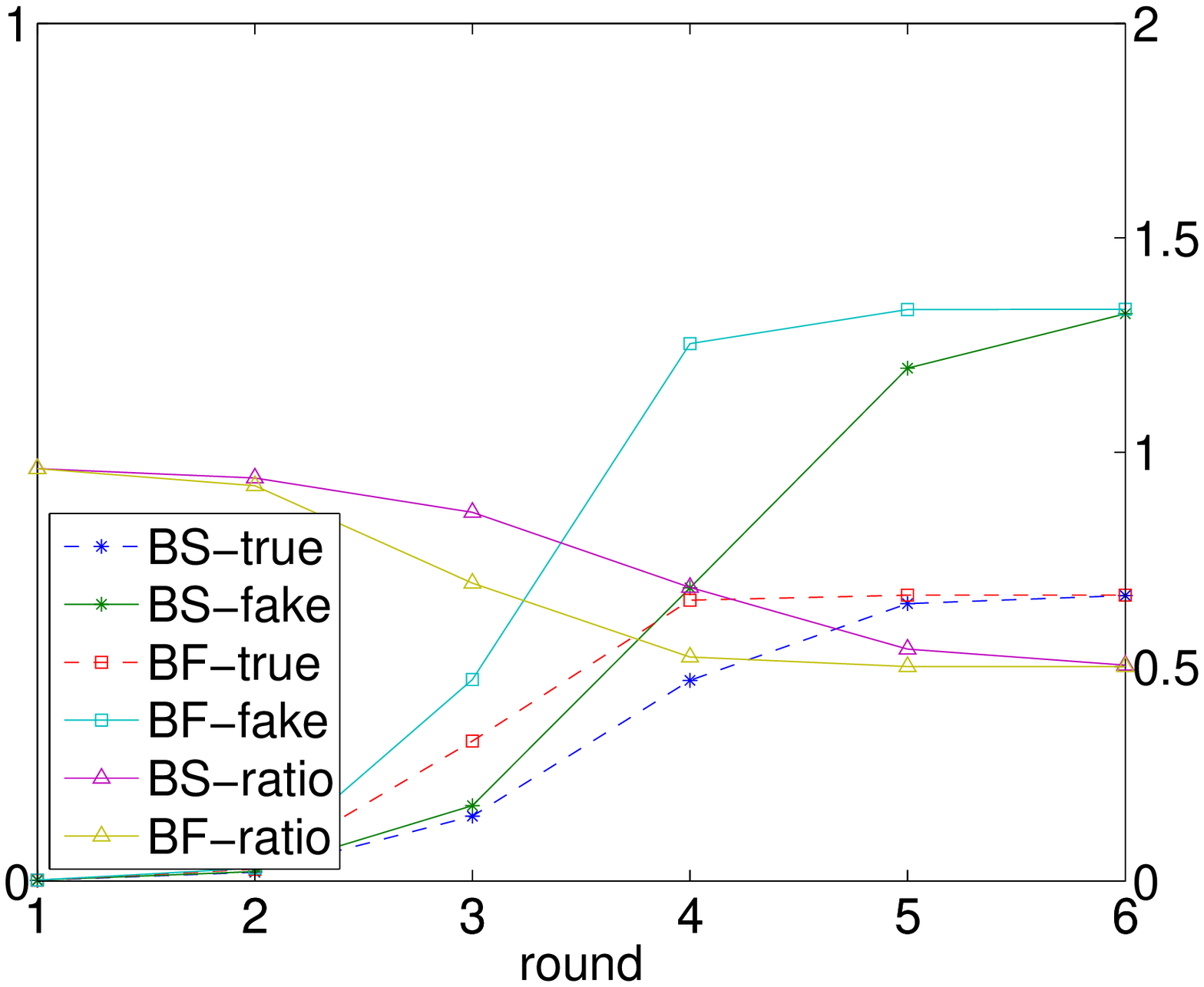, height=1.2in}
                  \setlength{\abovecaptionskip}{0pt}
                 \caption{$\alpha=0.5, \beta=1.0$}
                  \label{fig:pl2-2-2}
         \end{subfigure}
         \hfill         
         \begin{subfigure}[b]{0.22\textwidth}
                 \centering
                 \epsfig{file=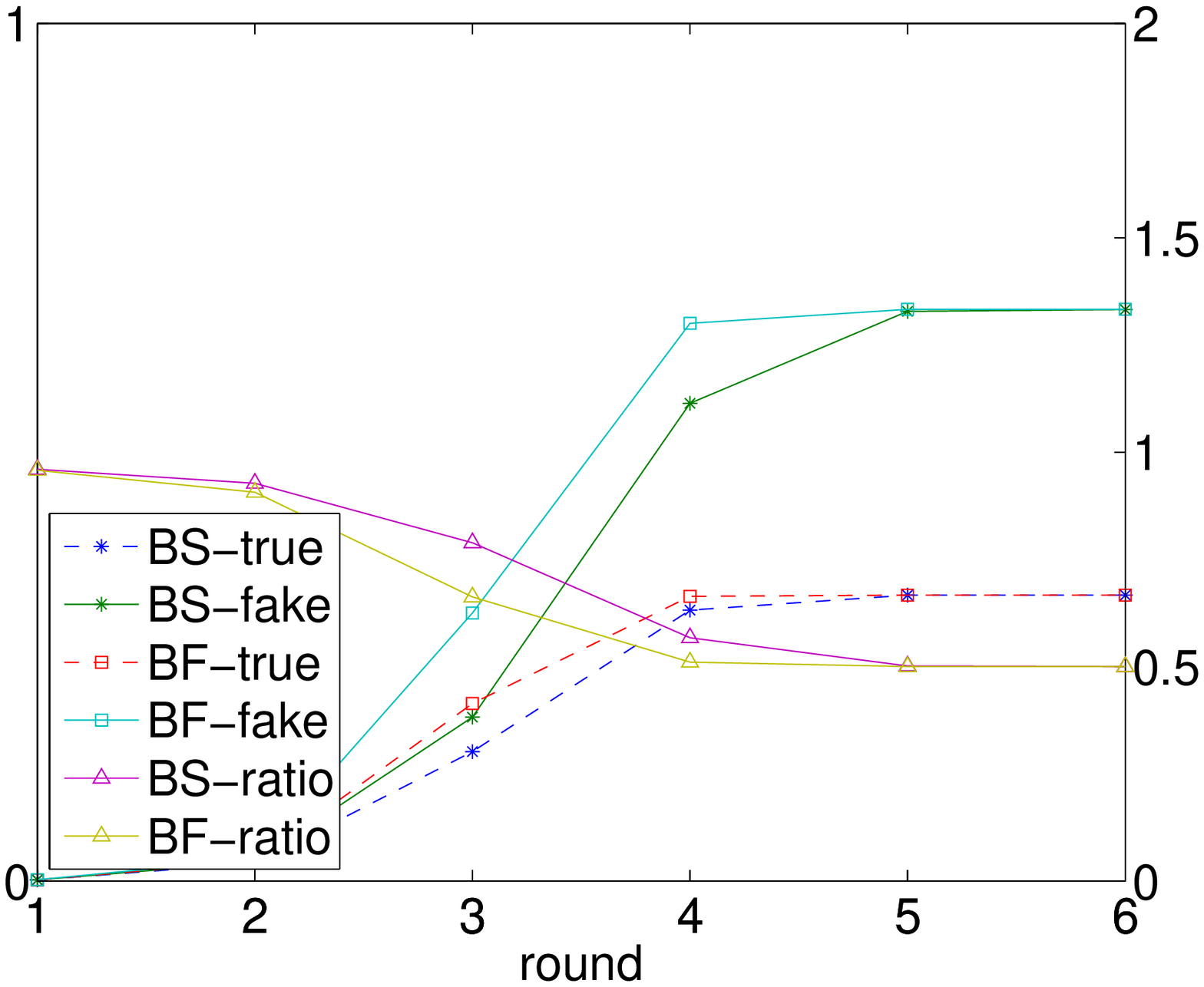, height=1.2in}
                 \setlength{\abovecaptionskip}{0pt}
                 \caption{$\alpha=0.75, \beta=1.0$}
                 \label{fig:pl2-3-2}
         \end{subfigure}
         \hfill         
         \begin{subfigure}[b]{0.22\textwidth}
                  \centering
                  \epsfig{file=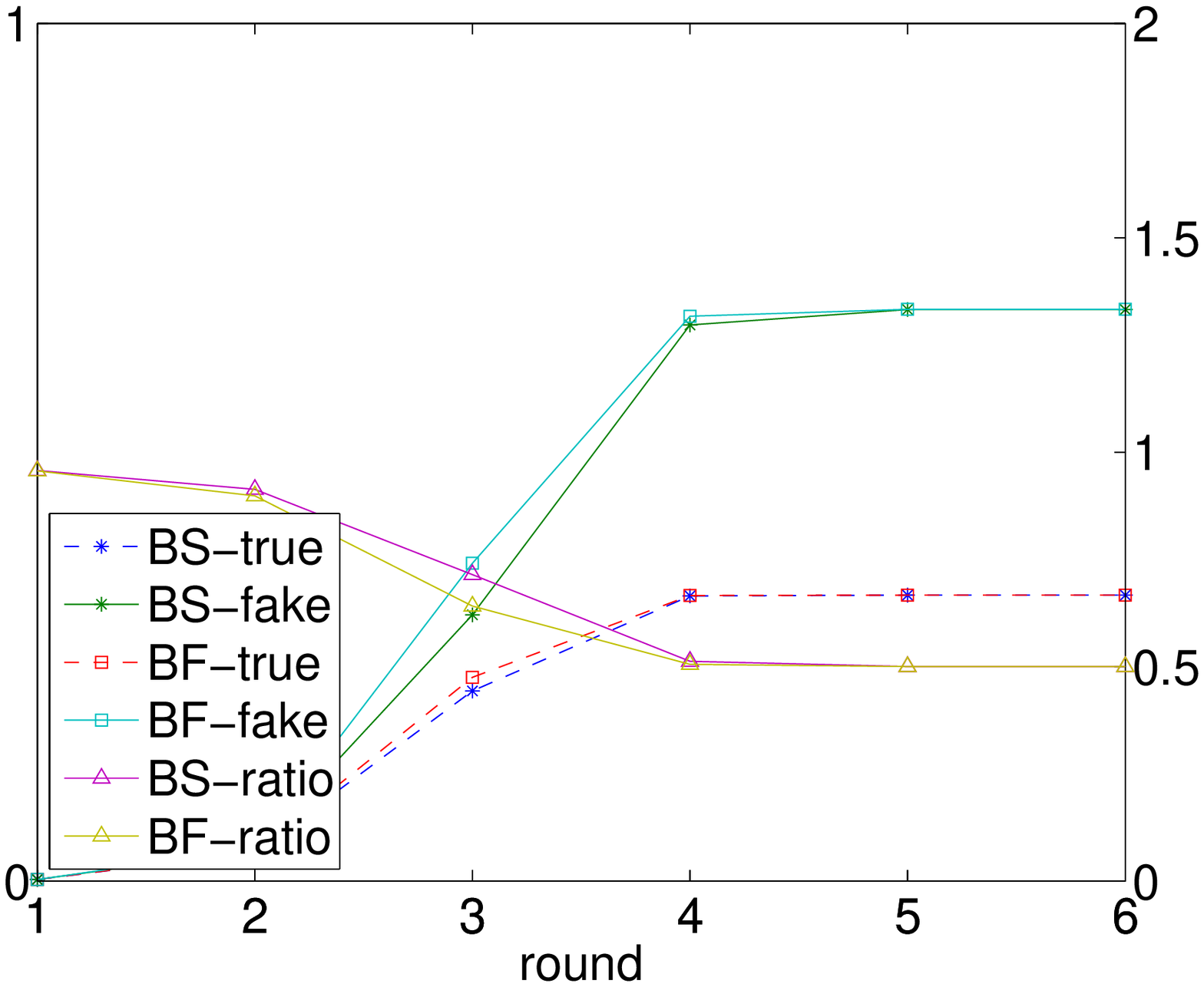, height=1.2in}
                  \setlength{\abovecaptionskip}{0pt}
                 \caption{$\alpha=1.0, \beta=1.0$}
                  \label{fig:pl2-4-2}
         \end{subfigure}        	
     \caption{Normalized number of true/fake links and link ratios on PL2}
     \label{fig:pl2}
 \end{figure*}

We investigate the message volume by the total number of true/fake links at all nodes after each round $t=1..Diam(G)$. These values are normalized by dividing them by $N.M.(1+2\beta)$. We also estimate the inference attacks by the ratio between the number of true links and the number of fake links. Fig. \ref{fig:er2} and Fig. \ref{fig:pl2} show two-y-axis charts. The left y-axis is for the normalized number of links. The right y-axis is for the ratios.

Several observations can be made clearly from Figures \ref{fig:er2} and \ref{fig:pl2}. First, the number of true/fake links increases exponentially and converges fast as all nodes reach the round at $Diam(G)$. For $\alpha = 0.25$, Baseline does not converge because not all links are propagated to all nodes. Bloom filter scheme produces higher number of true/fake links, especially at $\alpha=0.25, 0.5$. For larger values of $\alpha$, the two schemes almost coincide. Second, the ratio of true links over fake links decreases round by round and converges to $\frac{1}{2\beta}$. In early rounds, the ratios are lower than $\frac{1}{\beta}$. Higher the ratio, higher inference risk of true links. Clearly, Bloom Filter scheme reduces the risk better than Baseline for $\alpha=0.25, 0.5$ in later rounds.

\begin{figure*}
 	\centering
         \begin{subfigure}[b]{0.22\textwidth}
                 \centering
                 \epsfig{file=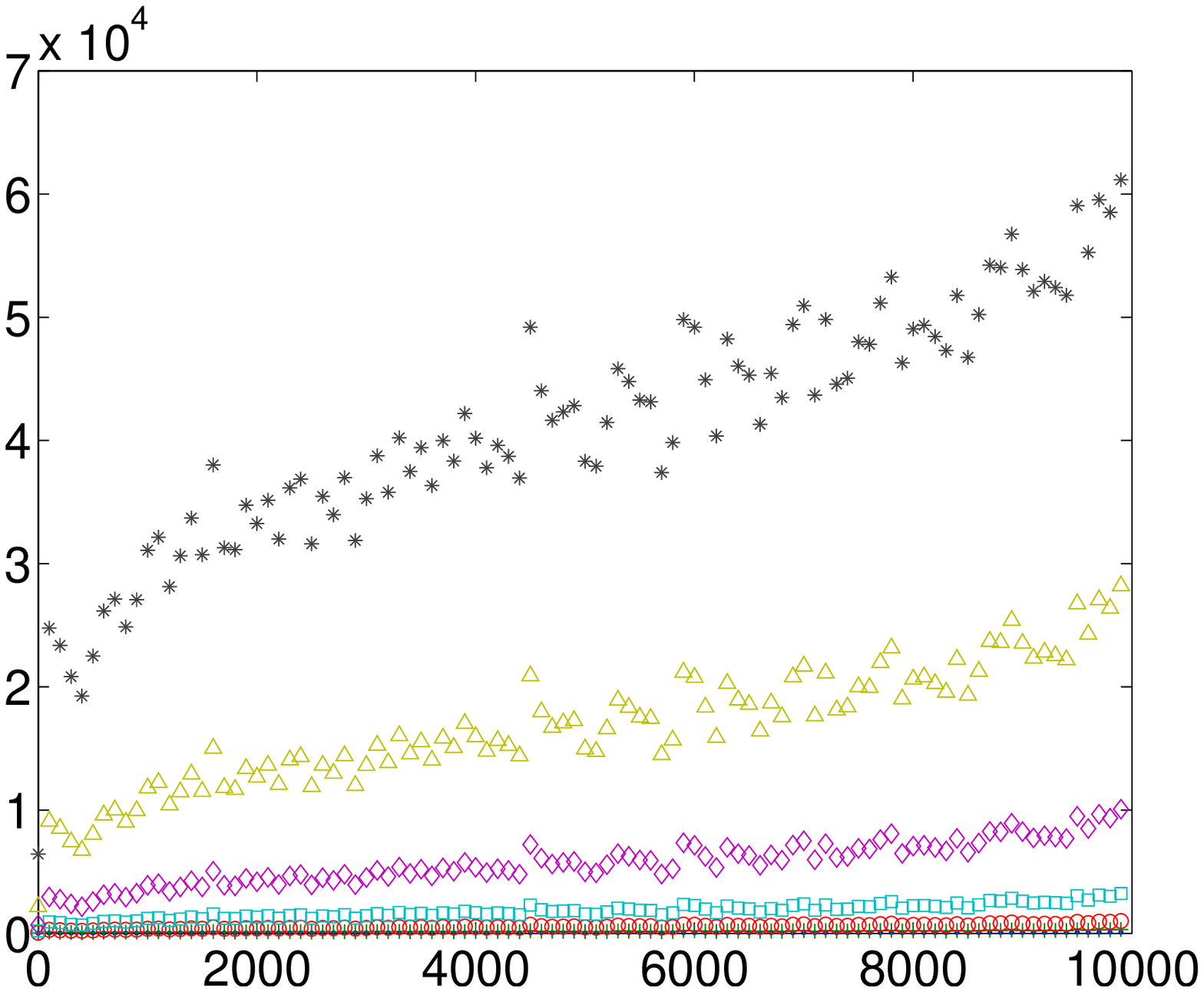, height=1.2in}
                 \setlength{\abovecaptionskip}{0pt}
                 \caption{$\alpha=0.25, \beta=0.5$, ER2}	
                 \label{fig:deg-er-1-1}
         \end{subfigure}
         \hfill
         \begin{subfigure}[b]{0.22\textwidth}
                 \centering
                 \epsfig{file=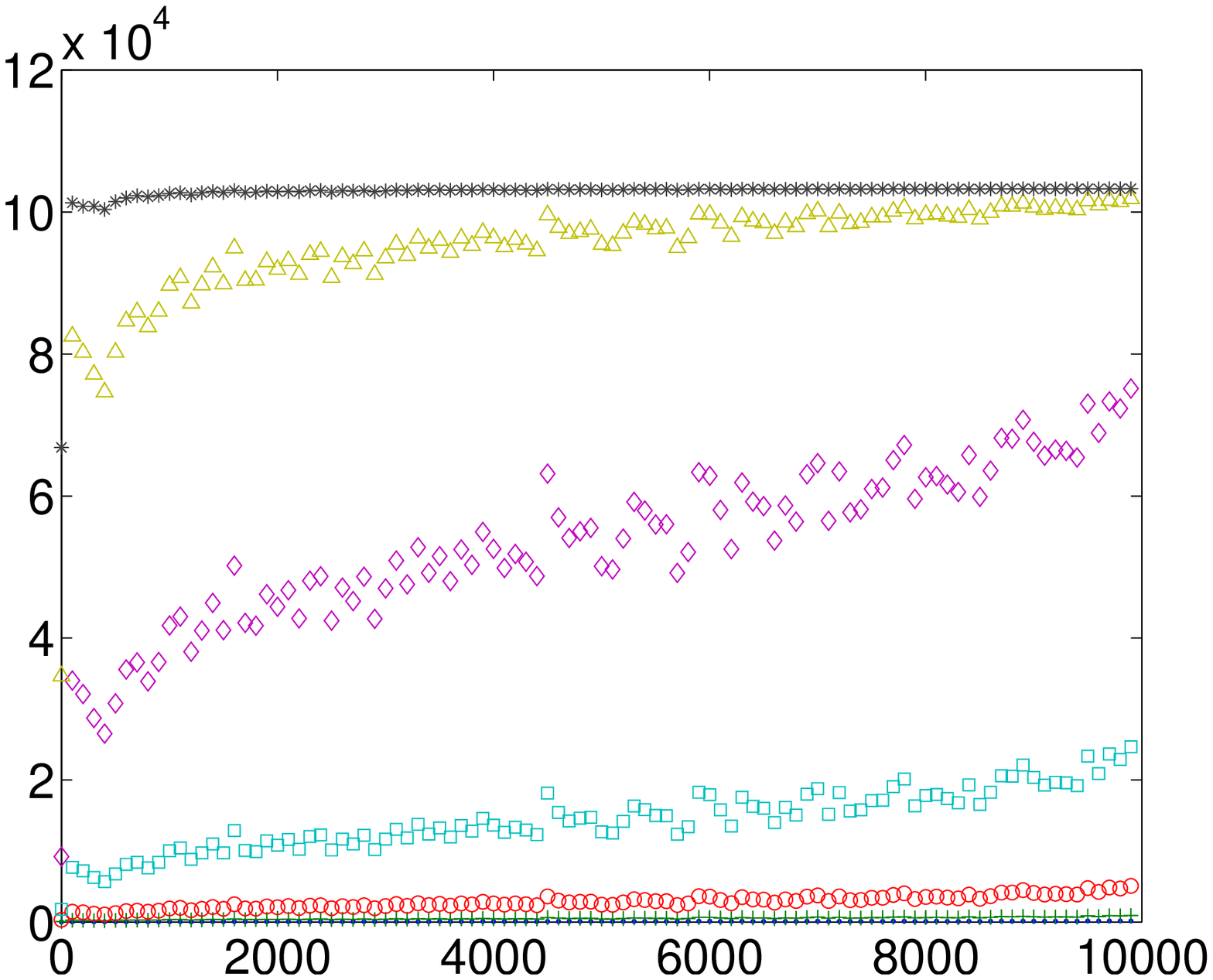, height=1.2in}
                 \setlength{\abovecaptionskip}{0pt}
                 \caption{$\alpha=0.5, \beta=0.5$, ER2}
                 \label{fig:deg-er-1-2}
         \end{subfigure}
         \hfill         
         \begin{subfigure}[b]{0.22\textwidth}
                 \centering
                 \epsfig{file=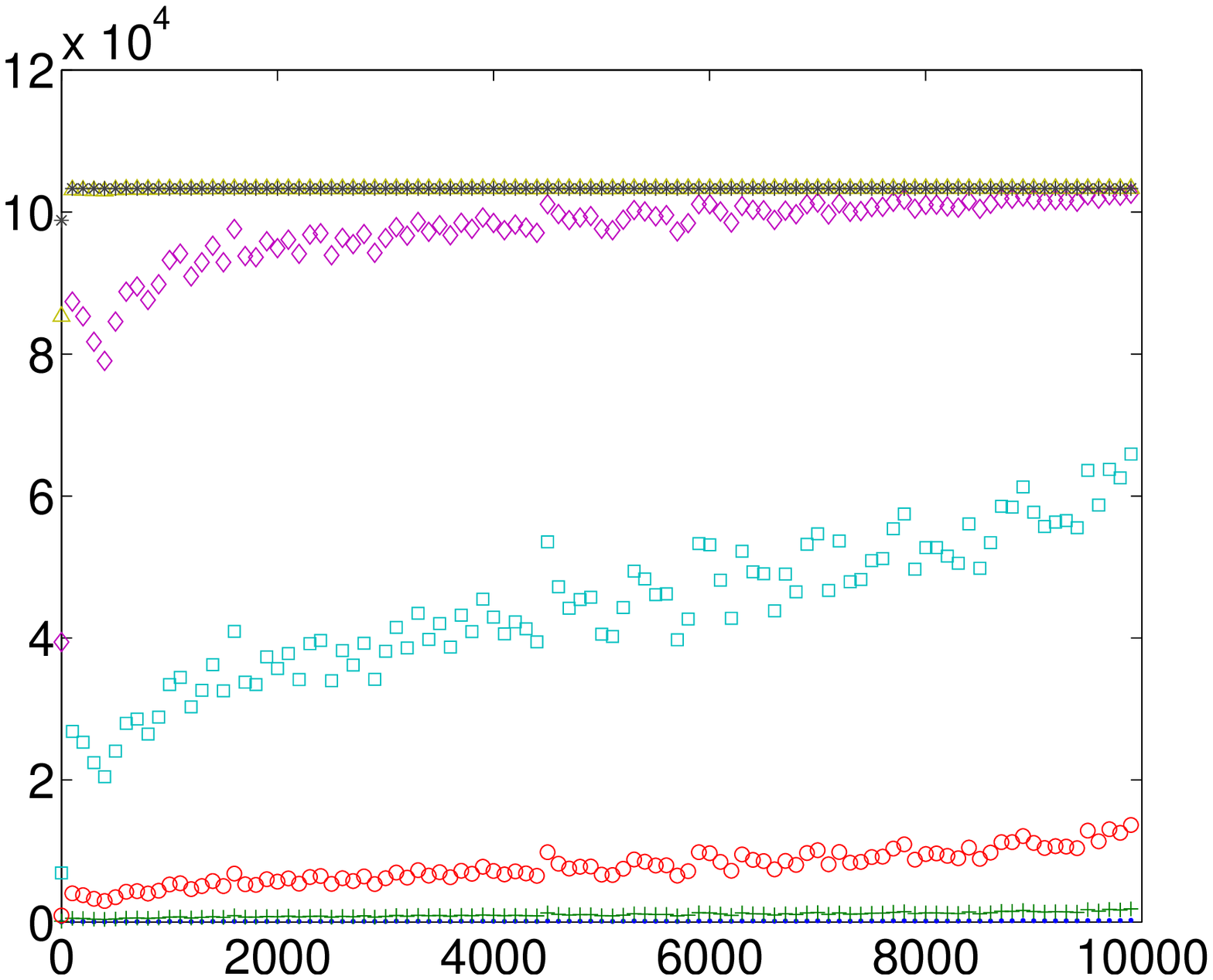, height=1.2in}
                 \setlength{\abovecaptionskip}{0pt}
                 \caption{$\alpha=0.75, \beta=0.5$, ER2}
                 \label{fig:deg-er-2-1}
         \end{subfigure}
         \hfill         
         \begin{subfigure}[b]{0.22\textwidth}
                  \centering
                  \epsfig{file=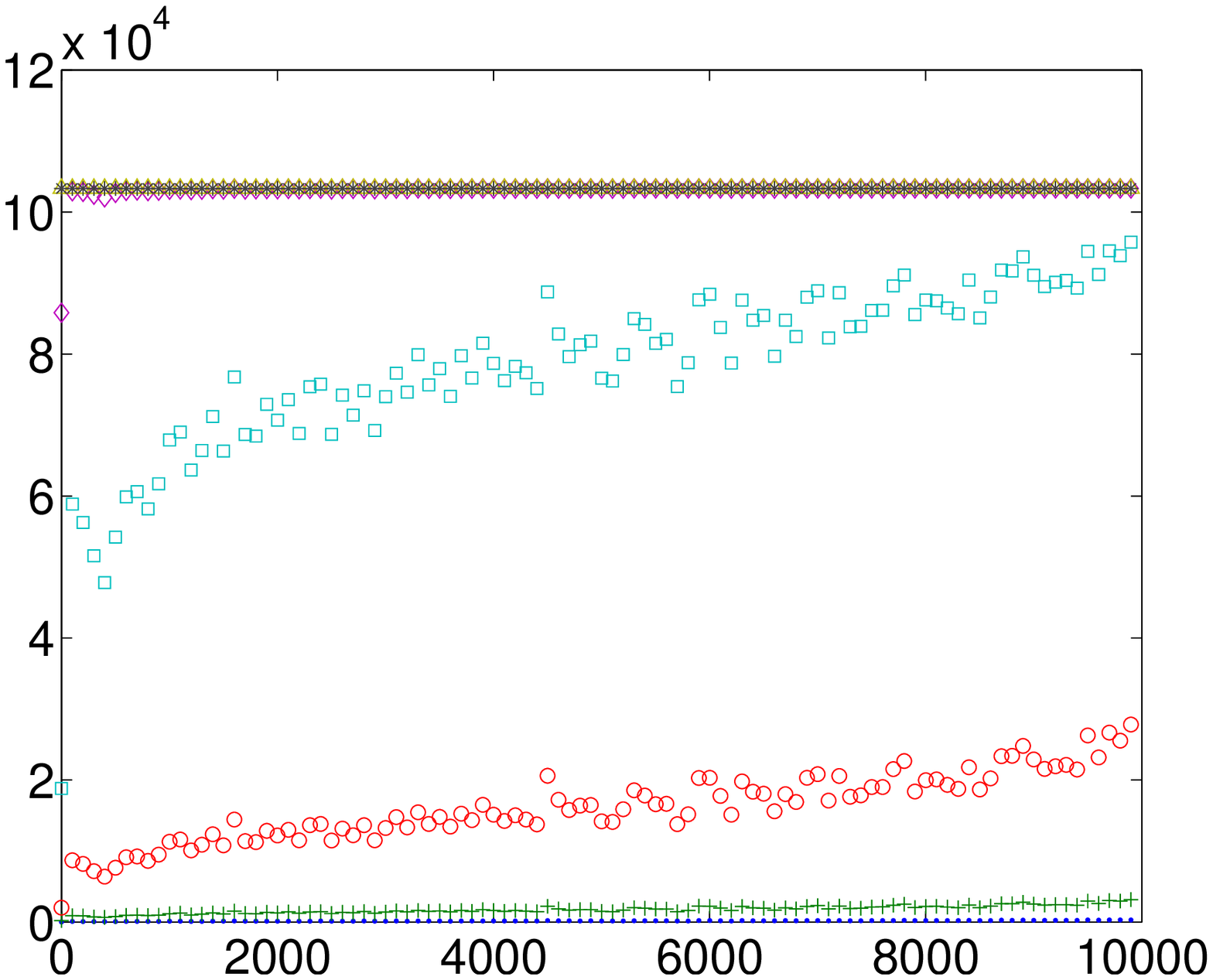, height=1.2in}
                  \setlength{\abovecaptionskip}{0pt}
                 \caption{$\alpha=1.0, \beta=0.5$, ER2}
                  \label{fig:deg-er-2-2}
         \end{subfigure}
         
         \begin{subfigure}[b]{0.22\textwidth}
                 \centering
                 \epsfig{file=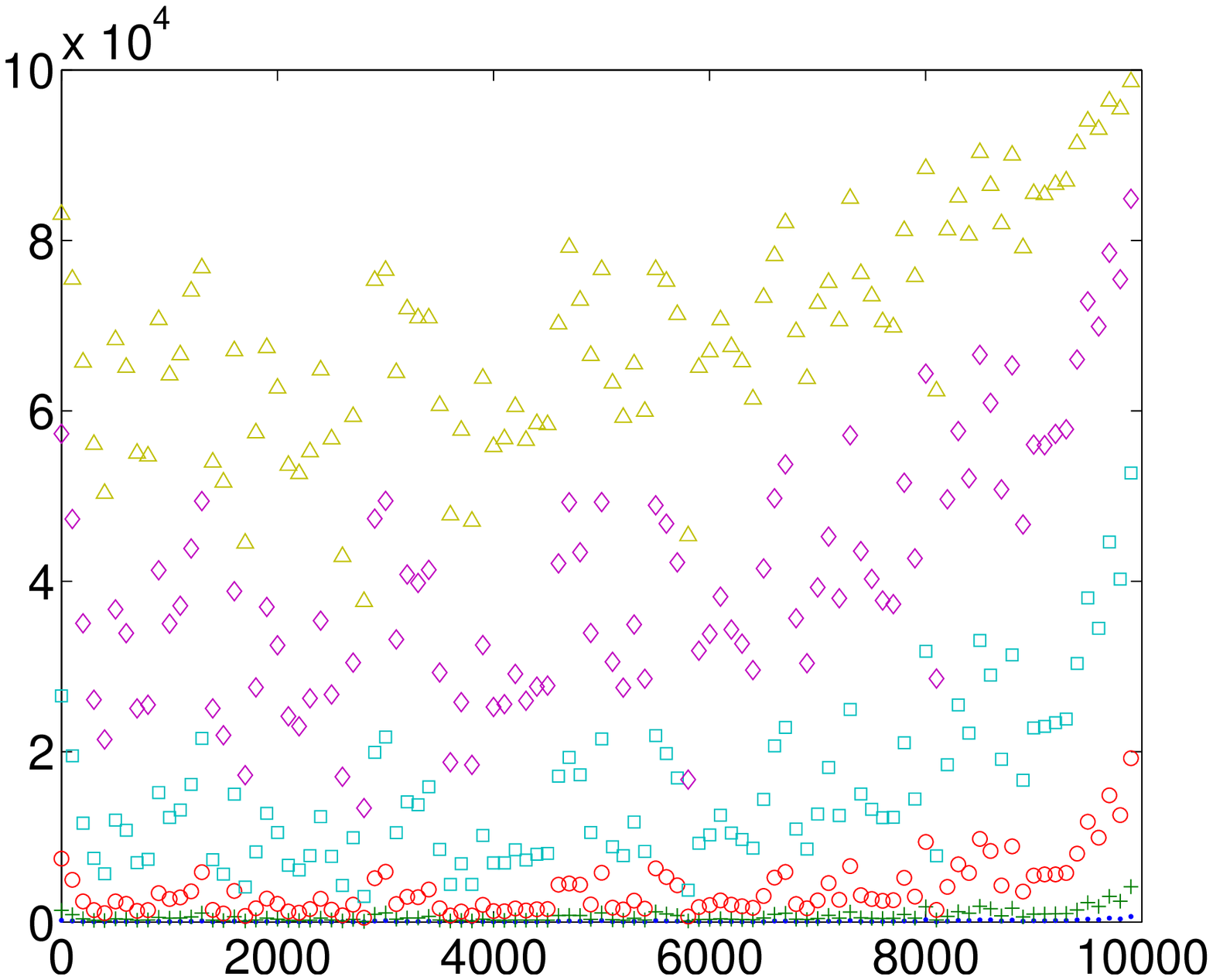, height=1.2in}
                 \setlength{\abovecaptionskip}{0pt}
                 \caption{$\alpha=0.25, \beta=0.5$, PL2}	
                 \label{fig:deg-pl-1-1}
         \end{subfigure}
         \hfill
         \begin{subfigure}[b]{0.22\textwidth}
                 \centering
                 \epsfig{file=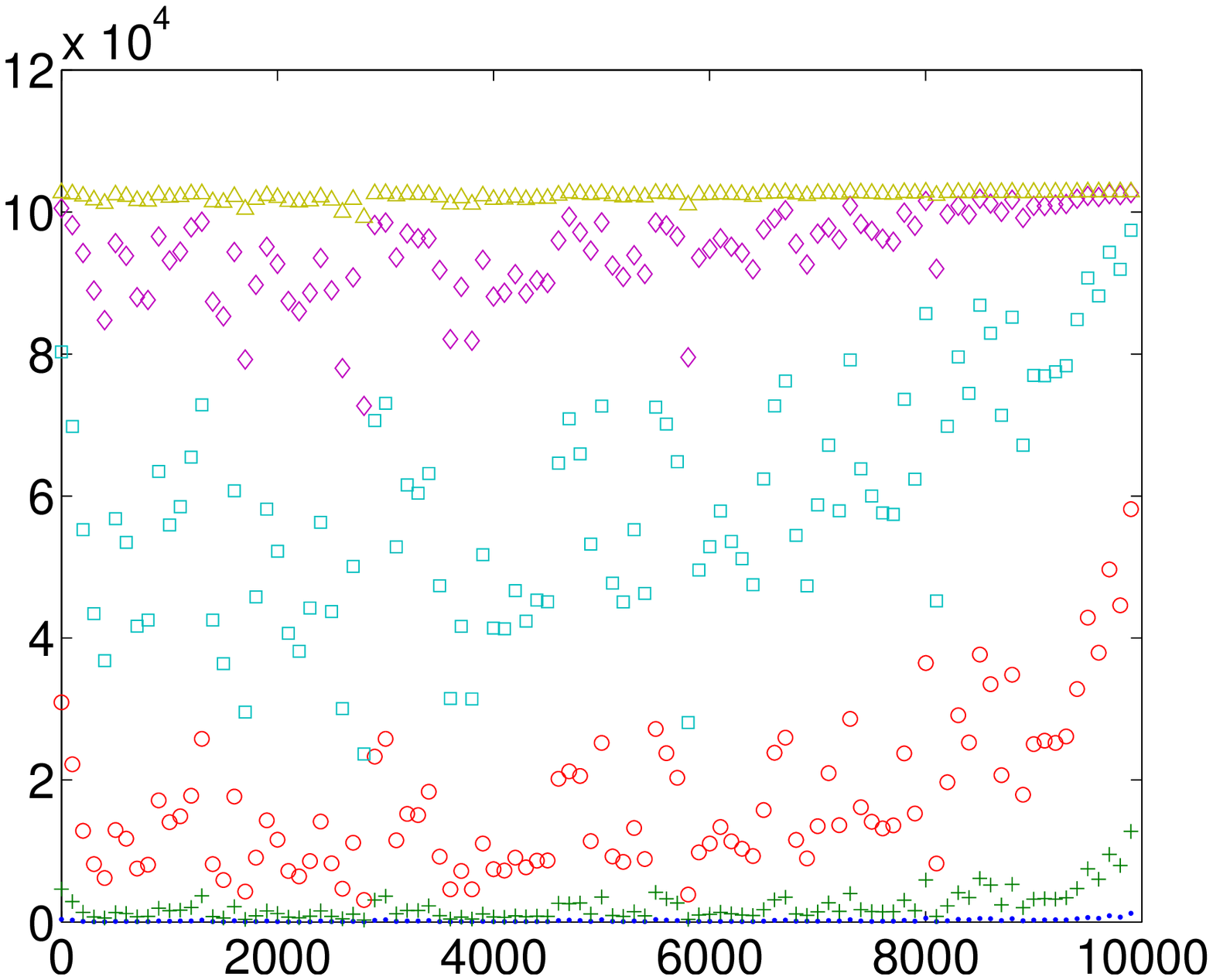, height=1.2in}
                 \setlength{\abovecaptionskip}{0pt}
                 \caption{$\alpha=0.5, \beta=0.5$, PL2}
                 \label{fig:deg-pl-1-2}
         \end{subfigure}
         \hfill         
         \begin{subfigure}[b]{0.22\textwidth}
                 \centering
                 \epsfig{file=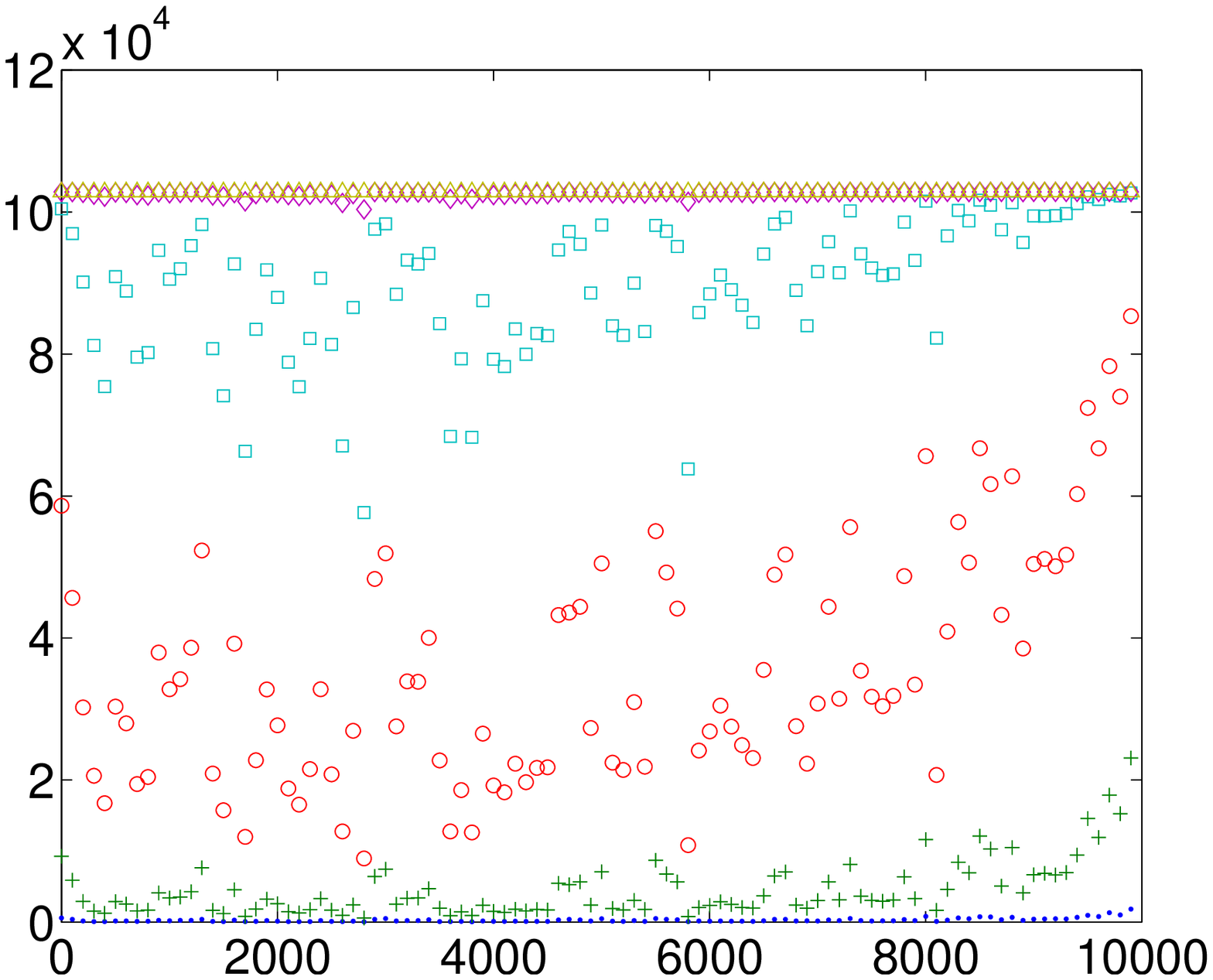, height=1.2in}
                 \setlength{\abovecaptionskip}{0pt}
                 \caption{$\alpha=0.75, \beta=0.5$, PL2}
                 \label{fig:deg-pl-2-1}
         \end{subfigure}
         \hfill         
         \begin{subfigure}[b]{0.22\textwidth}
                  \centering
                  \epsfig{file=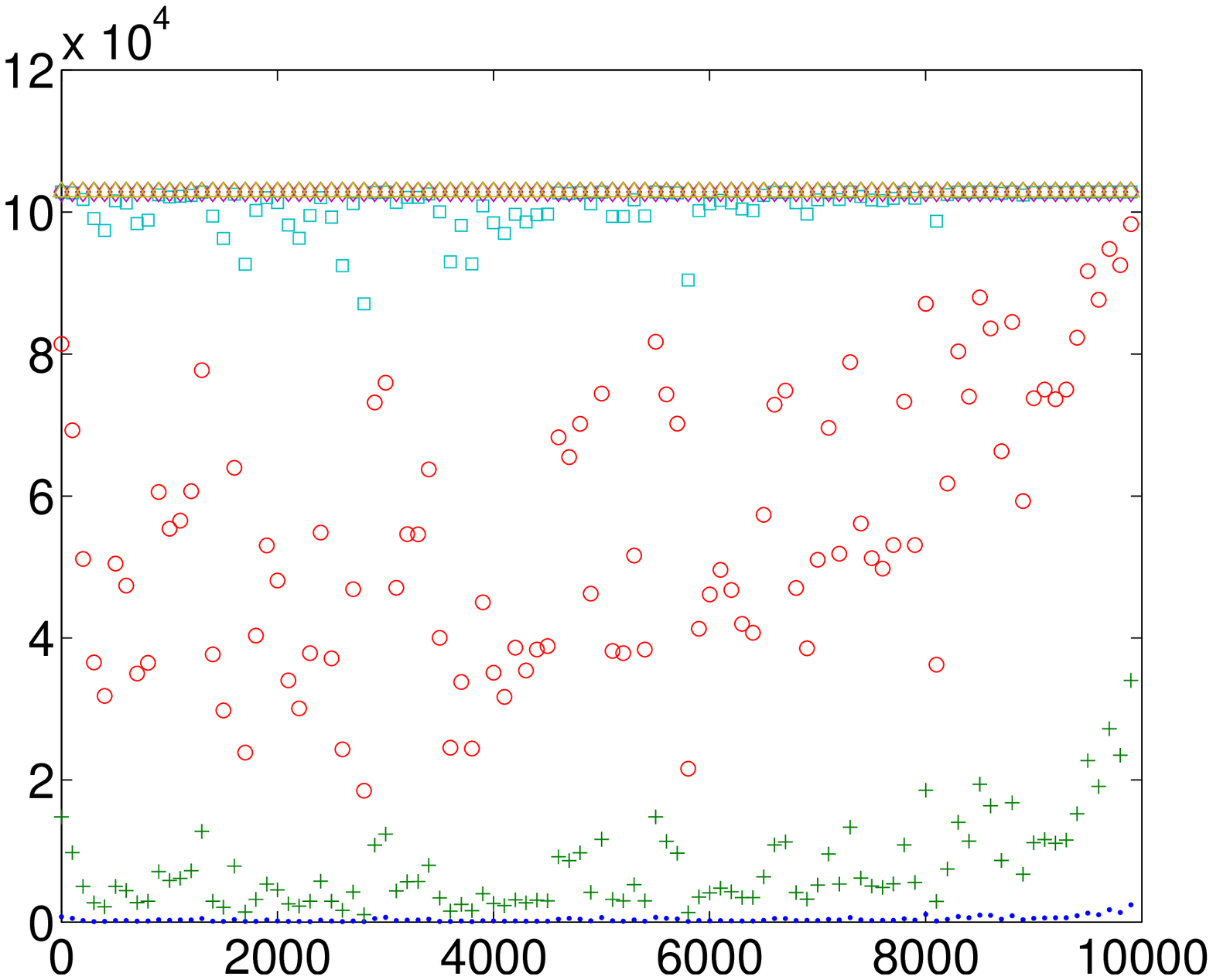, height=1.2in}
                  \setlength{\abovecaptionskip}{0pt}
                 \caption{$\alpha=1.0, \beta=0.5$, PL2}
                  \label{fig:deg-pl-2-2}
         \end{subfigure}        	
     \caption{Number of links at sampled nodes ($t=1(.), t=2(+), t=3(\circ), t=4(\square), t=5(\diamond), t=6(\triangle), t=7(*)$)}
     \label{fig:deg-volume}
 \end{figure*}

Fig. \ref{fig:deg-volume} displays the distribution of link volume collected at sample nodes. We sort $V$ by degree and take 100 sample nodes. ER graphs which are commonly called \textit{homogeneous} graphs show nearly uniform distributions for various values of $(\alpha, \beta)$. On the contrary, PL graphs are \textit{heterogeneous} ones and sample nodes exhibit much more random distributions.

\begin{figure*}
 	\centering
         \begin{subfigure}[b]{0.22\textwidth}
                 \centering
                 \epsfig{file=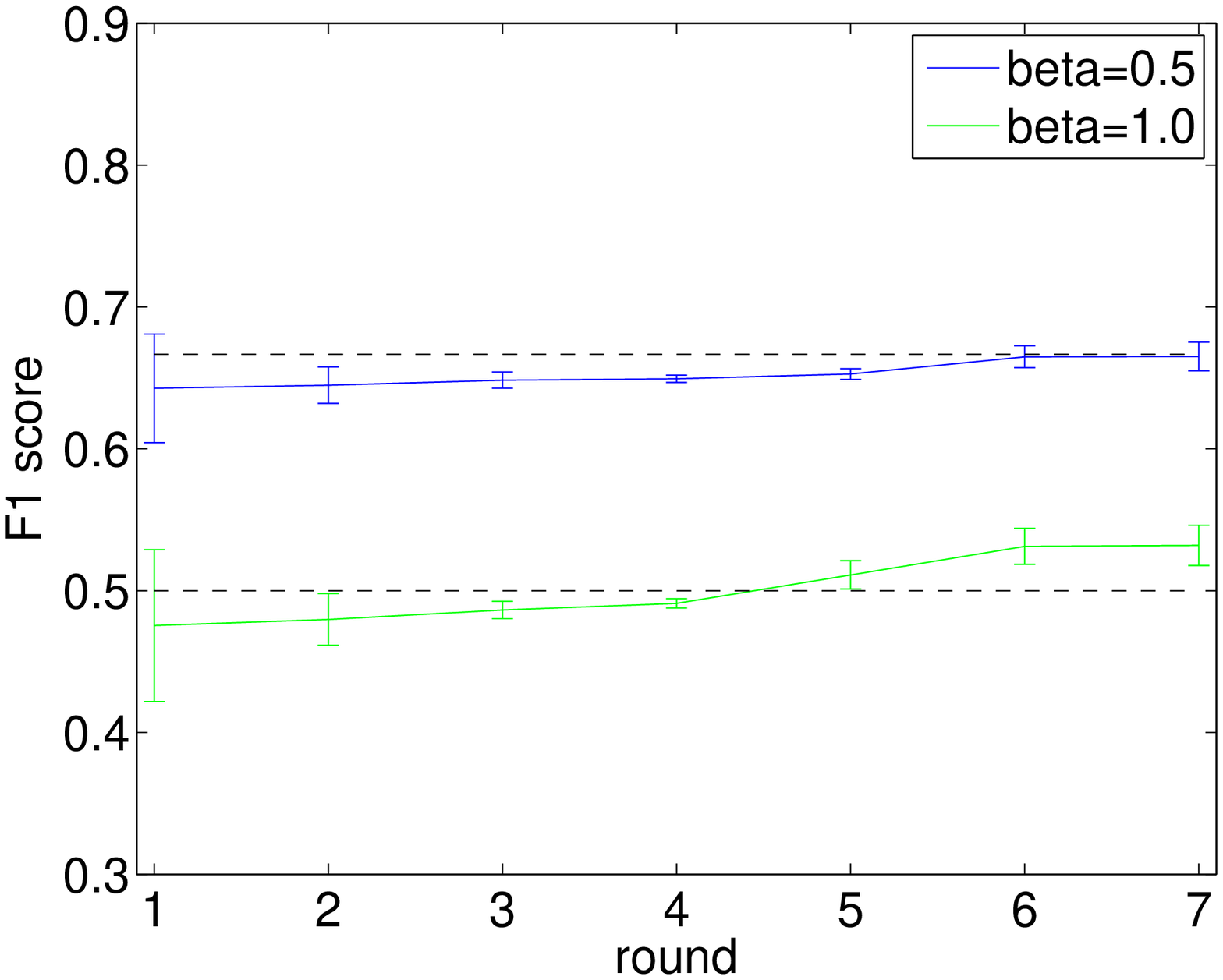, height=1.2in}
                 \setlength{\abovecaptionskip}{0pt}
                 \caption{ER2, $\alpha=0.5$}	
                 \label{fig:attack-er2-1}
         \end{subfigure}
         \hfill
         \begin{subfigure}[b]{0.22\textwidth}
                 \centering
                 \epsfig{file=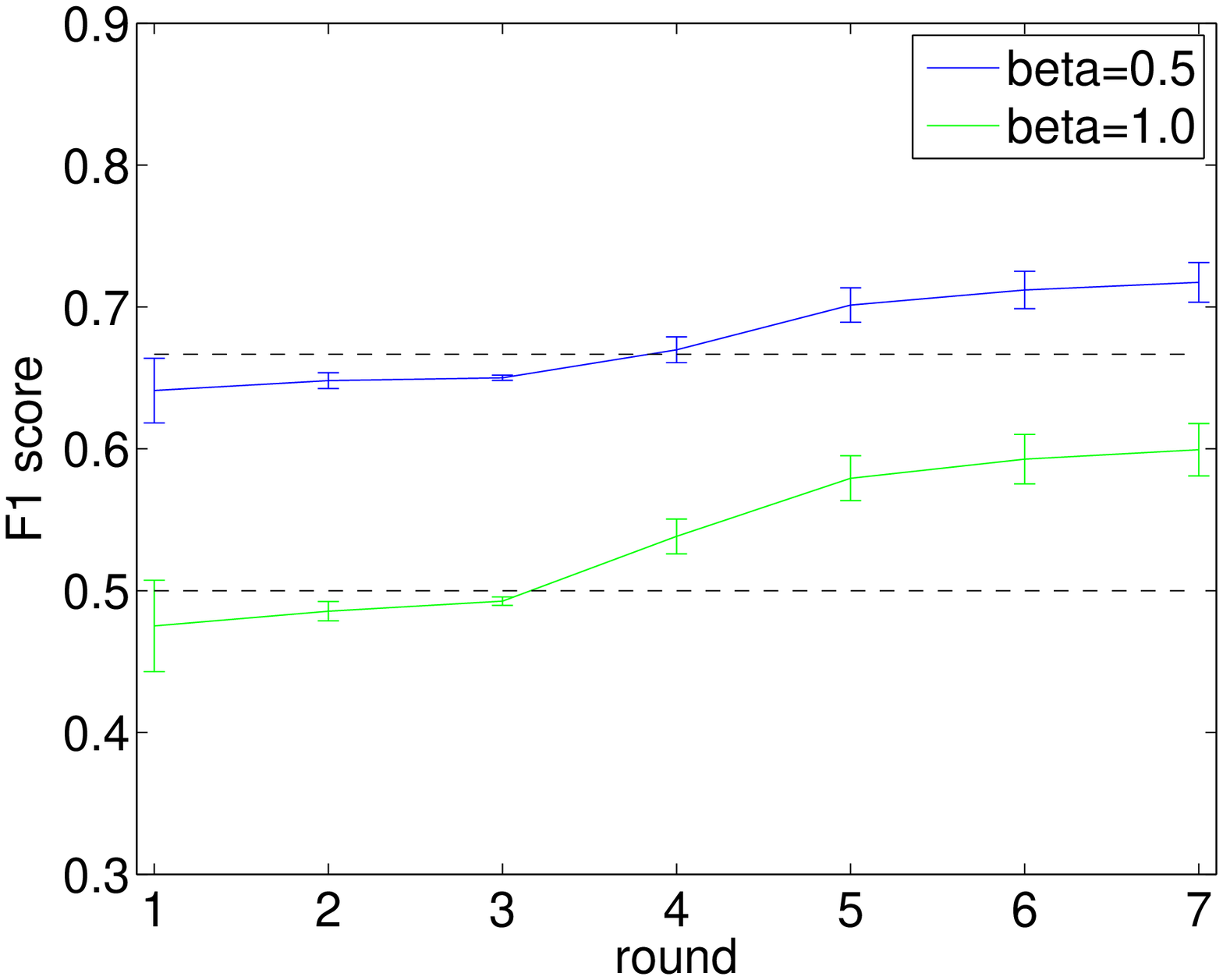, height=1.2in}
                 \setlength{\abovecaptionskip}{0pt}
                 \caption{ER2, $\alpha=1.0$}
                 \label{fig:attack-er2-2}
         \end{subfigure}
         \hfill         
         \begin{subfigure}[b]{0.22\textwidth}
                 \centering
                 \epsfig{file=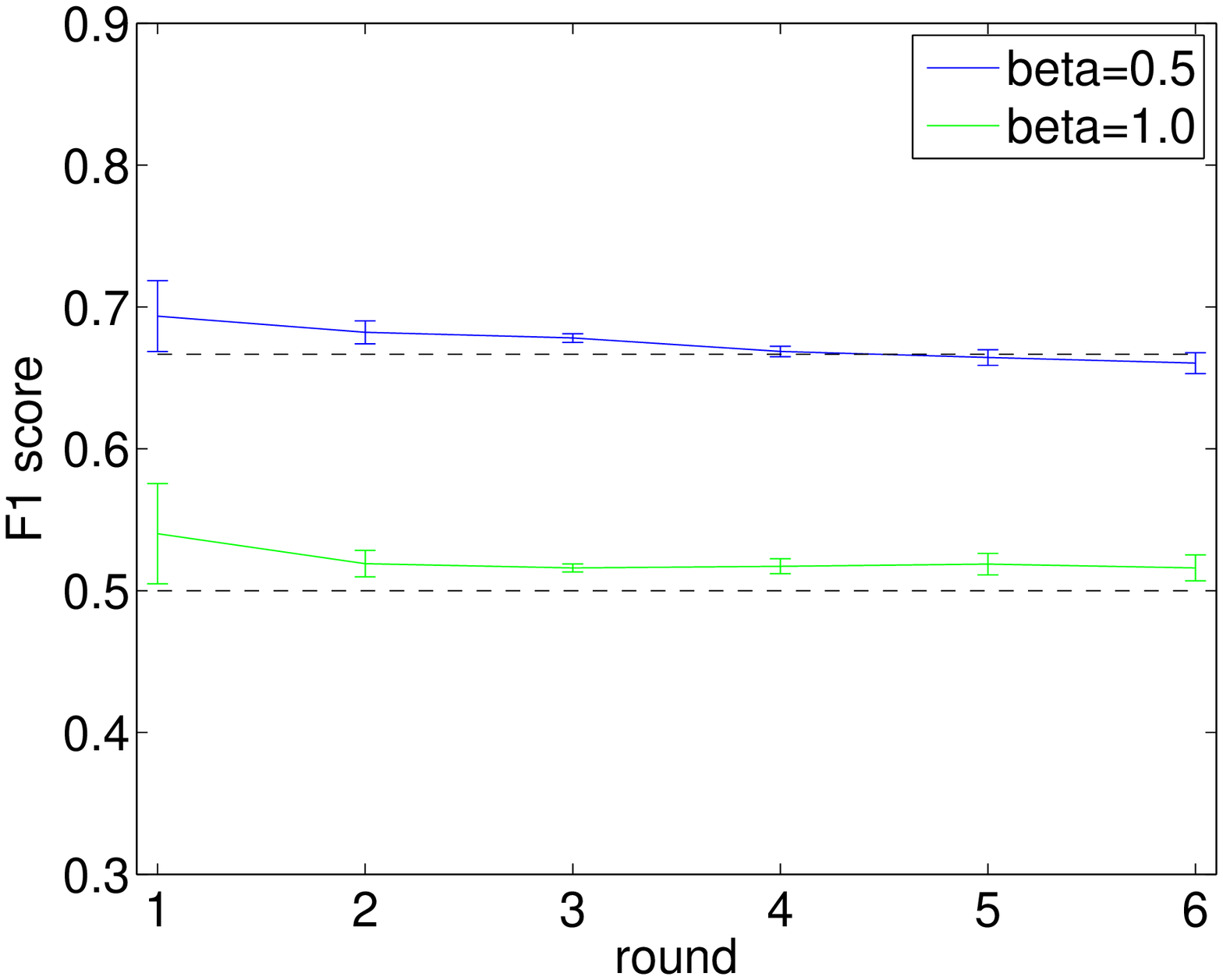, height=1.2in}
                 \setlength{\abovecaptionskip}{0pt}
                 \caption{PL2, $\alpha=0.5$}	
                 \label{fig:attack-pl2-1}
         \end{subfigure}
         \hfill         
         \begin{subfigure}[b]{0.22\textwidth}
                  \centering
                 \epsfig{file=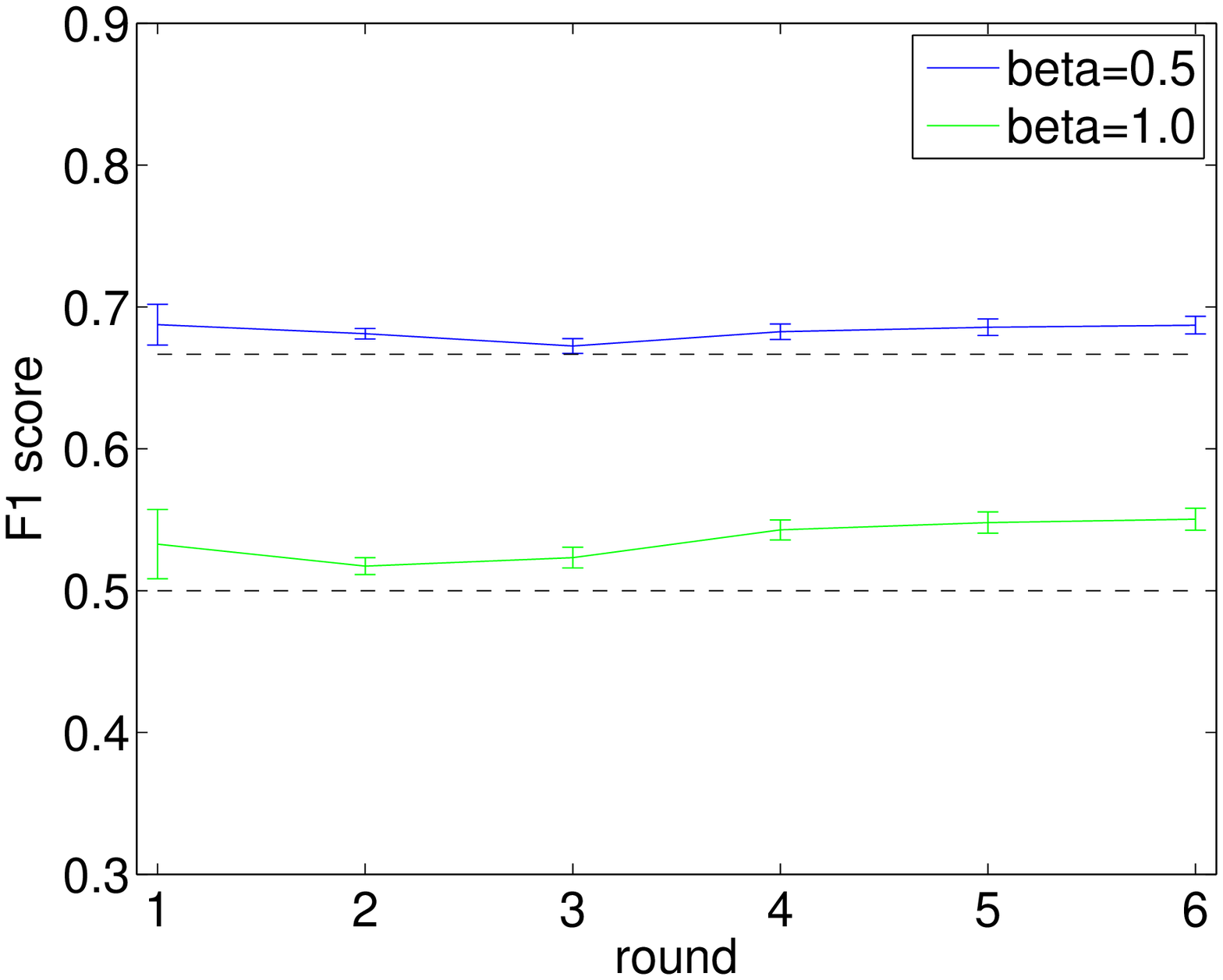, height=1.2in}
                 \setlength{\abovecaptionskip}{0pt}
                 \caption{PL2, $\alpha=1.0$}
                 \label{fig:attack-pl2-2}
         \end{subfigure}
     \caption{Inference attacks}
     \label{fig:attack}
 \end{figure*}

The inference attack on Baseline scheme (Section \ref{subsec:baseline-priv}) is shown in Fig. \ref{fig:attack}. The average F1 scores for two values of $\beta$ are plotted at different rounds of Baseline protocol. We observe that the scores are quite close to the theoretical values $1/(1+\beta)$ (see the dashed lines). On ER2 graph, the inference attack is more effective at latter rounds and for larger $\alpha$ while this is not clear on PL2.

\subsection{Bloom Filter Scheme}
\label{subsec:eval-bf}
\begin{figure*}
 	\centering
         \begin{subfigure}[b]{0.22\textwidth}
                 \centering
                 \epsfig{file=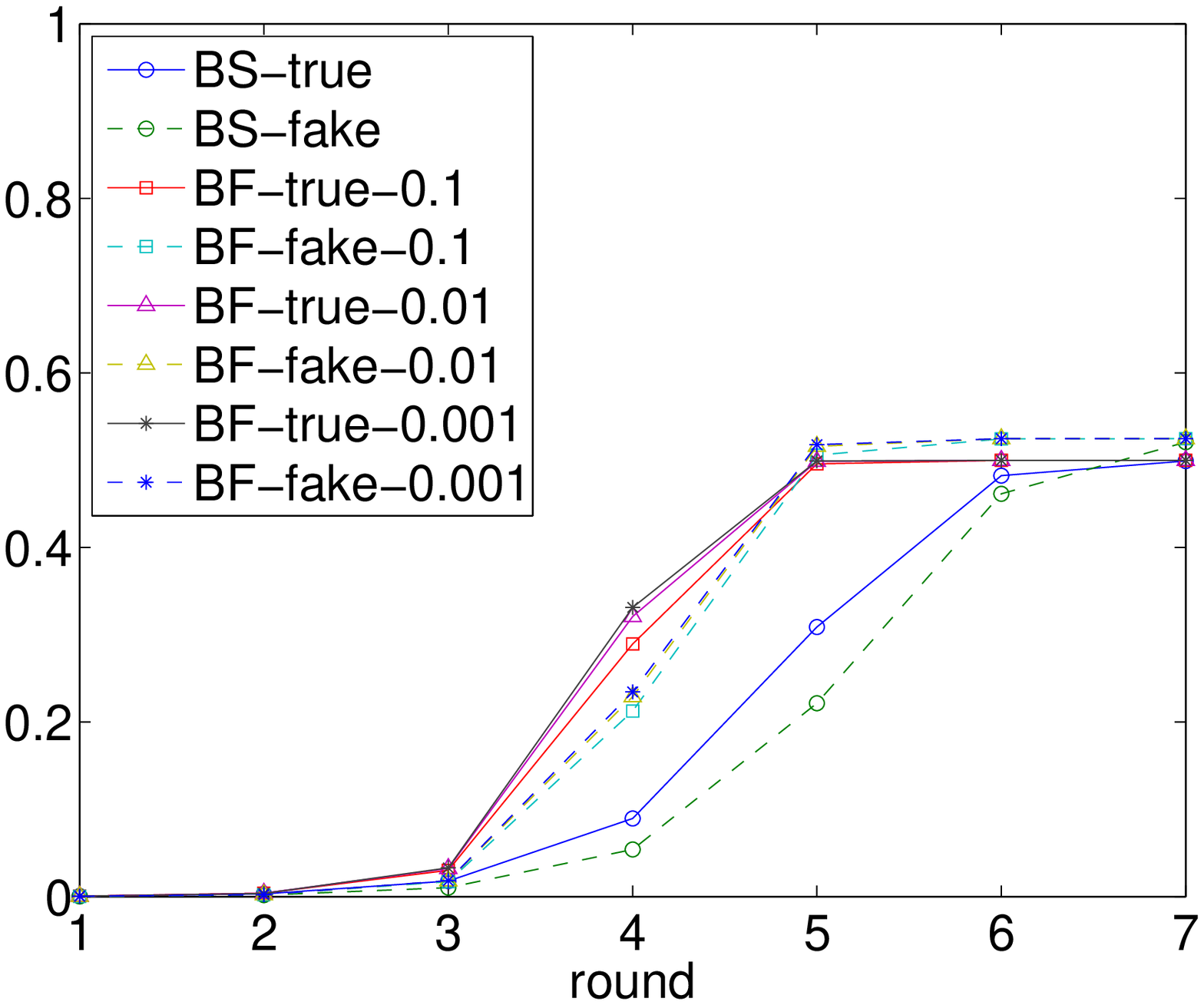, height=1.2in}
                 \setlength{\abovecaptionskip}{0pt}
                 \caption{$\alpha=0.5, \beta=0.5$, ER2}	
                 \label{fig:fp-er-1-1}
         \end{subfigure}
         \hfill
         \begin{subfigure}[b]{0.22\textwidth}
                 \centering
                 \epsfig{file=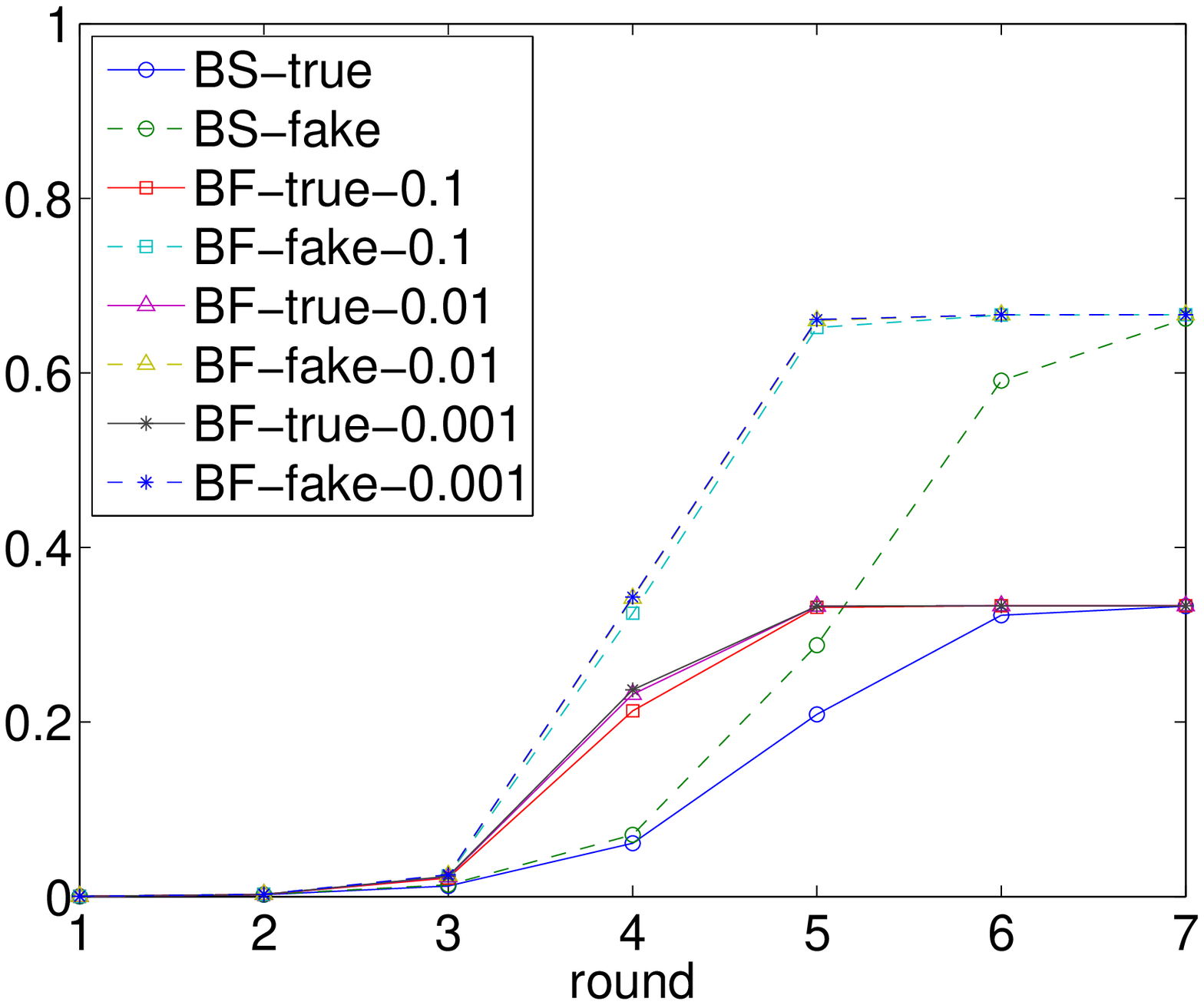, height=1.2in}
                 \setlength{\abovecaptionskip}{0pt}
                 \caption{$\alpha=0.5, \beta=1.0$, ER2}
                 \label{fig:fp-er-1-2}
         \end{subfigure}
         \hfill         
         \begin{subfigure}[b]{0.22\textwidth}
                 \centering
                 \epsfig{file=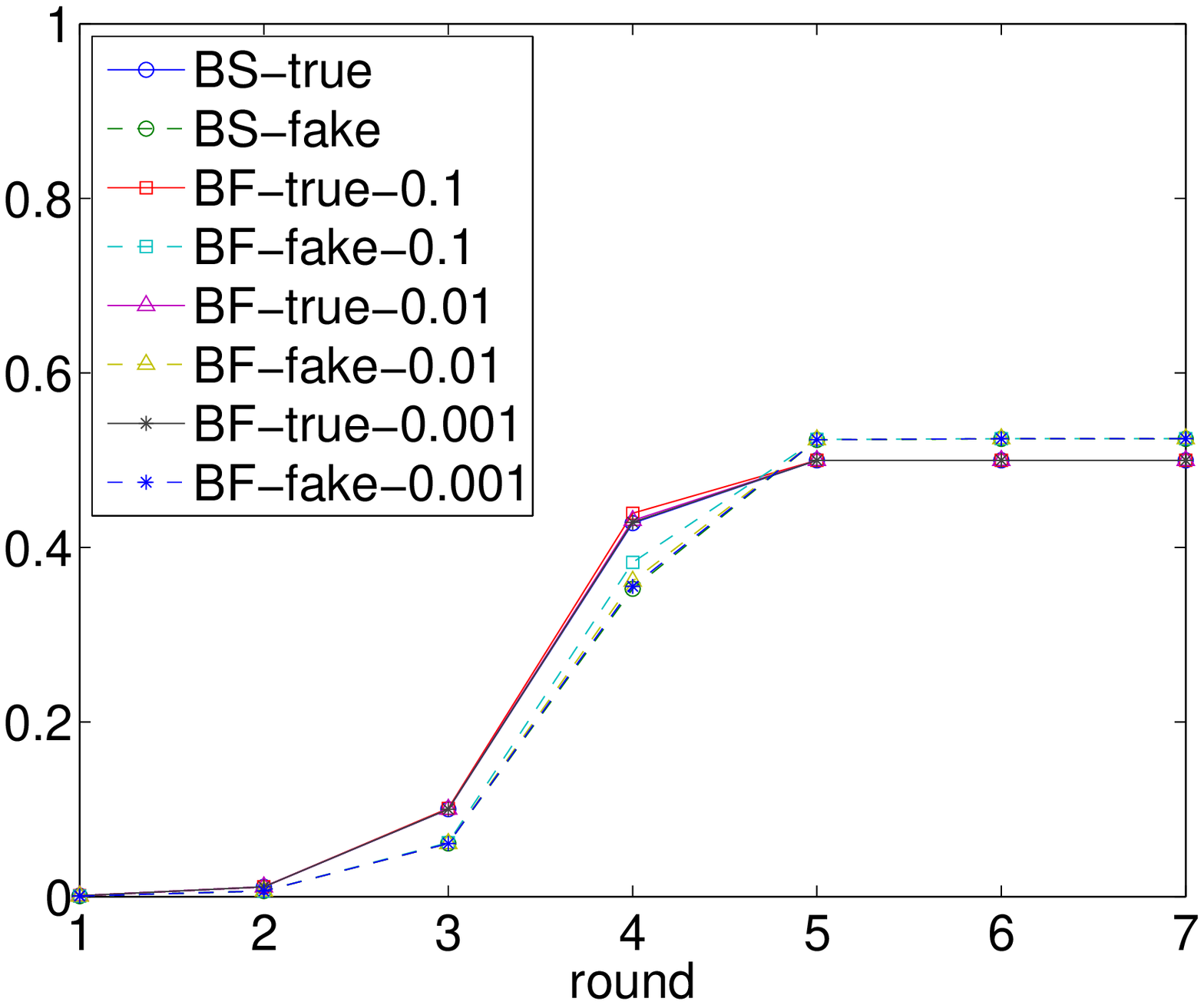, height=1.2in}
                 \setlength{\abovecaptionskip}{0pt}
                 \caption{$\alpha=1.0, \beta=0.5$, ER2}
                 \label{fig:fp-er-2-1}
         \end{subfigure}
         \hfill         
         \begin{subfigure}[b]{0.22\textwidth}
                  \centering
                  \epsfig{file=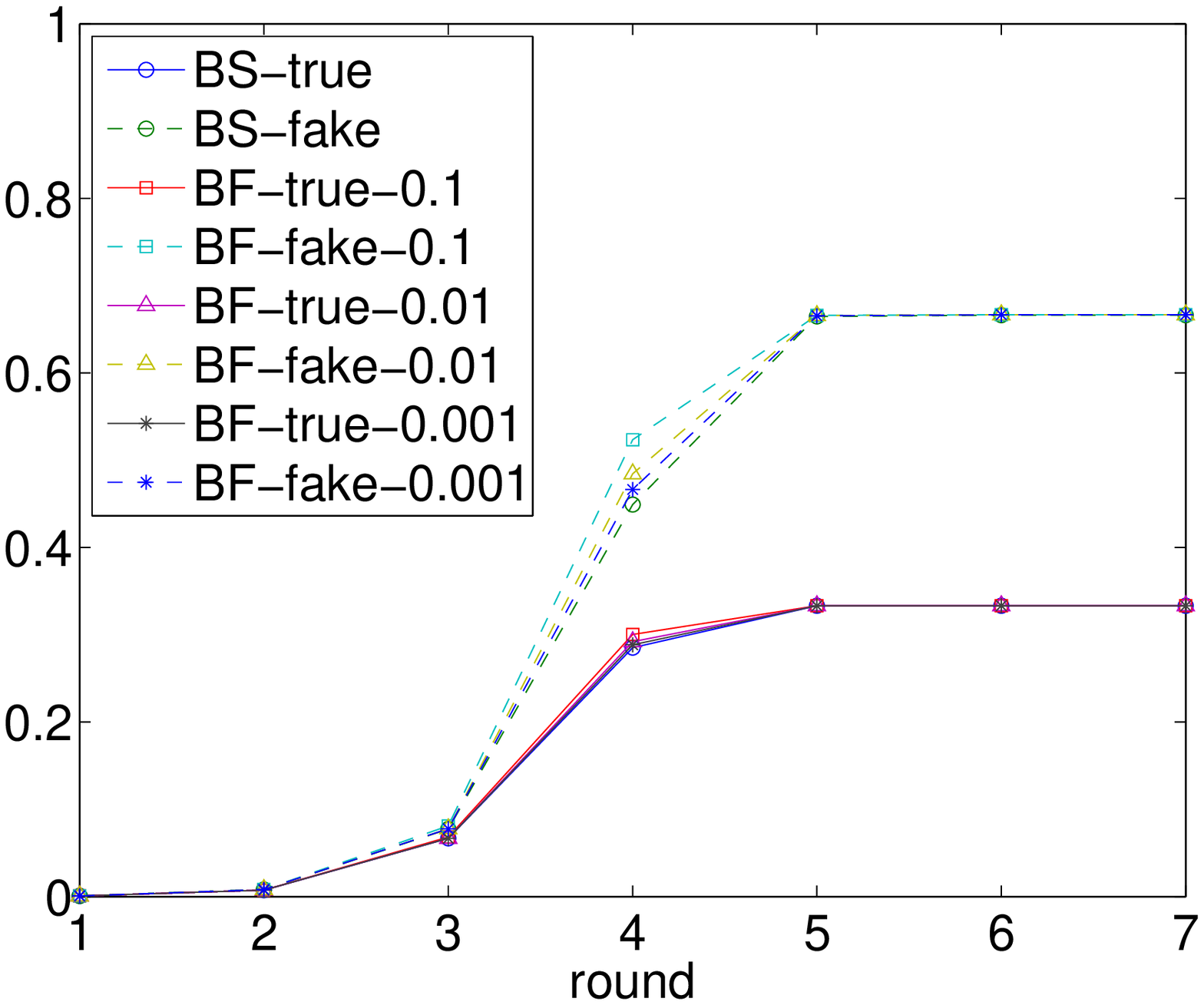, height=1.2in}
                  \setlength{\abovecaptionskip}{0pt}
                 \caption{$\alpha=1.0, \beta=1.0$, ER2}
                  \label{fig:fp-er-2-2}
         \end{subfigure}
         
         \begin{subfigure}[b]{0.22\textwidth}
                 \centering
                 \epsfig{file=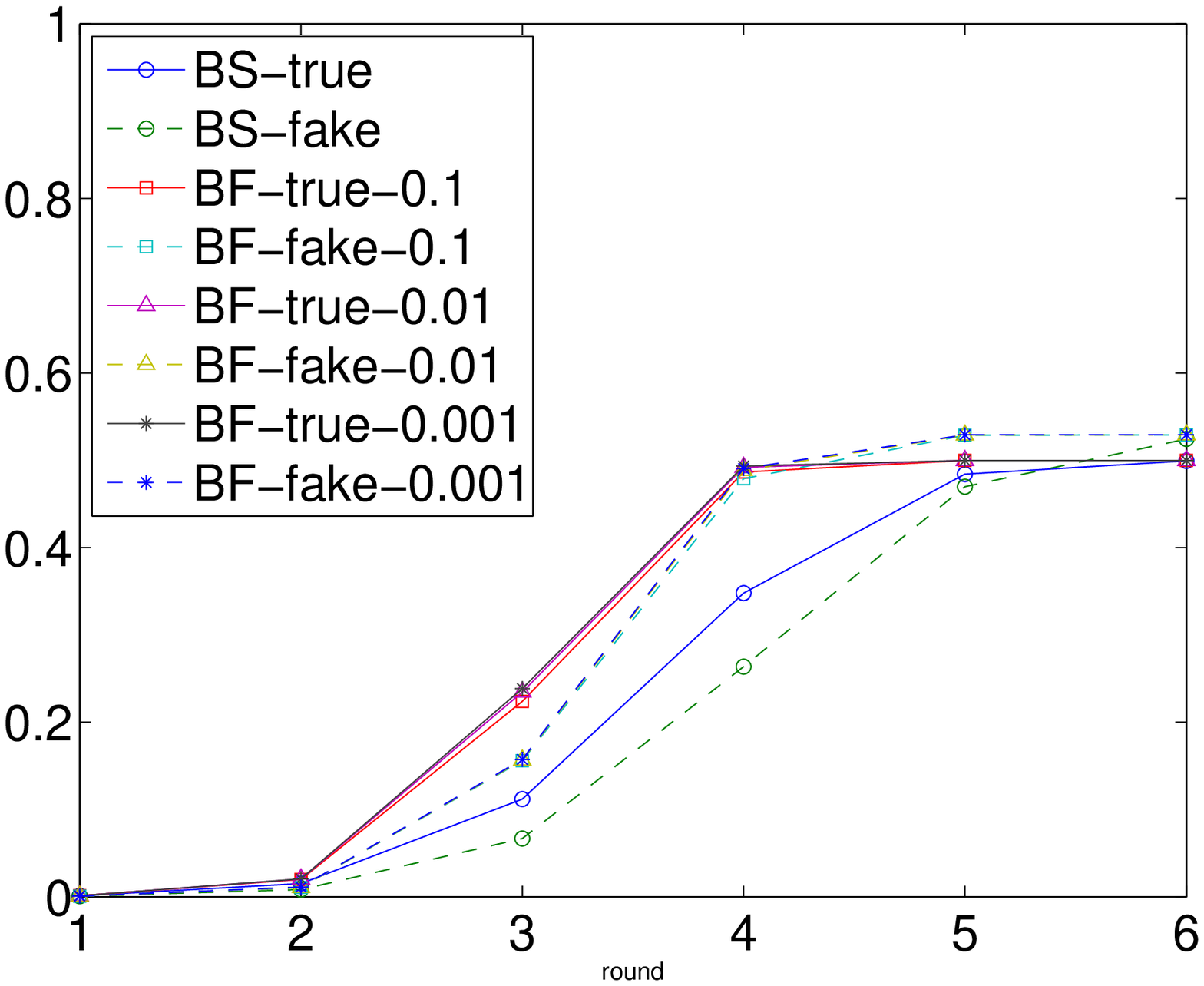, height=1.2in}
                 \setlength{\abovecaptionskip}{0pt}
                 \caption{$\alpha=0.5, \beta=0.5$, PL2}	
                 \label{fig:fp-pl-1-1}
         \end{subfigure}
         \hfill
         \begin{subfigure}[b]{0.22\textwidth}
                 \centering
                 \epsfig{file=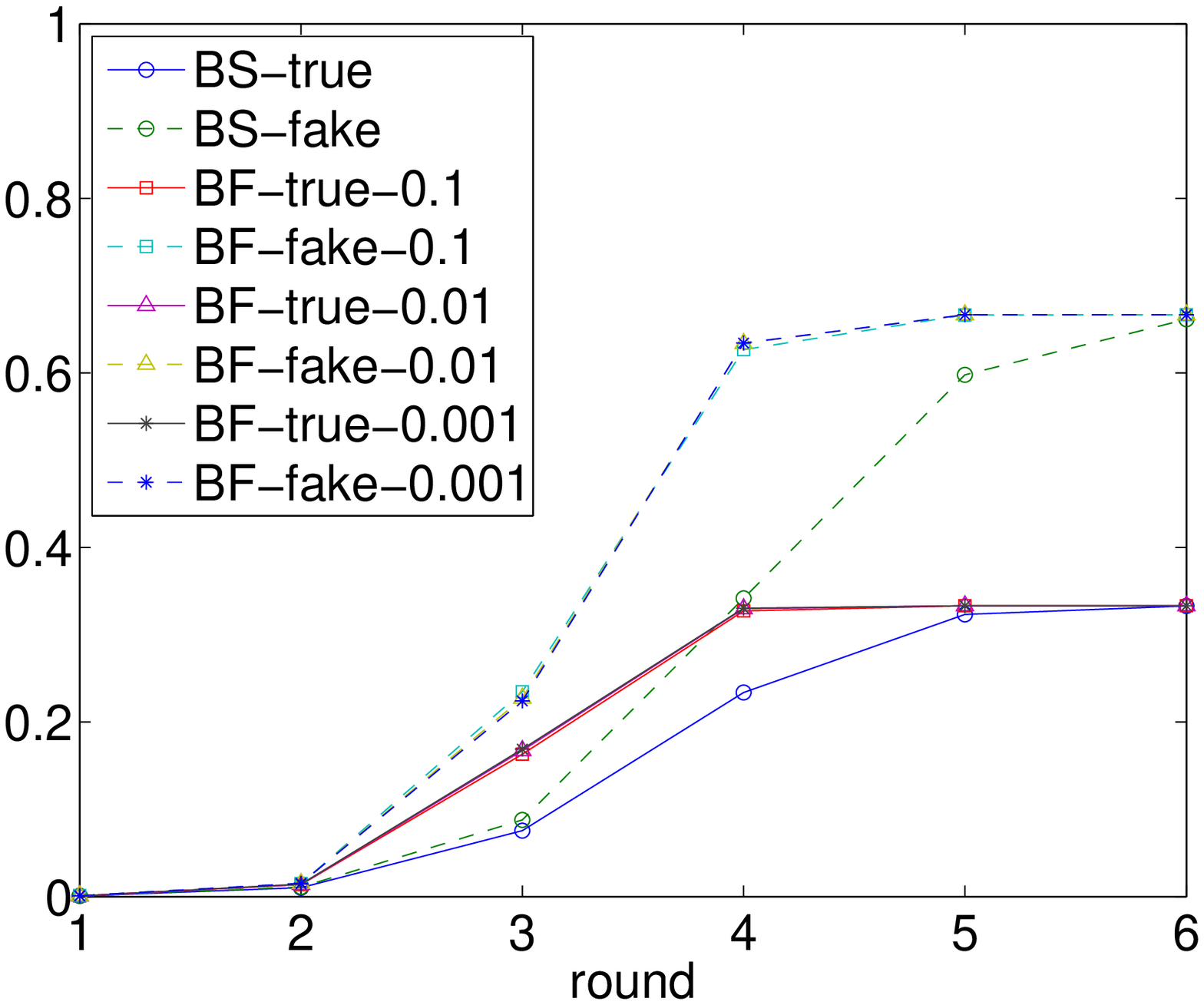, height=1.2in}
                 \setlength{\abovecaptionskip}{0pt}
                 \caption{$\alpha=0.5, \beta=1.0$, PL2}
                 \label{fig:fp-pl-1-2}
         \end{subfigure}
         \hfill         
         \begin{subfigure}[b]{0.22\textwidth}
                 \centering
                 \epsfig{file=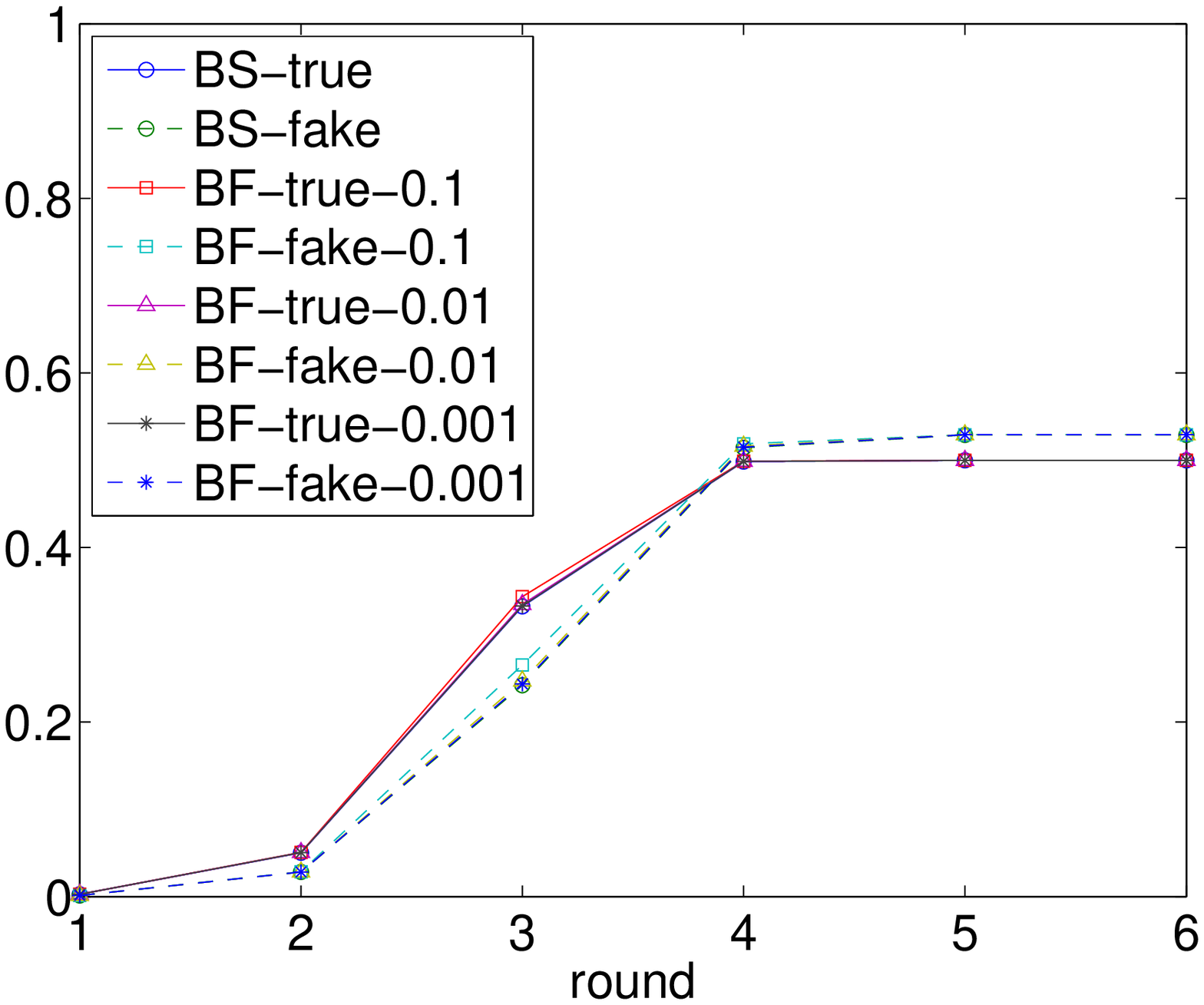, height=1.2in}
                 \setlength{\abovecaptionskip}{0pt}
                 \caption{$\alpha=1.0, \beta=0.5$, PL2}
                 \label{fig:fp-pl-2-1}
         \end{subfigure}
         \hfill         
         \begin{subfigure}[b]{0.22\textwidth}
                  \centering
                  \epsfig{file=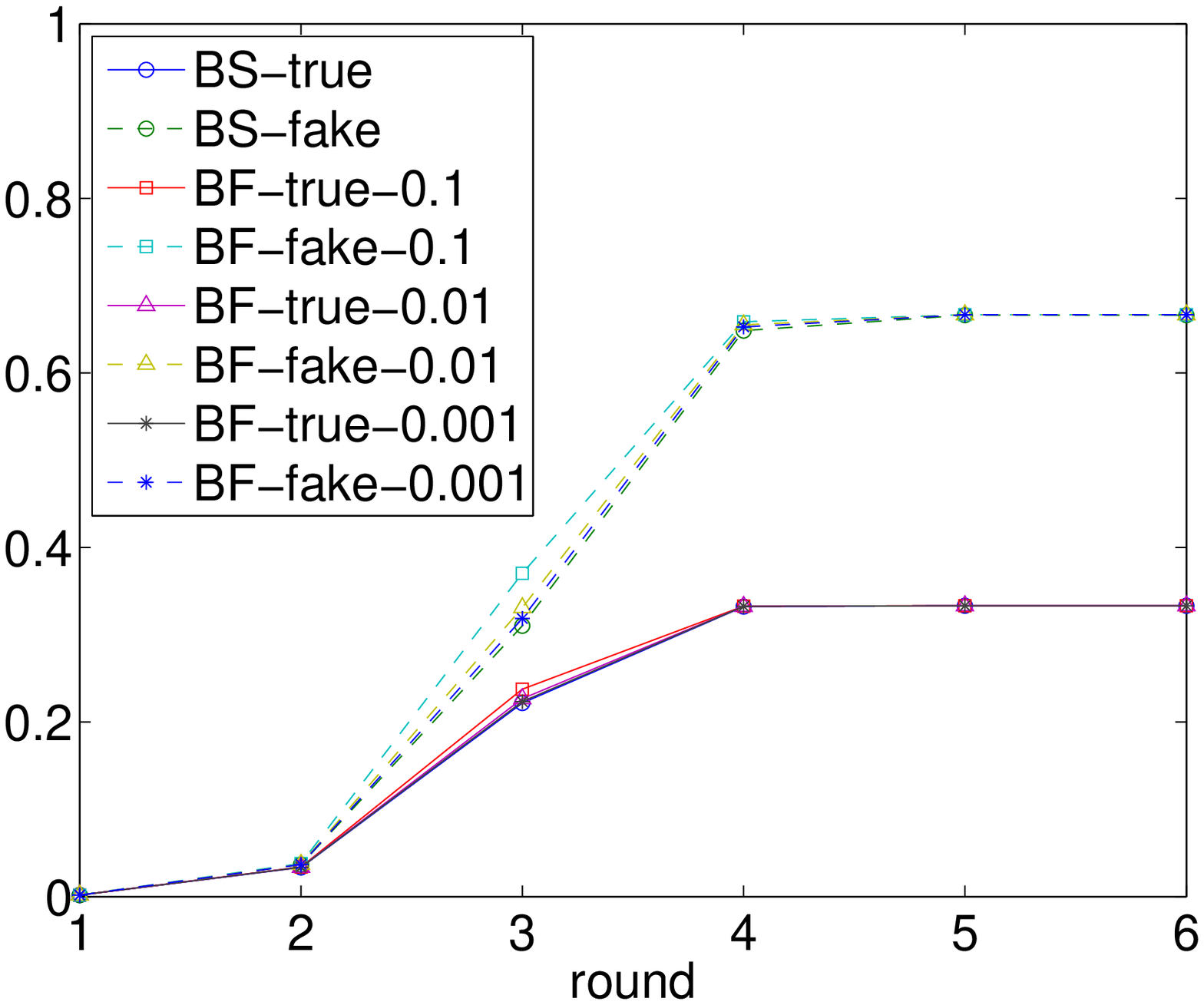, height=1.2in}
                  \setlength{\abovecaptionskip}{0pt}
                 \caption{$\alpha=1.0, \beta=1.0$, PL2}
                  \label{fig:fp-pl-2-2}
         \end{subfigure}        	
     \caption{Normalized number of true/fake links by different false positive rates}
     \label{fig:fp-volume}
 \end{figure*}

In this section, we examine the performance of Bloom Filter scheme. We set the false positive rate of Bloom Filter to 0.1, 0.01 and 0.001 (the number of hash functions $k$ is 4,7 and 10 respectively). Fig. \ref{fig:fp-volume} displays the normalized number of true/fake links by Baseline and Bloom Filter with different false positive rates. We find that lower false positive rates make no improvement for $\alpha=0.25, 0.5$. Bit Erasure (Algorithm \ref{algo-bit-erasure}) causes this effect. Lower $\alpha$ means more bits to be erased in Bloom filters. Consequently, the number of false positive links and false negative links is amplified round by round for small $\alpha$.

\begin{figure*}
 	\centering
         \begin{subfigure}[b]{0.22\textwidth}
                 \centering
                 \epsfig{file=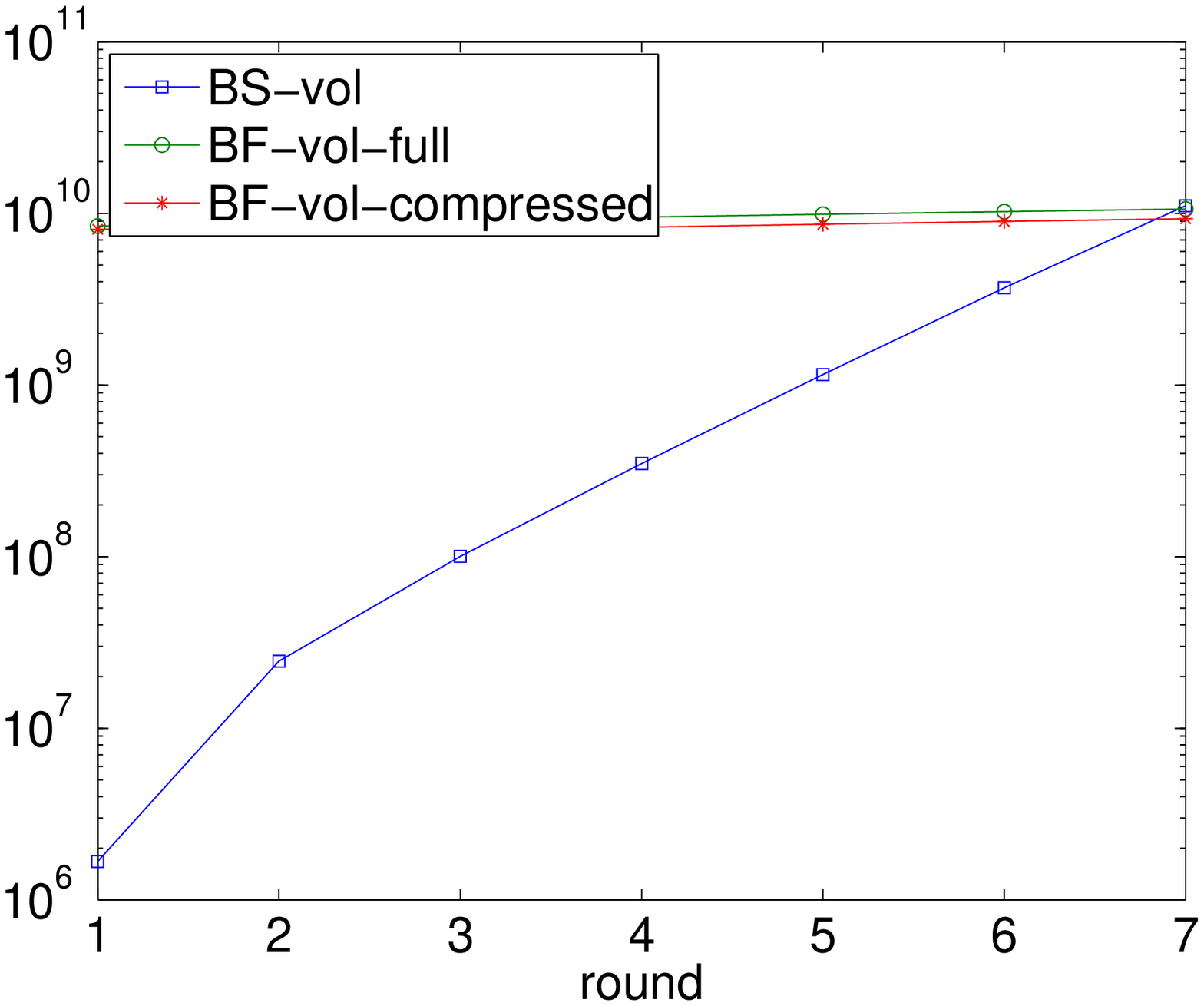, height=1.2in}
                 \setlength{\abovecaptionskip}{0pt}
                 \caption{$\alpha=0.25, \beta=0.5$, ER2}	
                 \label{fig:comp-er-1-1}
         \end{subfigure}
         \hfill
         \begin{subfigure}[b]{0.22\textwidth}
                 \centering
                 \epsfig{file=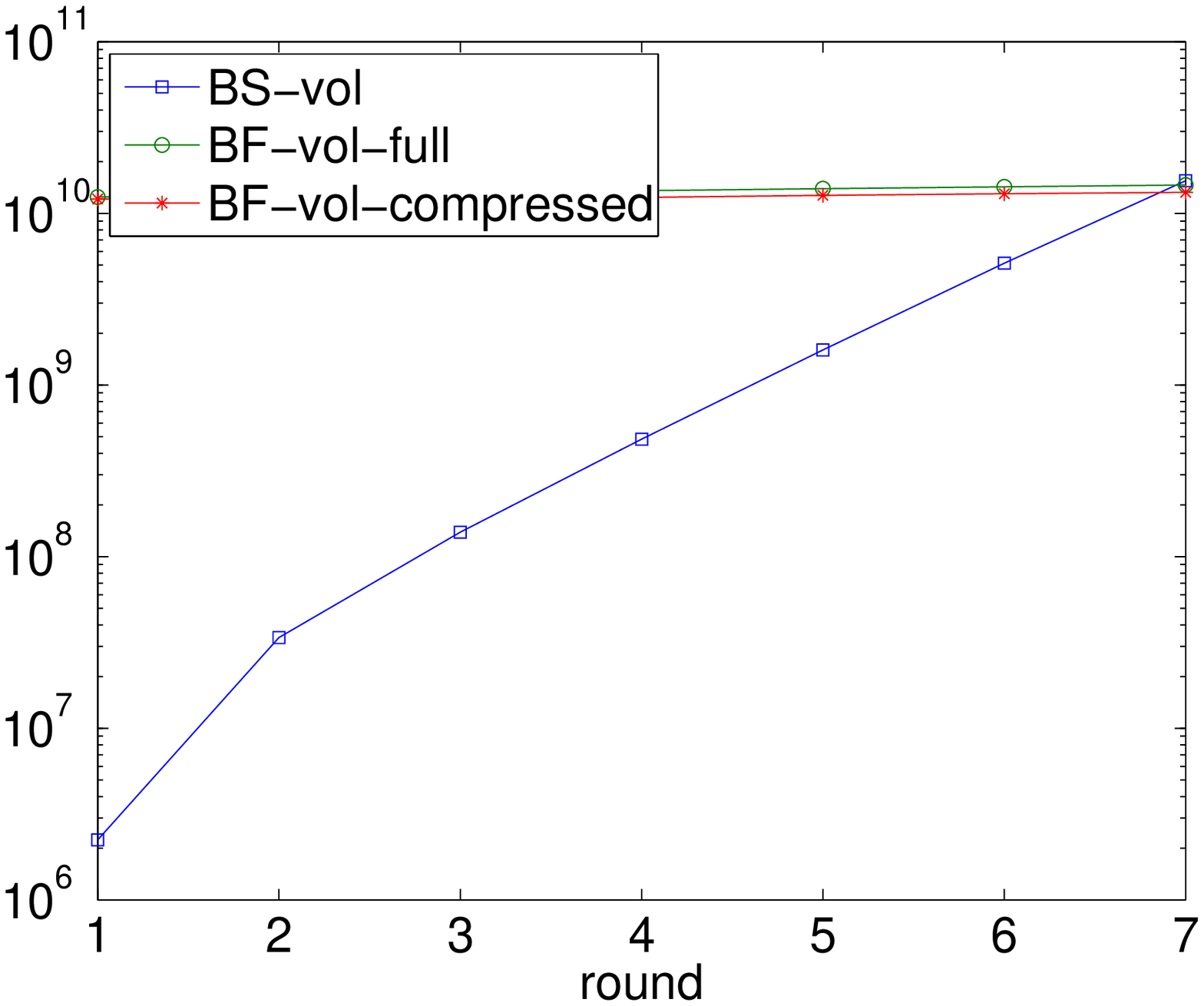, height=1.2in}
                 \setlength{\abovecaptionskip}{0pt}
                 \caption{$\alpha=0.25, \beta=1.0$, ER2}
                 \label{fig:comp-er-1-2}
         \end{subfigure}
         \hfill         
         \begin{subfigure}[b]{0.22\textwidth}
                 \centering
                 \epsfig{file=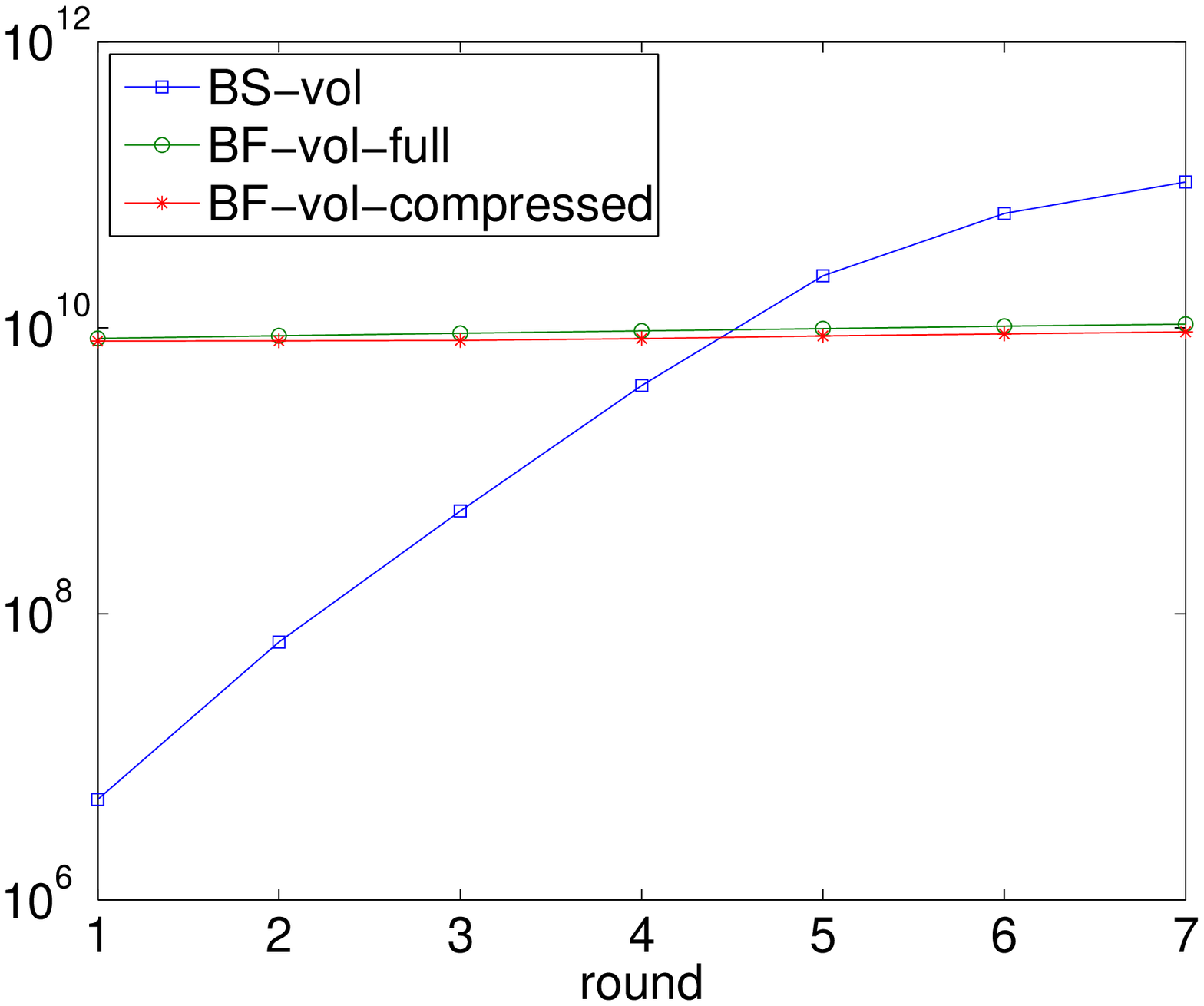, height=1.2in}
                 \setlength{\abovecaptionskip}{0pt}
                 \caption{$\alpha=0.75, \beta=0.5$, ER2}
                 \label{fig:comp-er-2-1}
         \end{subfigure}
         \hfill         
         \begin{subfigure}[b]{0.22\textwidth}
                  \centering
                  \epsfig{file=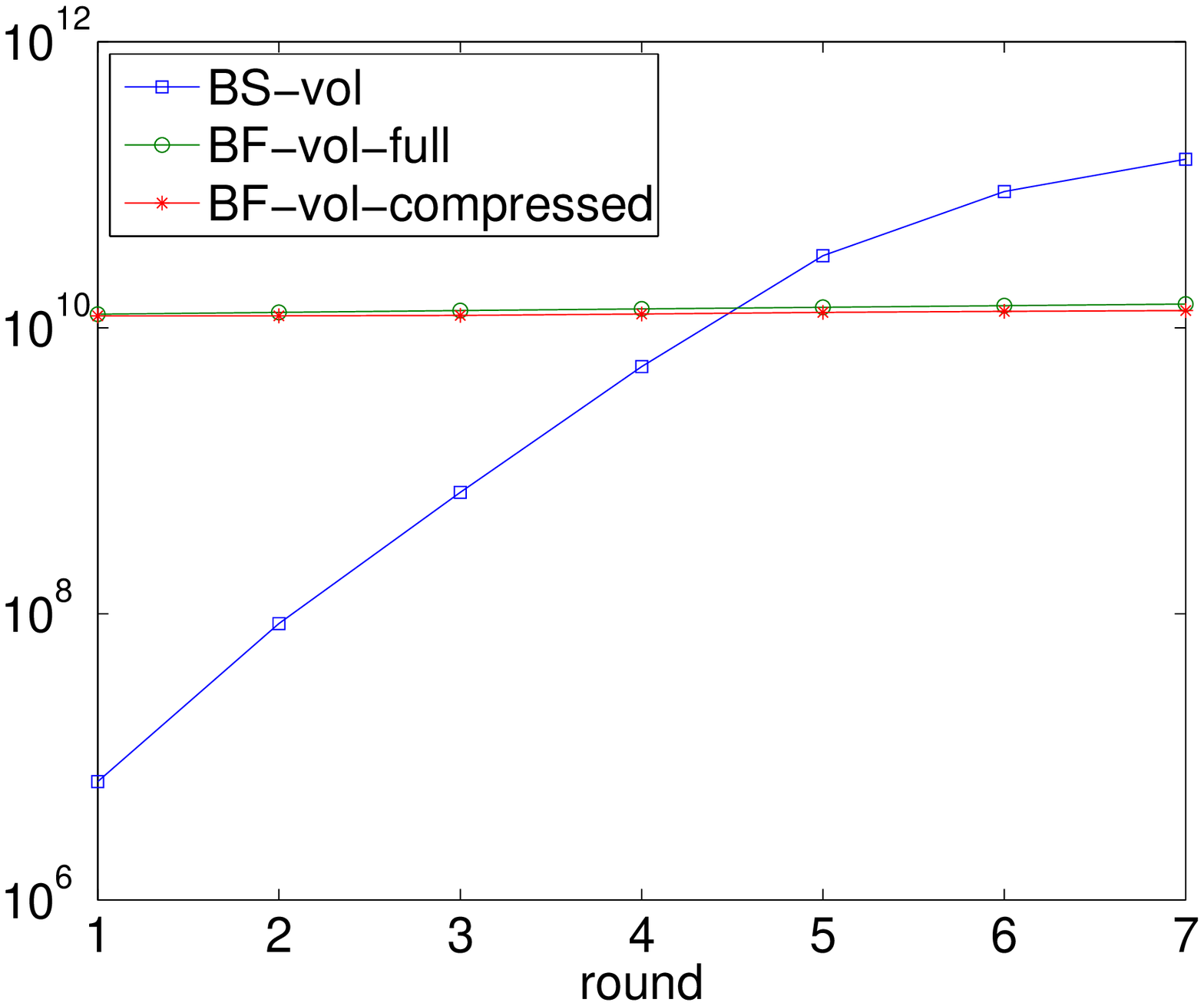, height=1.2in}
                  \setlength{\abovecaptionskip}{0pt}
                 \caption{$\alpha=0.75, \beta=1.0$, ER2}
                  \label{fig:comp-er-2-2}
         \end{subfigure}
         
         \begin{subfigure}[b]{0.22\textwidth}
                 \centering
                 \epsfig{file=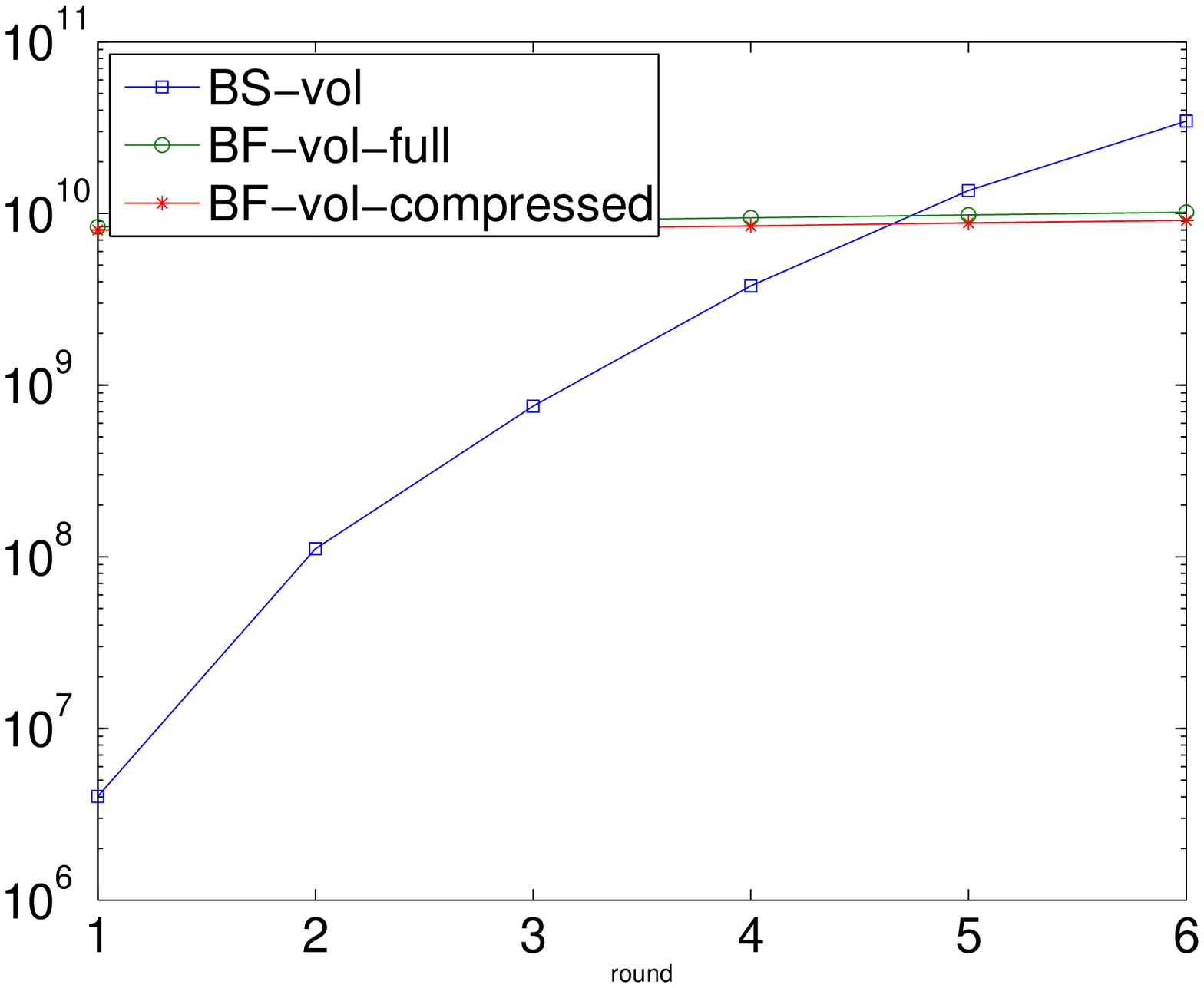, height=1.2in}
                 \setlength{\abovecaptionskip}{0pt}
                 \caption{$\alpha=0.25, \beta=0.5$, PL2}	
                 \label{fig:comp-pl-1-1}
         \end{subfigure}
         \hfill
         \begin{subfigure}[b]{0.22\textwidth}
                 \centering
                 \epsfig{file=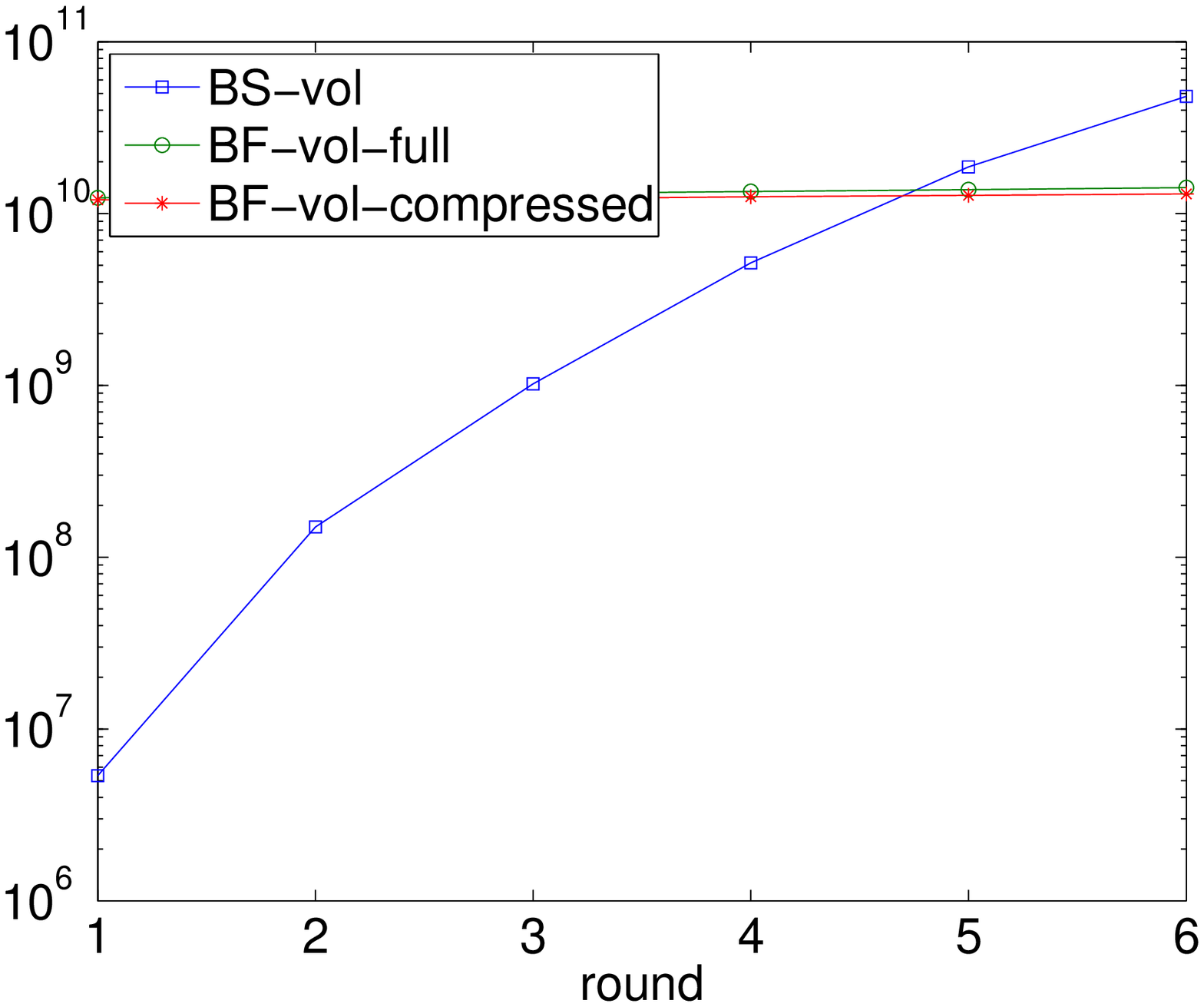, height=1.2in}
                 \setlength{\abovecaptionskip}{0pt}
                 \caption{$\alpha=0.25, \beta=1.0$, PL2}
                 \label{fig:comp-pl-1-2}
         \end{subfigure}
         \hfill         
         \begin{subfigure}[b]{0.22\textwidth}
                 \centering
                 \epsfig{file=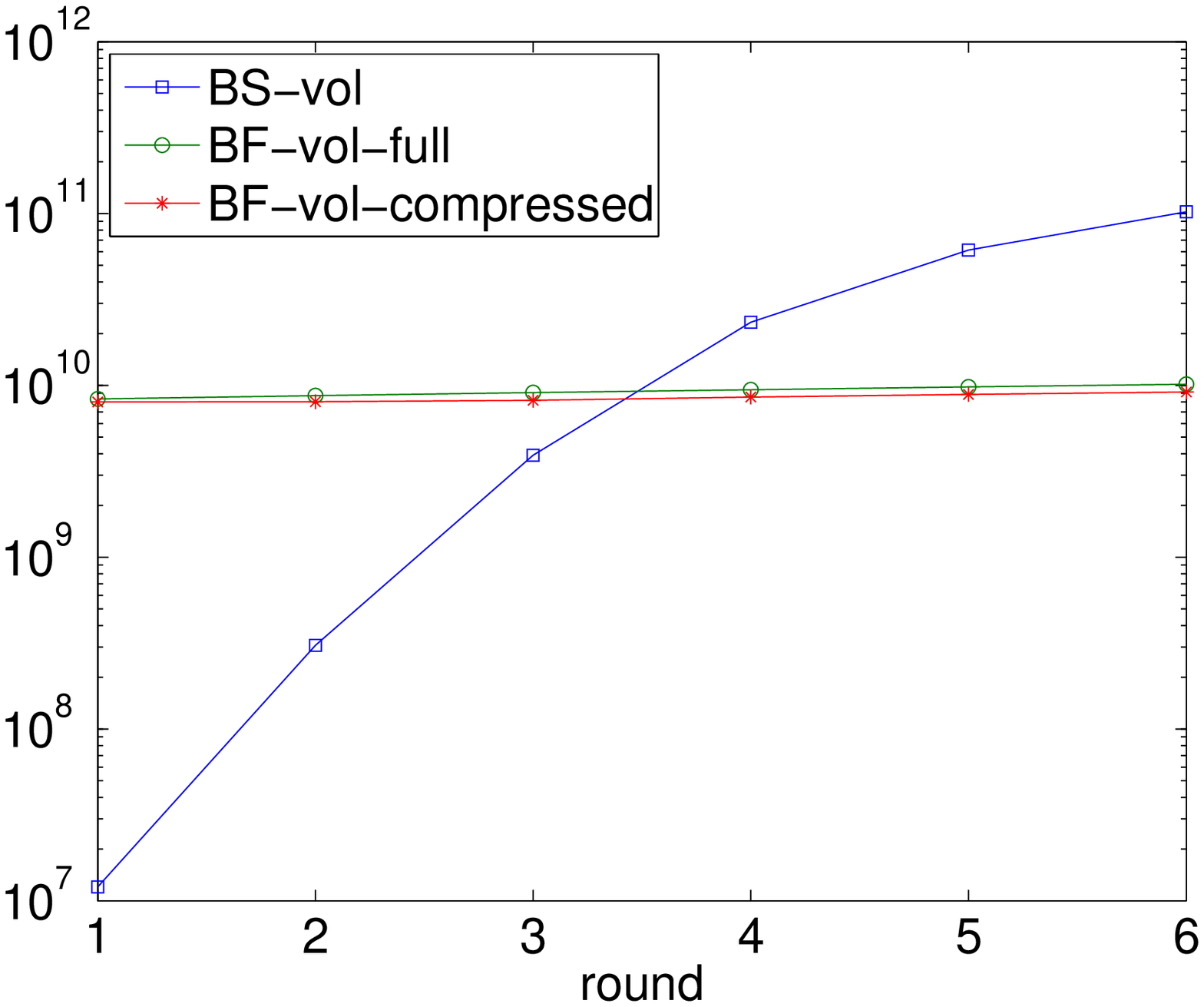, height=1.2in}
                 \setlength{\abovecaptionskip}{0pt}
                 \caption{$\alpha=0.75, \beta=0.5$, PL2}
                 \label{fig:comp-pl-2-1}
         \end{subfigure}
         \hfill         
         \begin{subfigure}[b]{0.22\textwidth}
                  \centering
                  \epsfig{file=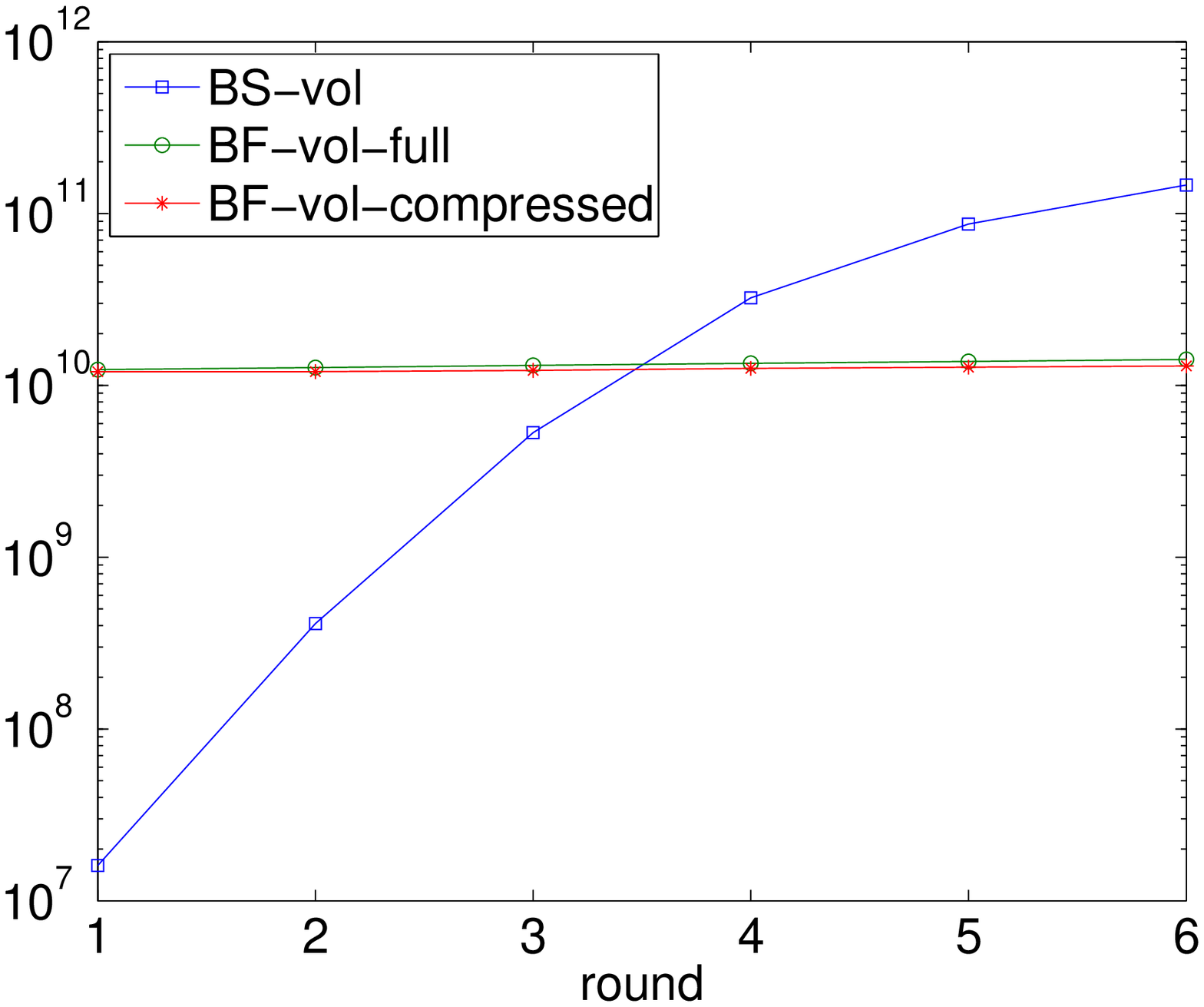, height=1.2in}
                  \setlength{\abovecaptionskip}{0pt}
                 \caption{$\alpha=0.75, \beta=1.0$, PL2}
                  \label{fig:comp-pl-2-2}
         \end{subfigure}        	
     \caption{Communication complexity. Y-axis is the total number of bytes transmitted among nodes (log-scale)}
     \label{fig:comp-volume}
 \end{figure*}

\begin{figure*}
 	\centering
         \begin{subfigure}[b]{0.22\textwidth}
                 \centering
                 \epsfig{file=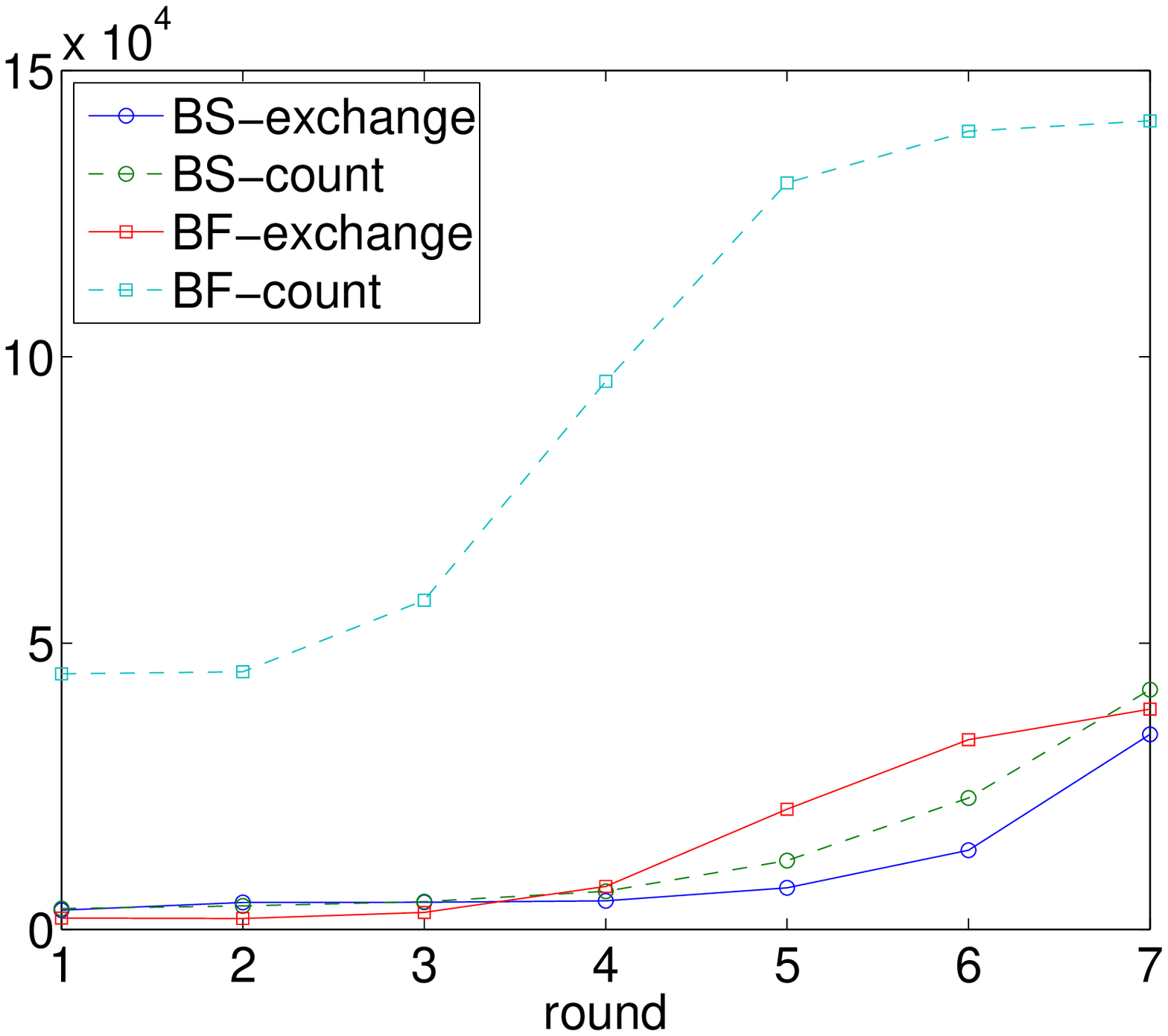, height=1.2in}
                 \setlength{\abovecaptionskip}{0pt}
                 \caption{$\alpha=0.25, \beta=0.5$, ER2}	
                 \label{fig:runtime-er-1-1}
         \end{subfigure}
         \hfill
         \begin{subfigure}[b]{0.22\textwidth}
                 \centering
                 \epsfig{file=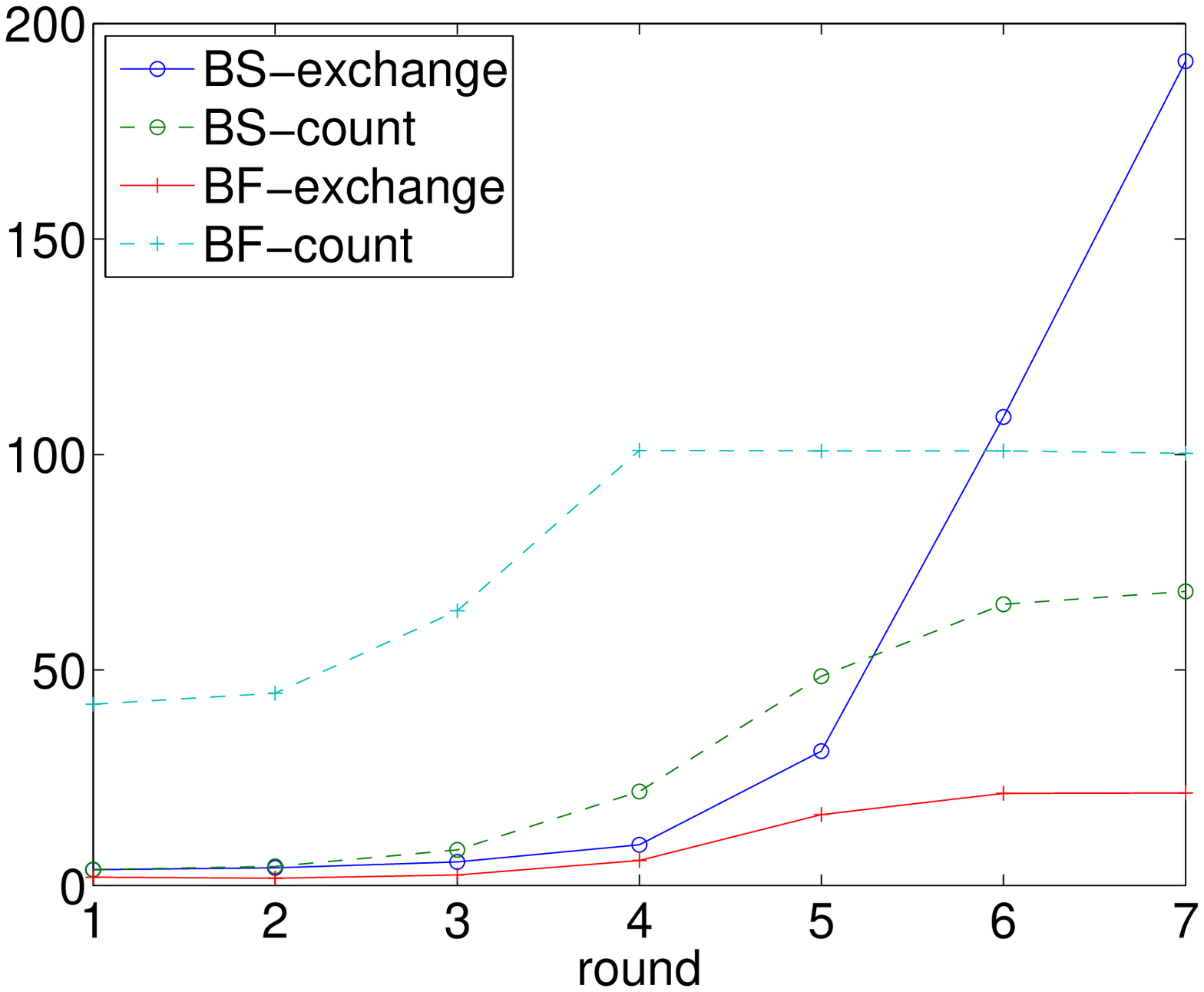, height=1.2in}
                 \setlength{\abovecaptionskip}{0pt}
                 \caption{$\alpha=0.5, \beta=0.5$, ER2}
                 \label{fig:runtime-er-1-2}
         \end{subfigure}
         \hfill         
         \begin{subfigure}[b]{0.22\textwidth}
                 \centering
                 \epsfig{file=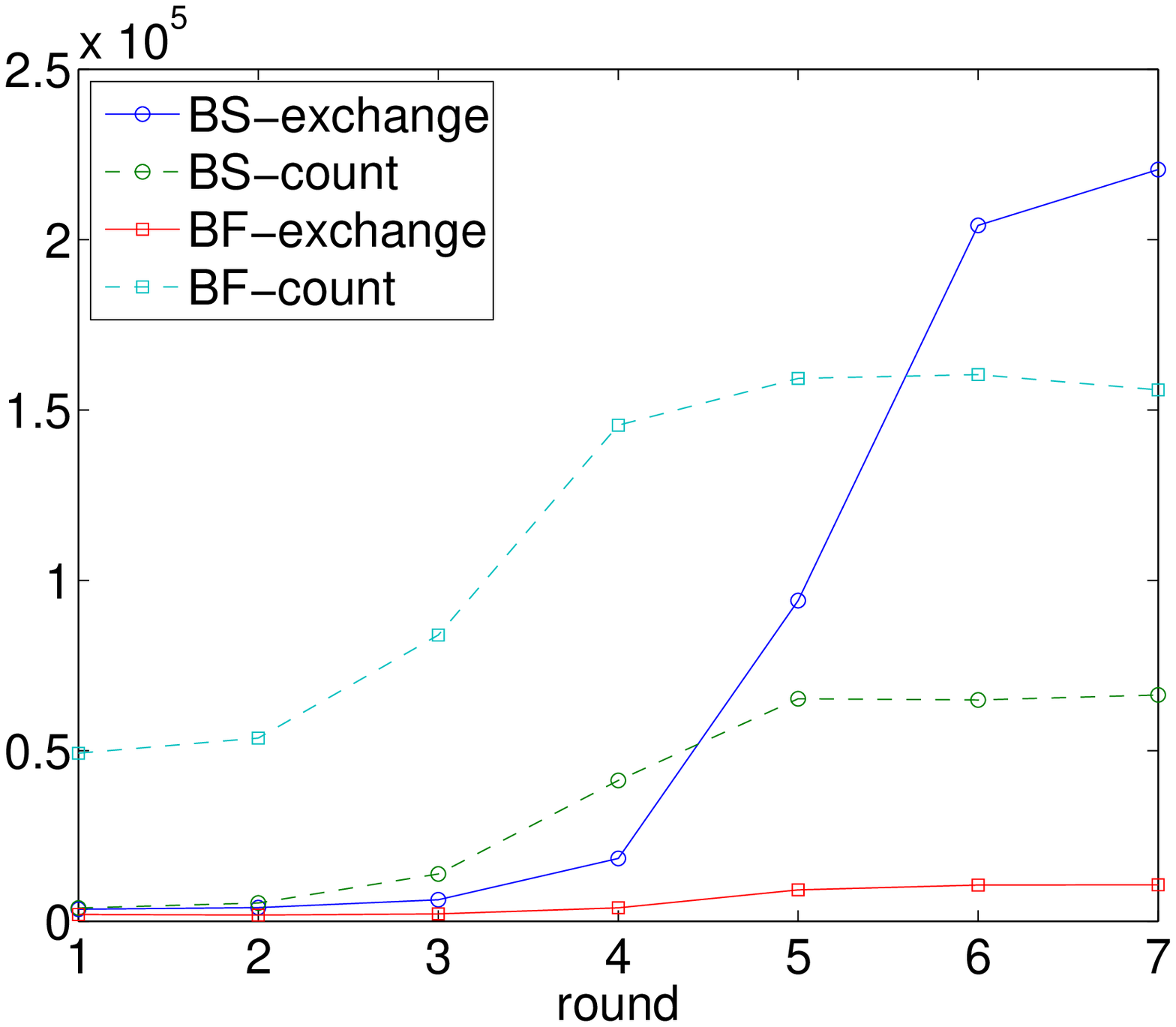, height=1.2in}
                 \setlength{\abovecaptionskip}{0pt}
                 \caption{$\alpha=0.75, \beta=0.5$, ER2}
                 \label{fig:runtime-er-2-1}
         \end{subfigure}
         \hfill         
         \begin{subfigure}[b]{0.22\textwidth}
                  \centering
                  \epsfig{file=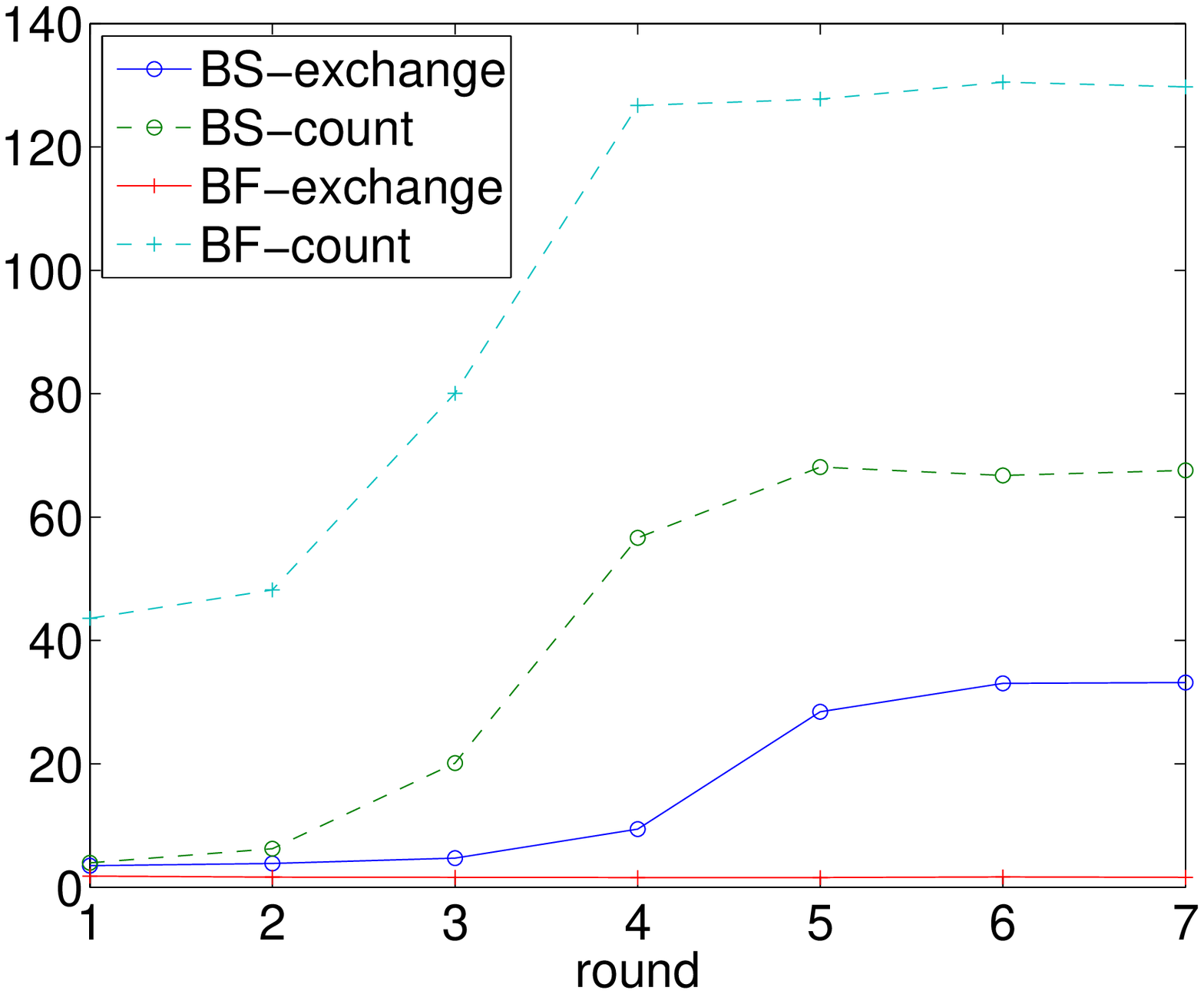, height=1.2in}
                  \setlength{\abovecaptionskip}{0pt}
                 \caption{$\alpha=1.0, \beta=0.5$, ER2}
                  \label{fig:runtime-er-2-2}
         \end{subfigure}
         
         \begin{subfigure}[b]{0.22\textwidth}
                 \centering
                 \epsfig{file=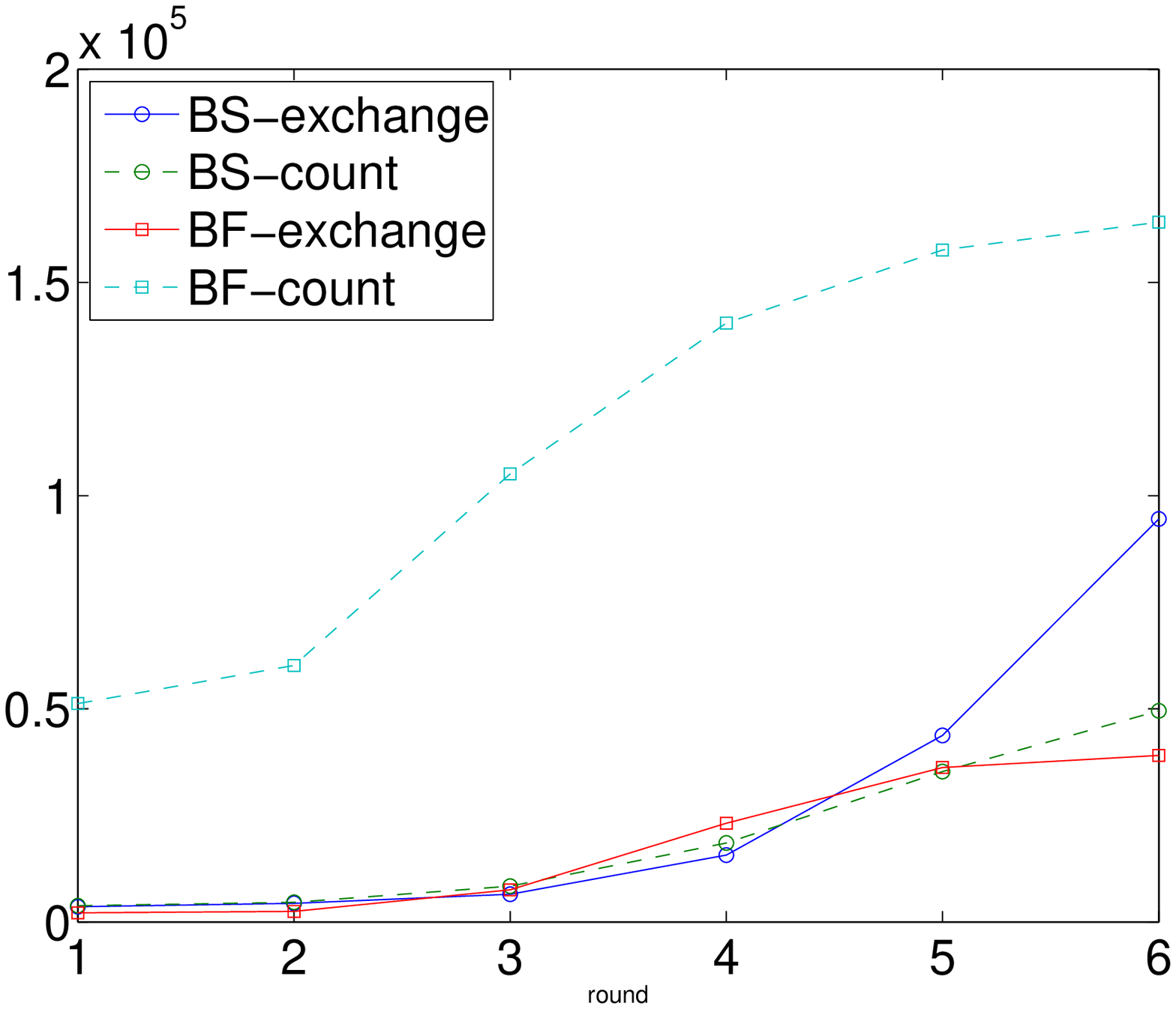, height=1.2in}
                 \setlength{\abovecaptionskip}{0pt}
                 \caption{$\alpha=0.25, \beta=0.5$, PL2}	
                 \label{fig:runtime-pl-1-1}
         \end{subfigure}
         \hfill
         \begin{subfigure}[b]{0.22\textwidth}
                 \centering
                 \epsfig{file=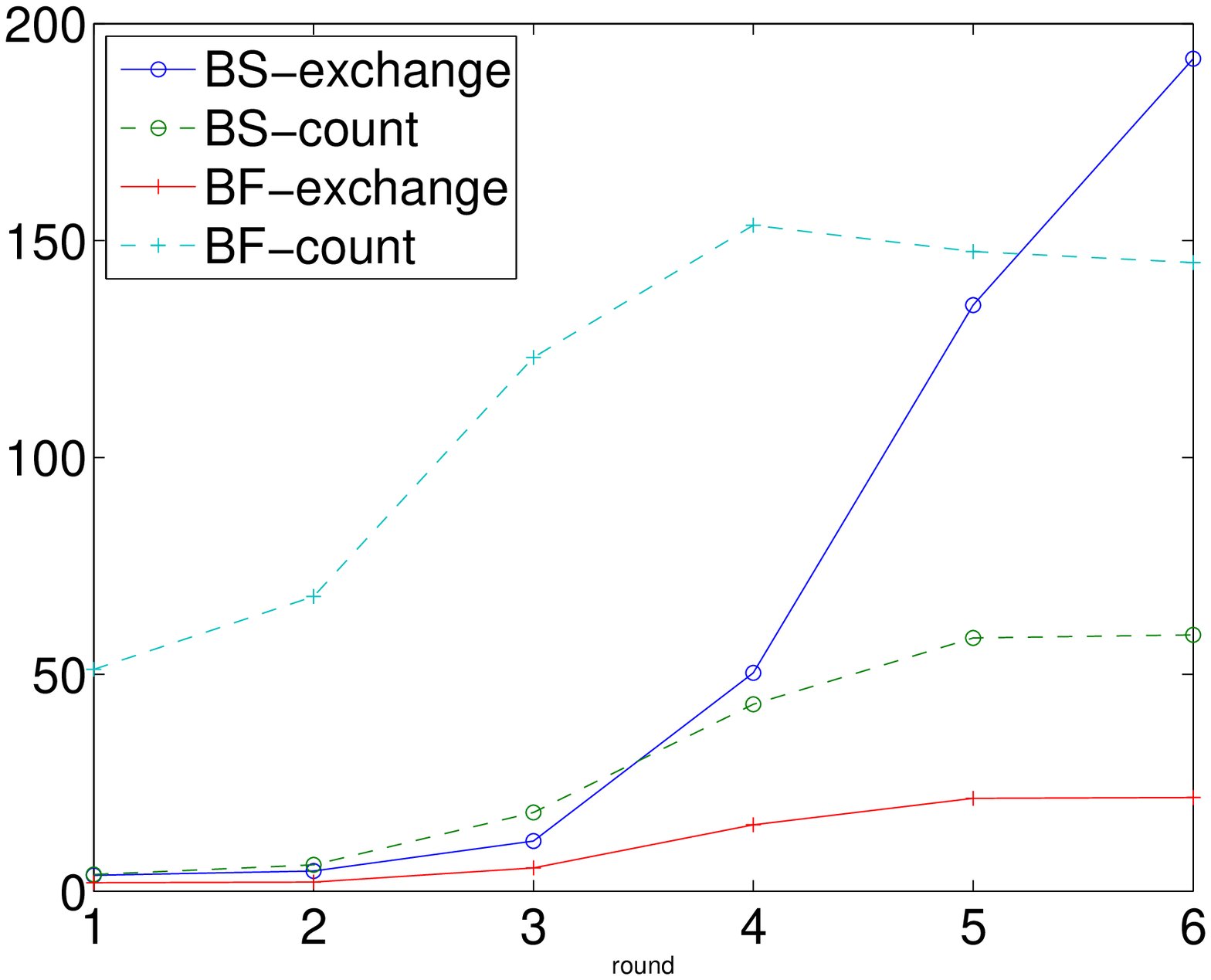, height=1.2in}
                 \setlength{\abovecaptionskip}{0pt}
                 \caption{$\alpha=0.5, \beta=0.5$, PL2}
                 \label{fig:runtime-pl-1-2}
         \end{subfigure}
         \hfill         
         \begin{subfigure}[b]{0.22\textwidth}
                 \centering
                 \epsfig{file=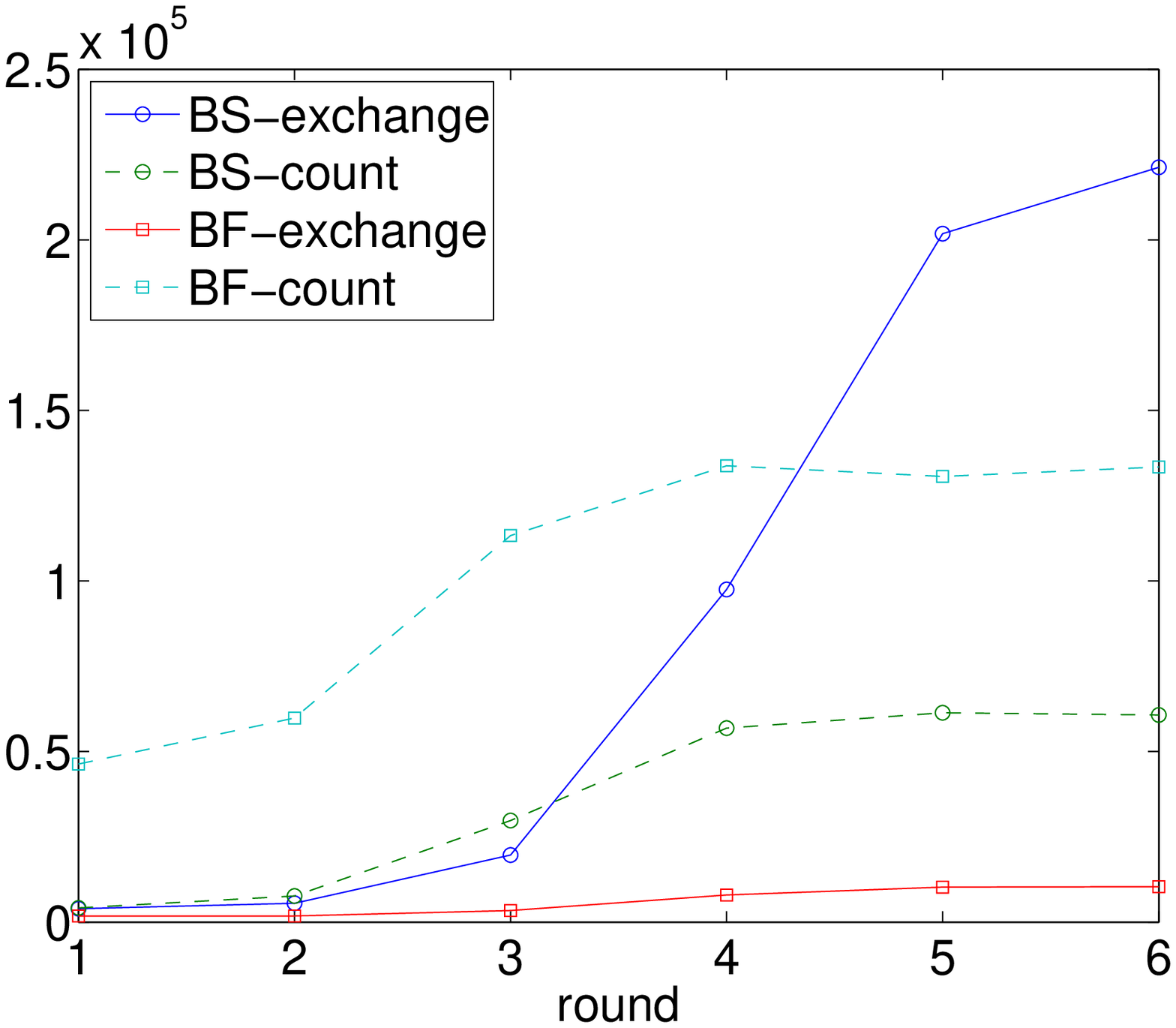, height=1.2in}
                 \setlength{\abovecaptionskip}{0pt}
                 \caption{$\alpha=0.75, \beta=0.5$, PL2}
                 \label{fig:runtime-pl-2-1}
         \end{subfigure}
         \hfill         
         \begin{subfigure}[b]{0.22\textwidth}
                  \centering
                  \epsfig{file=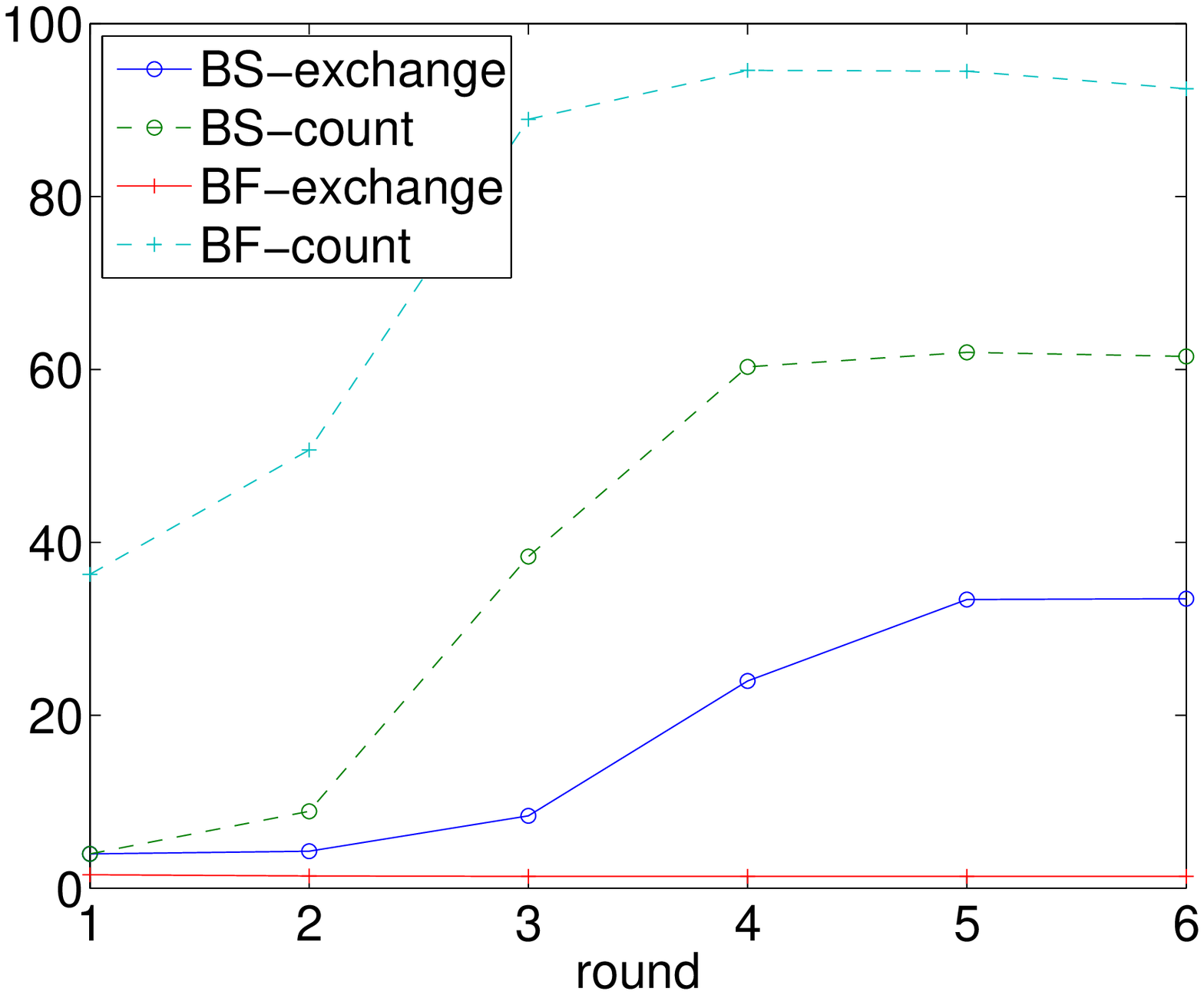, height=1.2in}
                  \setlength{\abovecaptionskip}{0pt}
                 \caption{$\alpha=1.0, \beta=0.5$, PL2}
                  \label{fig:runtime-pl-2-2}
         \end{subfigure}        	
     \caption{Total simulation runtime of all nodes (in millisecond)}
     \label{fig:runtime}
 \end{figure*}

We compare the communication complexity of Baseline and Bloom Filter schemes. Fig. \ref{fig:comp-volume} reports the number of bytes transmitted among nodes after each round in Baseline and Bloom Filter (with or without compression). 
Baseline scheme stores links in clear form, so it incurs exponential communication complexity. As discussed in Section \ref{subsec:bf-motivation}, we assume that each node ID cost 4 bytes and a link list of length $l$ may be stored compactly in $4l$ bytes. 
Bloom Filter uses constant-sized bit arrays, so its communication cost is constant too. However, each node running Bloom Filter scheme has to download the full noisy list of  $(1+2\beta)M$ links after the final round to find which links are contained in its bit array. The number of bytes in the download step is $4N(1+2\beta)M$ bytes for $N$ nodes. The communication cost of the download step dominates that of bit array exchange. Using bit array compression (Section \ref{subsec:bf-compress}), Bloom Filter scheme reduces the message size a little bit, especially at early rounds when a large part of bit arrays are zero bits. For $\alpha = 0.75$, Bloom Filter scheme saves the communication cost in the last three rounds in both ER2 and PL2. For $\alpha = 0.25$, it is worse than Baseline in all rounds (except the final round) on ER2 and in the first four rounds on PL2.

In Fig. \ref{fig:runtime}, we compare the runtime of Baseline and Bloom Filter simulations in a single PC. In each round, each node updates its link set (\texttt{count} operation) by aggregating noisy link lists from its neighbors. Then, each node prepares (\texttt{exchange} operation), for the next round, new noisy link lists sampled from its link set.
At $\alpha < 1$, the exchange operations cost an increasing time as more rounds are considered. Higher $\alpha$ makes the link sampling slower. Only at $\alpha = 1$, we have fast exchange operations. In particular, the exchange runtime of Bloom Filter scheme is constant for $\alpha = 1$ and is an increasing function of round for $\alpha < 1$ due to bit erasure operations. The count operation of Bloom Filter dominates that of Baseline because each node has to hash the full link set to recover its noisy link set at each round.

\begin{figure*}[t!]
 	\centering
         \begin{subfigure}[b]{0.22\textwidth}
                 \centering
                 \epsfig{file=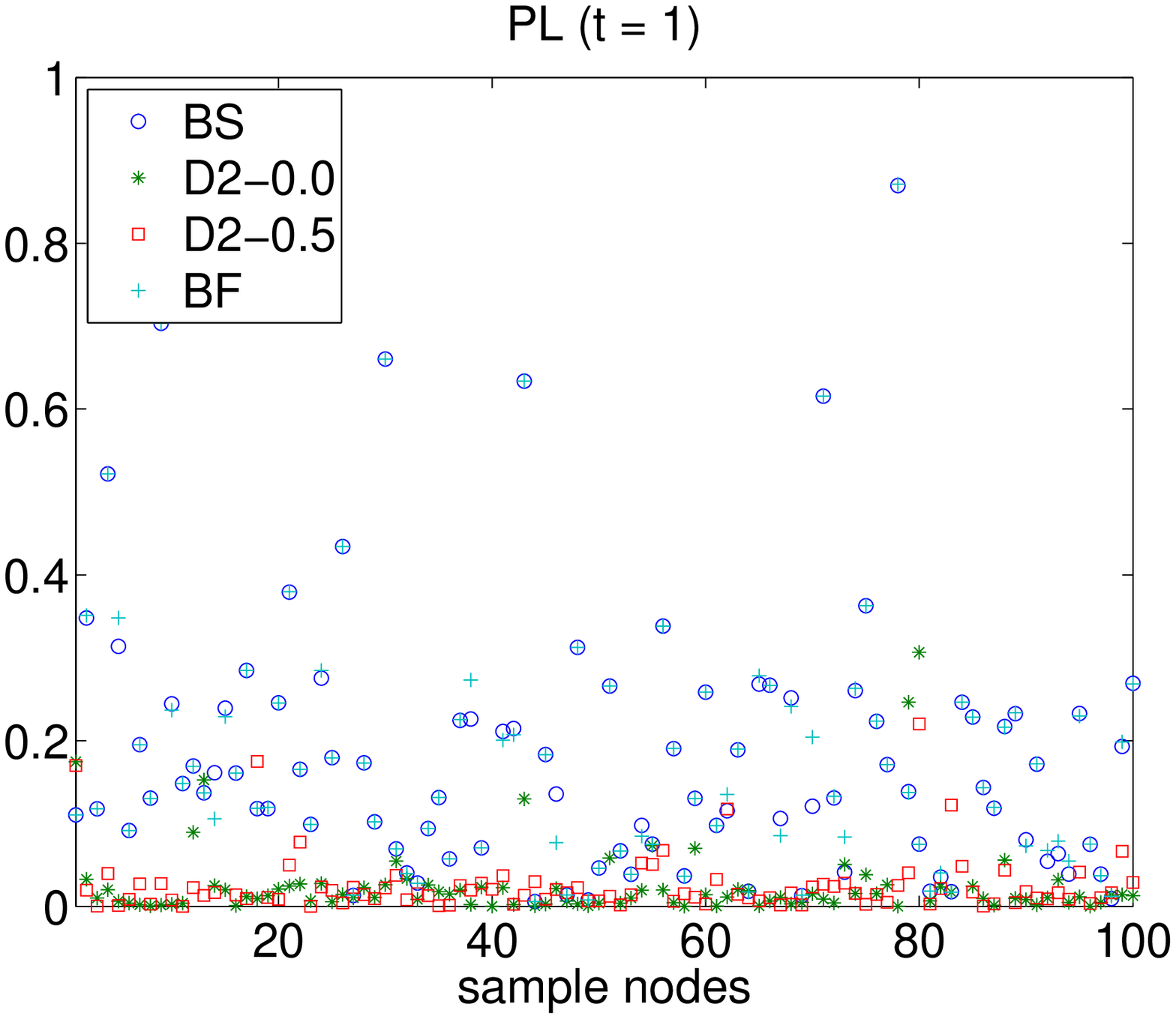, height=1.2in}
                 \setlength{\abovecaptionskip}{0pt}
                 \caption{}	
                 \label{fig:util-PL-1}
         \end{subfigure}
         \hfill
         \begin{subfigure}[b]{0.22\textwidth}
                 \centering
                 \epsfig{file=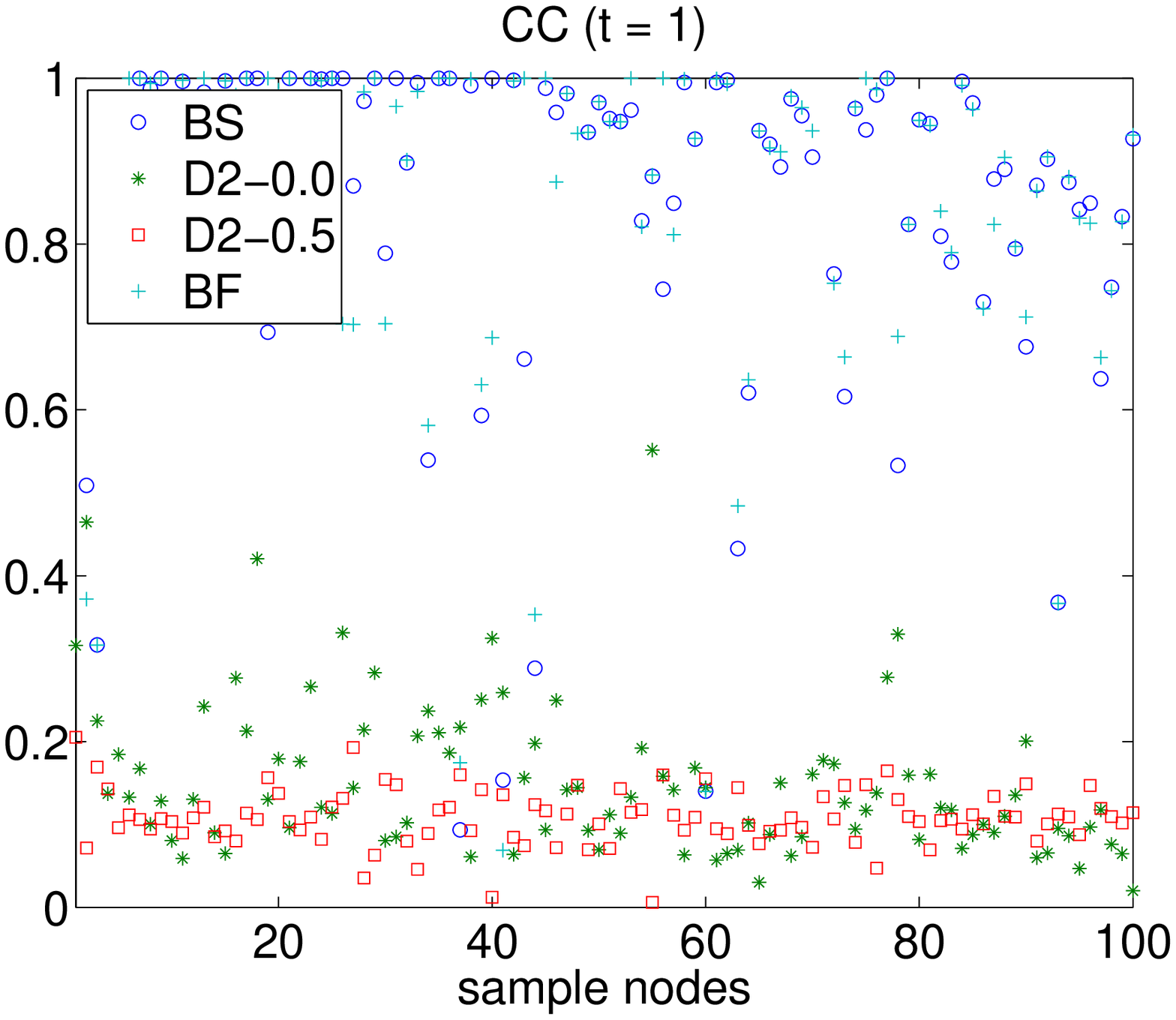, height=1.2in}
                 \setlength{\abovecaptionskip}{0pt}
                 \caption{}
                 \label{fig:util-CC-1}
         \end{subfigure}
         \hfill         
         \begin{subfigure}[b]{0.22\textwidth}
                 \centering
                 \epsfig{file=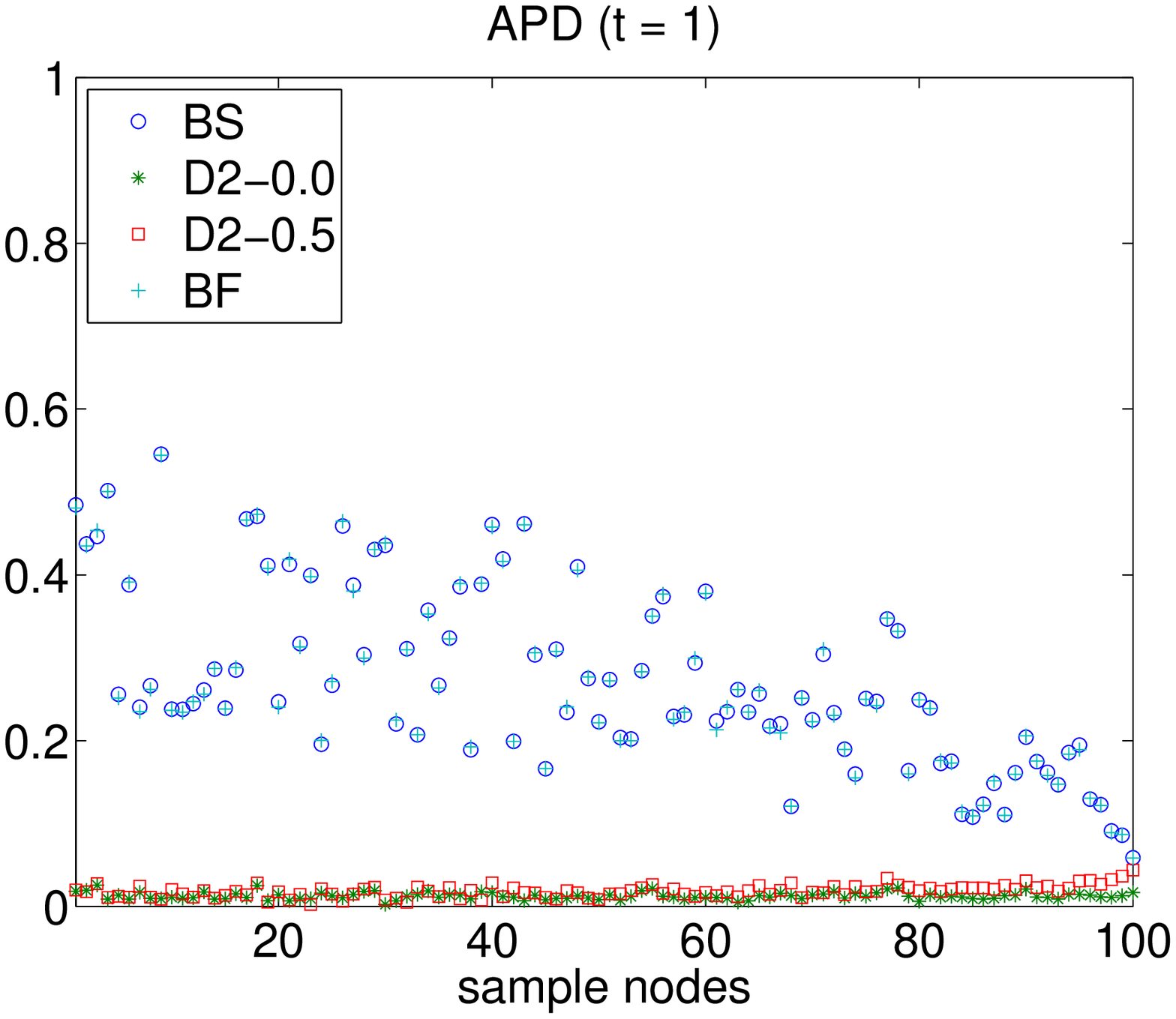, height=1.2in}
                 \setlength{\abovecaptionskip}{0pt}
                 \caption{}
                 \label{fig:util-APD-1}
         \end{subfigure}
         \hfill         
         \begin{subfigure}[b]{0.22\textwidth}
                  \centering
                  \epsfig{file=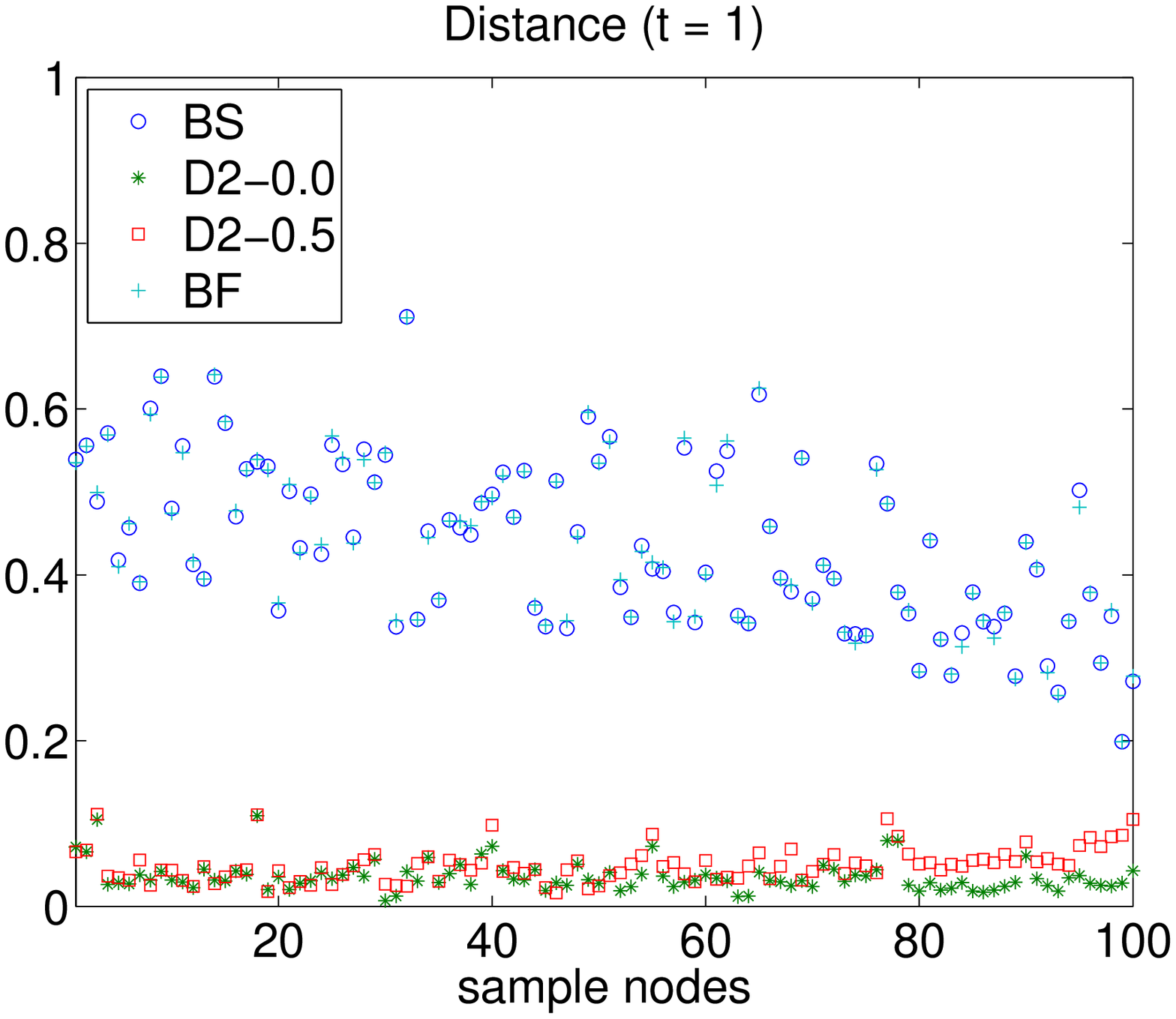, height=1.2in}
                  \setlength{\abovecaptionskip}{0pt}
                 \caption{}
                  \label{fig:util-Dist-1}
         \end{subfigure}        	

         \begin{subfigure}[b]{0.22\textwidth}
                 \centering
                 \epsfig{file=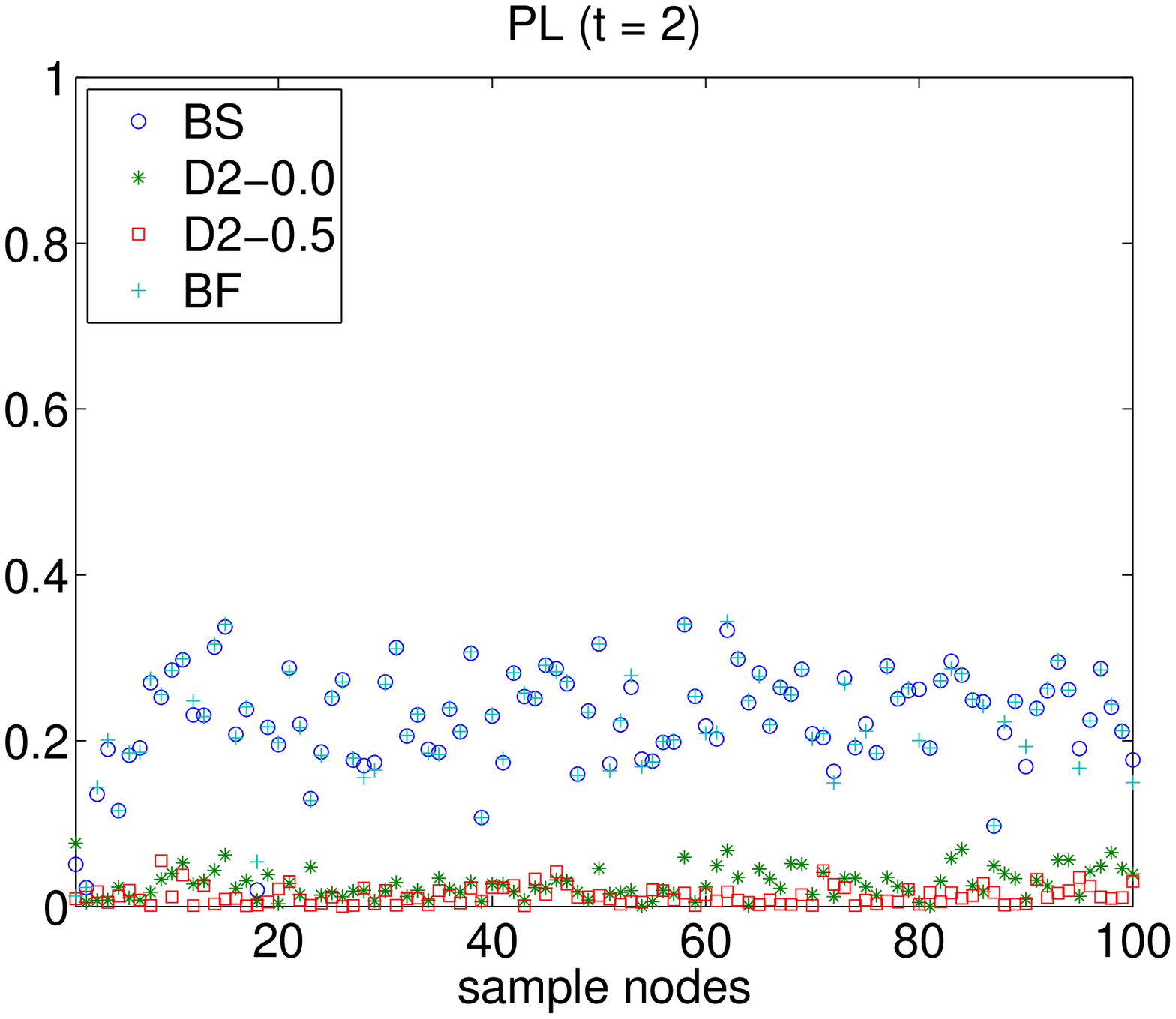, height=1.2in}
                 \setlength{\abovecaptionskip}{0pt}
                 \caption{}	
                 \label{fig:util-PL-2}
         \end{subfigure}
         \hfill
         \begin{subfigure}[b]{0.22\textwidth}
                 \centering
                 \epsfig{file=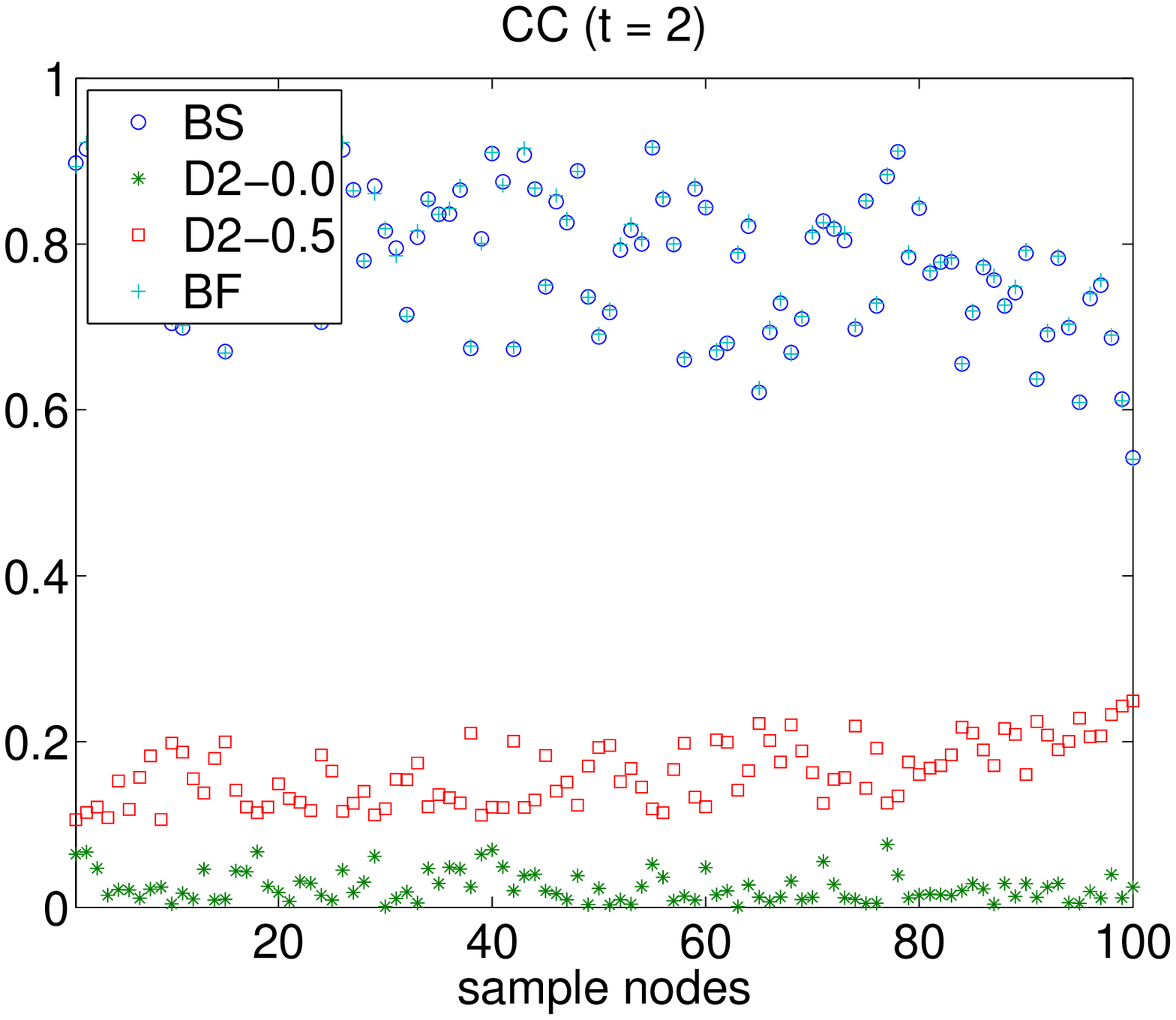, height=1.2in}
                 \setlength{\abovecaptionskip}{0pt}
                 \caption{}
                 \label{fig:util-CC-2}
         \end{subfigure}
         \hfill         
         \begin{subfigure}[b]{0.22\textwidth}
                 \centering
                 \epsfig{file=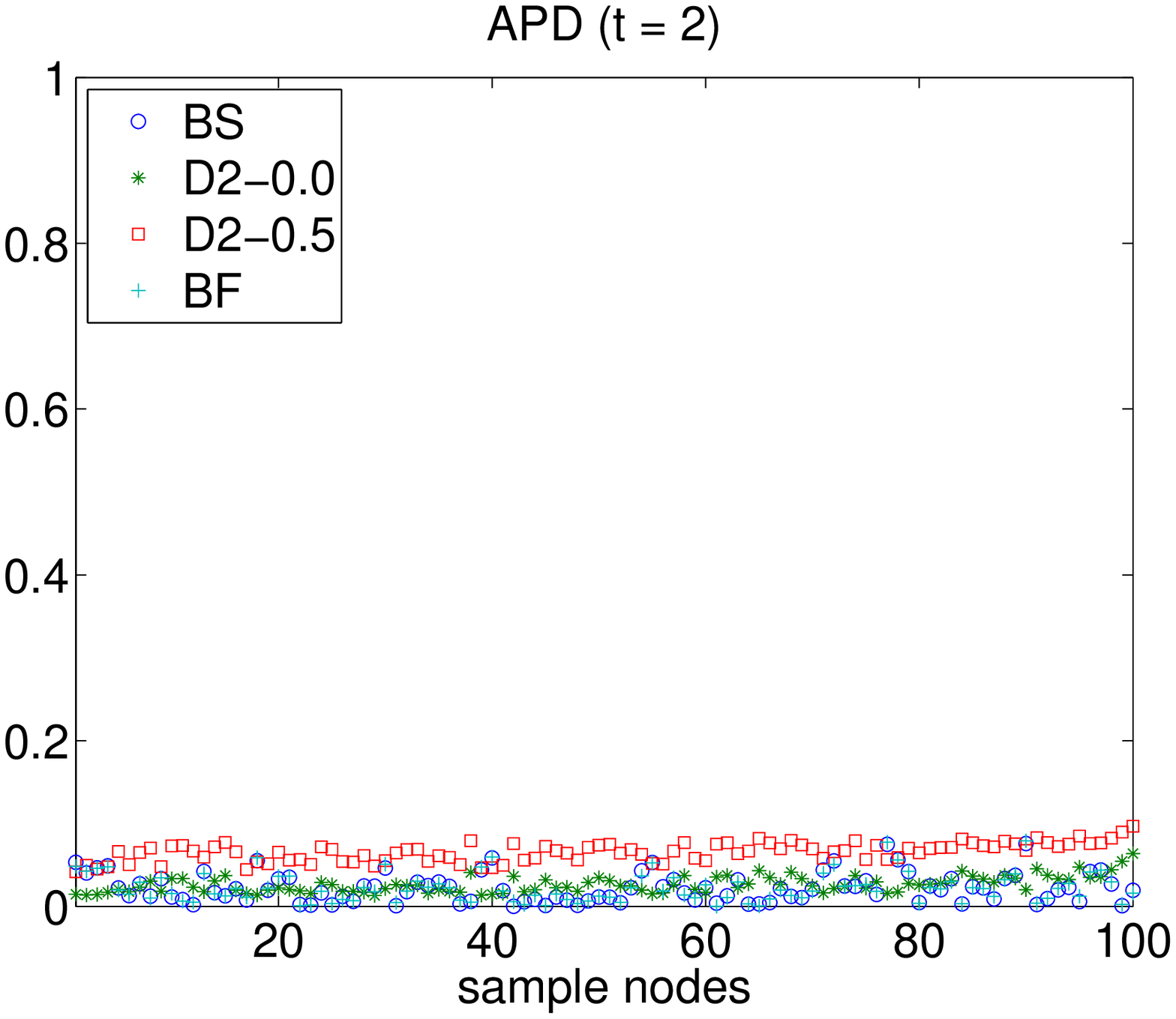, height=1.2in}
                 \setlength{\abovecaptionskip}{0pt}
                 \caption{}
                 \label{fig:util-APD-2}
         \end{subfigure}
         \hfill         
         \begin{subfigure}[b]{0.22\textwidth}
                  \centering
                  \epsfig{file=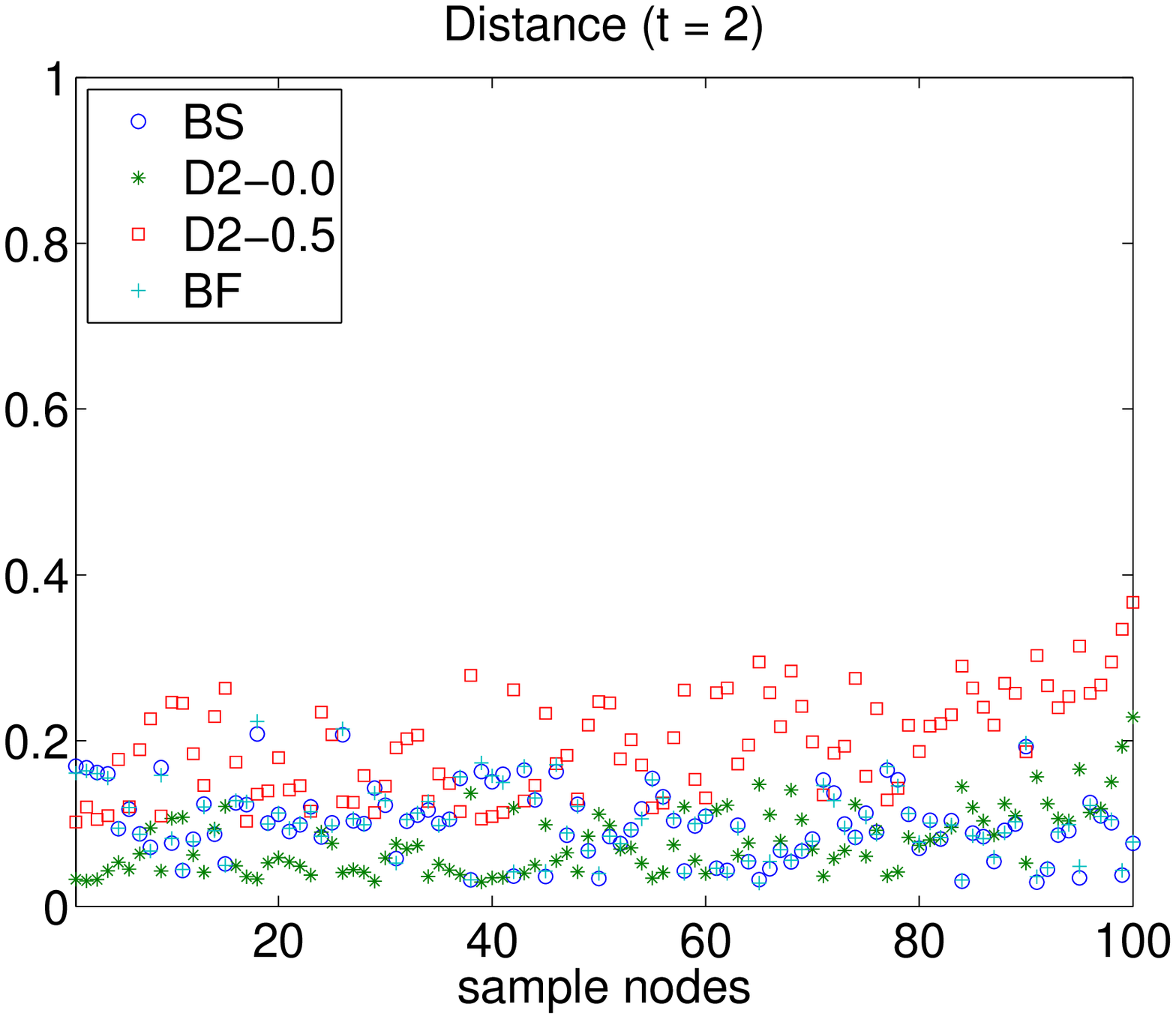, height=1.2in}
                  \setlength{\abovecaptionskip}{0pt}
                 \caption{}
                  \label{fig:util-Dist-2}
         \end{subfigure} 
                  
         \begin{subfigure}[b]{0.22\textwidth}
                 \centering
                 \epsfig{file=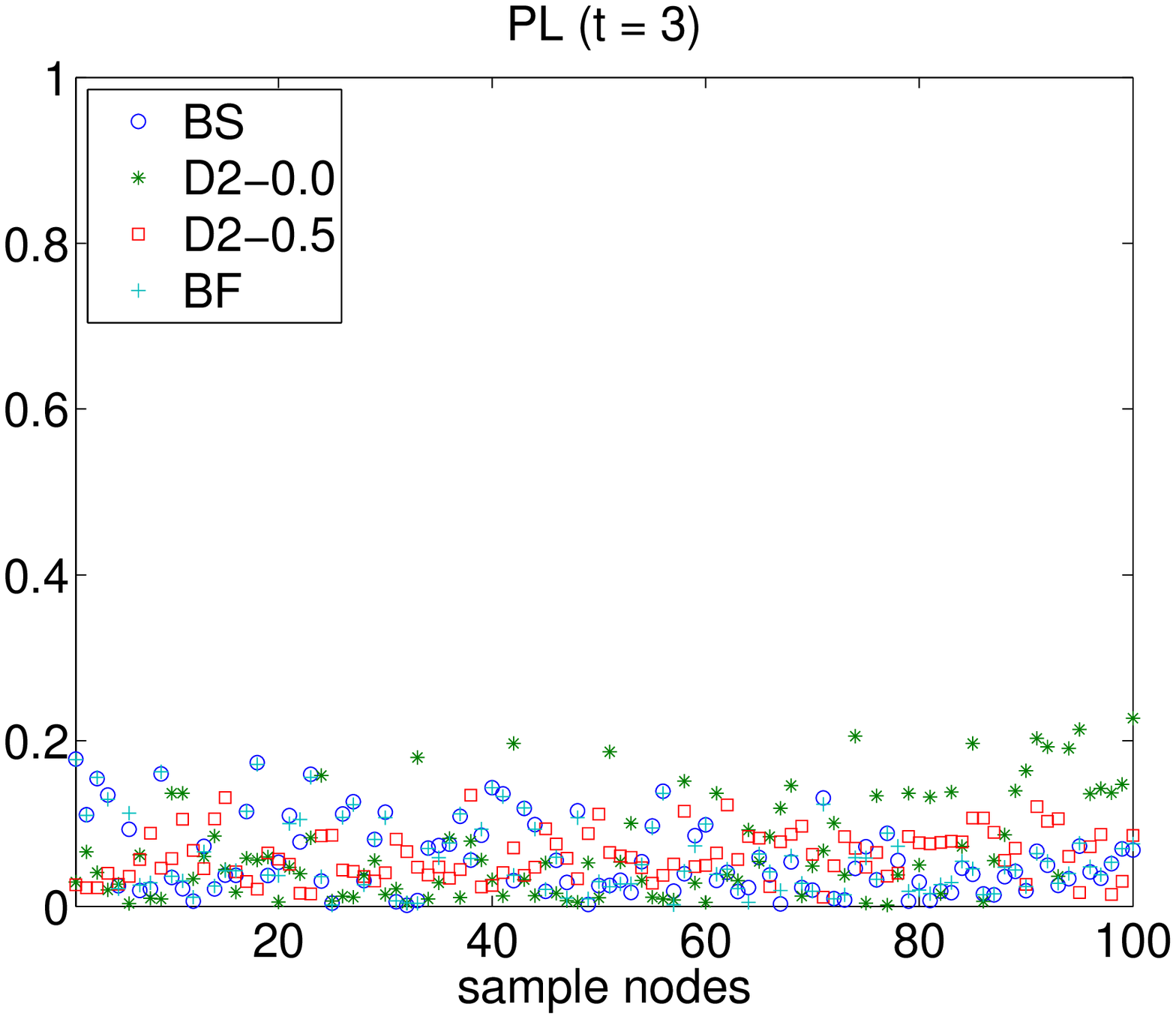, height=1.2in}
                 \setlength{\abovecaptionskip}{0pt}
                 \caption{}	
                 \label{fig:util-PL-3}
         \end{subfigure}
         \hfill
         \begin{subfigure}[b]{0.22\textwidth}
                 \centering
                 \epsfig{file=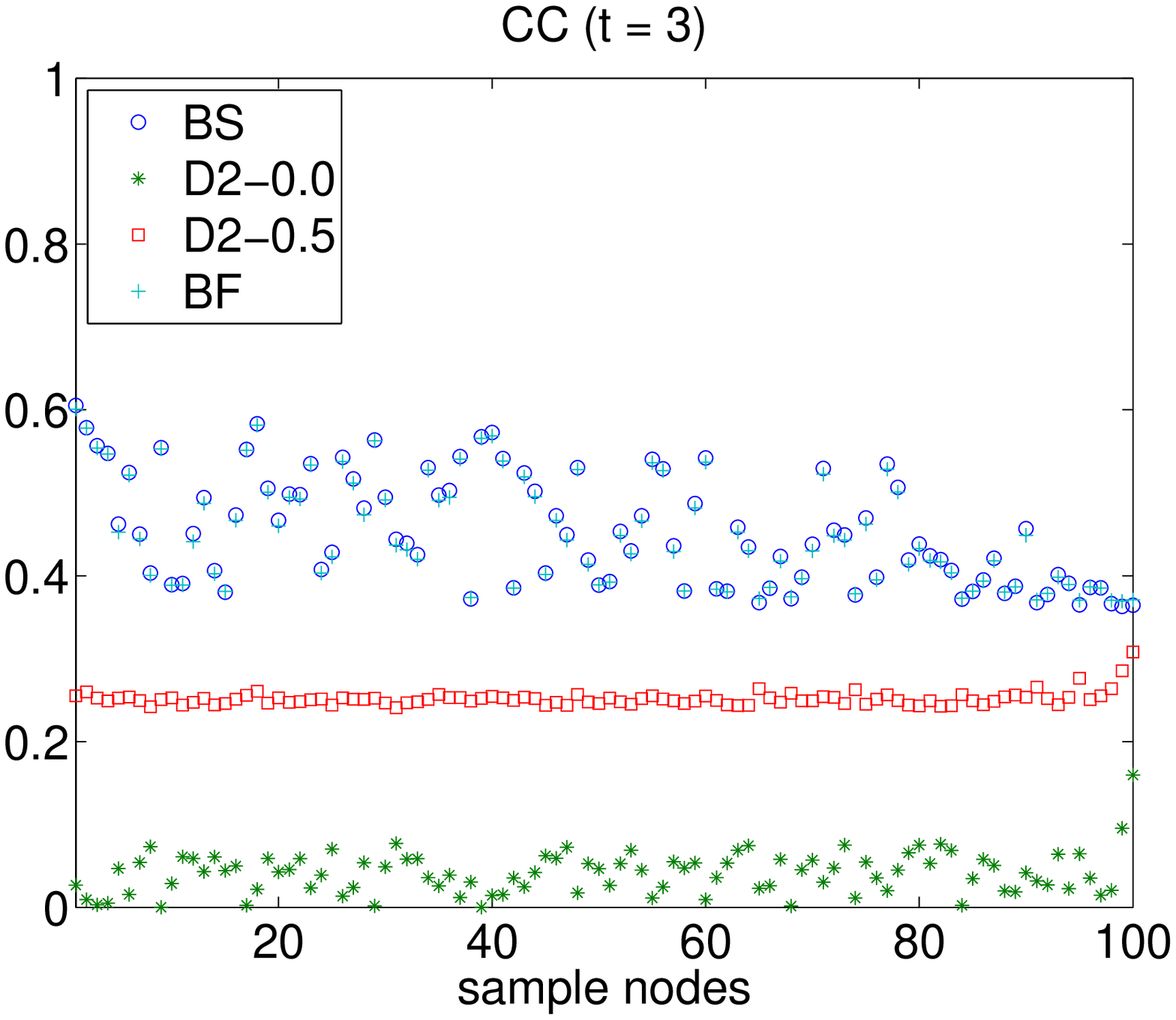, height=1.2in}
                 \setlength{\abovecaptionskip}{0pt}
                 \caption{}
                 \label{fig:util-CC-3}
         \end{subfigure}
         \hfill         
         \begin{subfigure}[b]{0.22\textwidth}
                 \centering
                 \epsfig{file=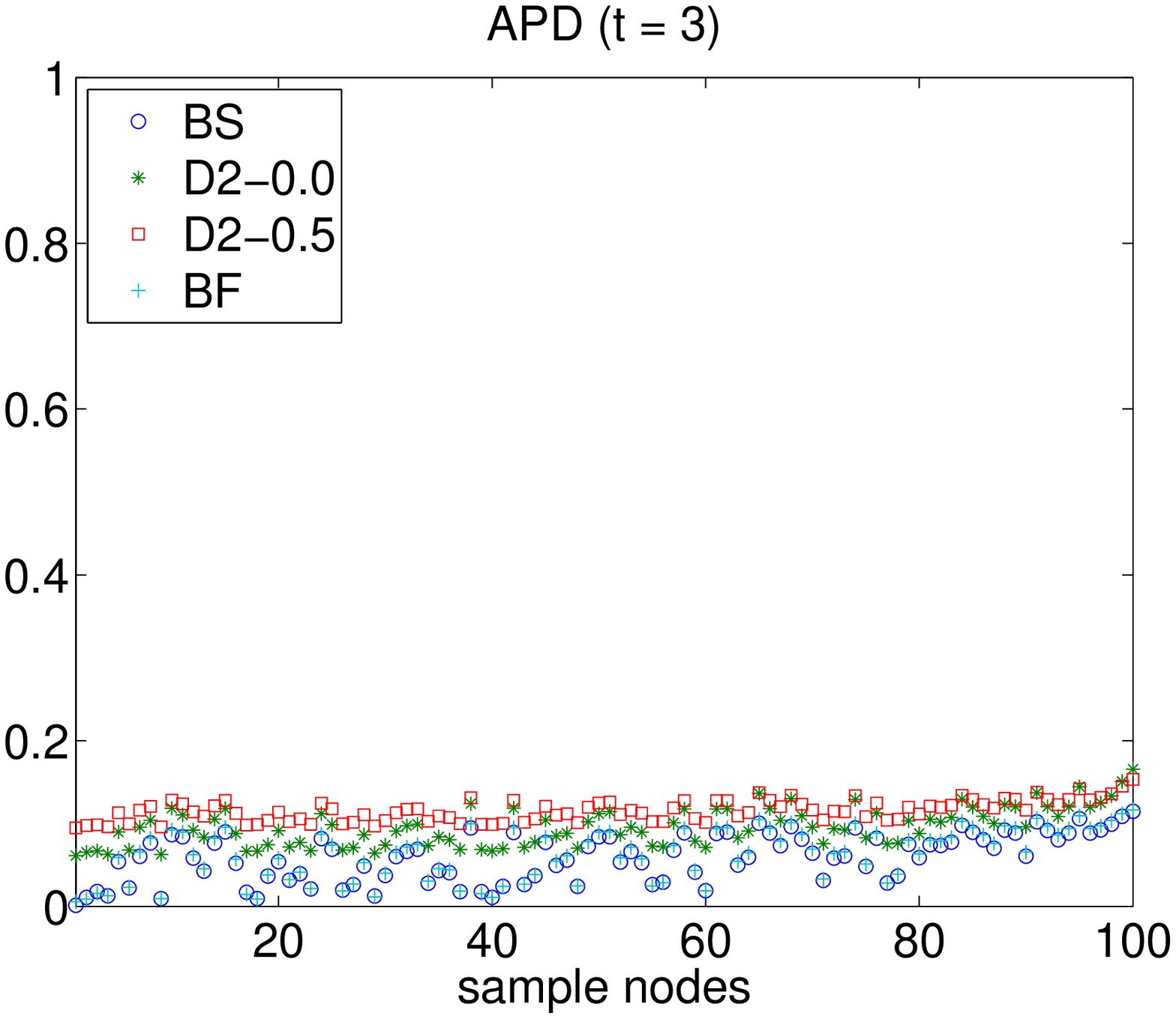, height=1.2in}
                 \setlength{\abovecaptionskip}{0pt}
                 \caption{}
                 \label{fig:util-APD-3}
         \end{subfigure}
         \hfill         
         \begin{subfigure}[b]{0.22\textwidth}
                  \centering
                  \epsfig{file=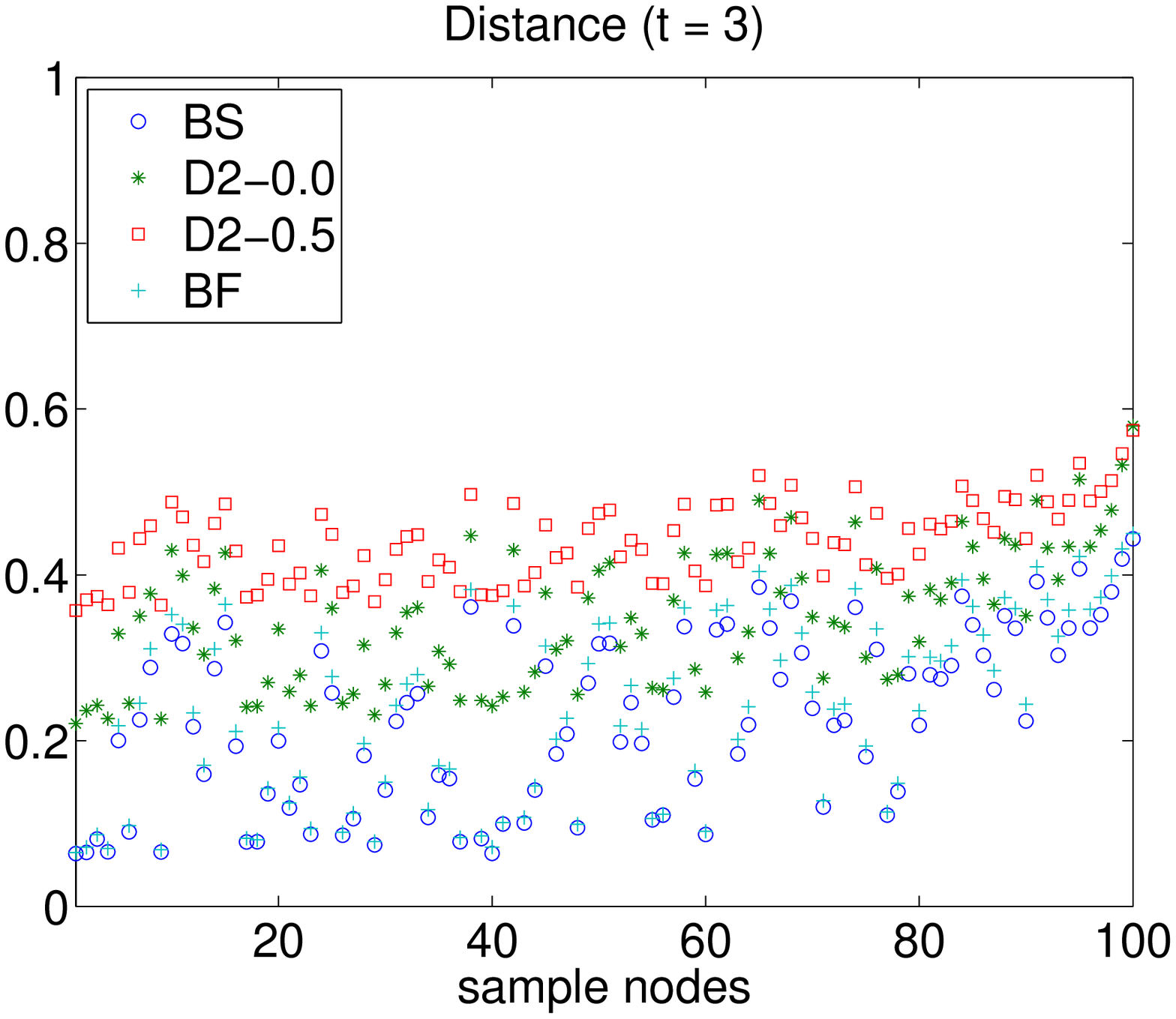, height=1.2in}
                  \setlength{\abovecaptionskip}{0pt}
                 \caption{}
                  \label{fig:util-Dist-3}
         \end{subfigure}         

         \begin{subfigure}[b]{0.22\textwidth}
                 \centering
                 \epsfig{file=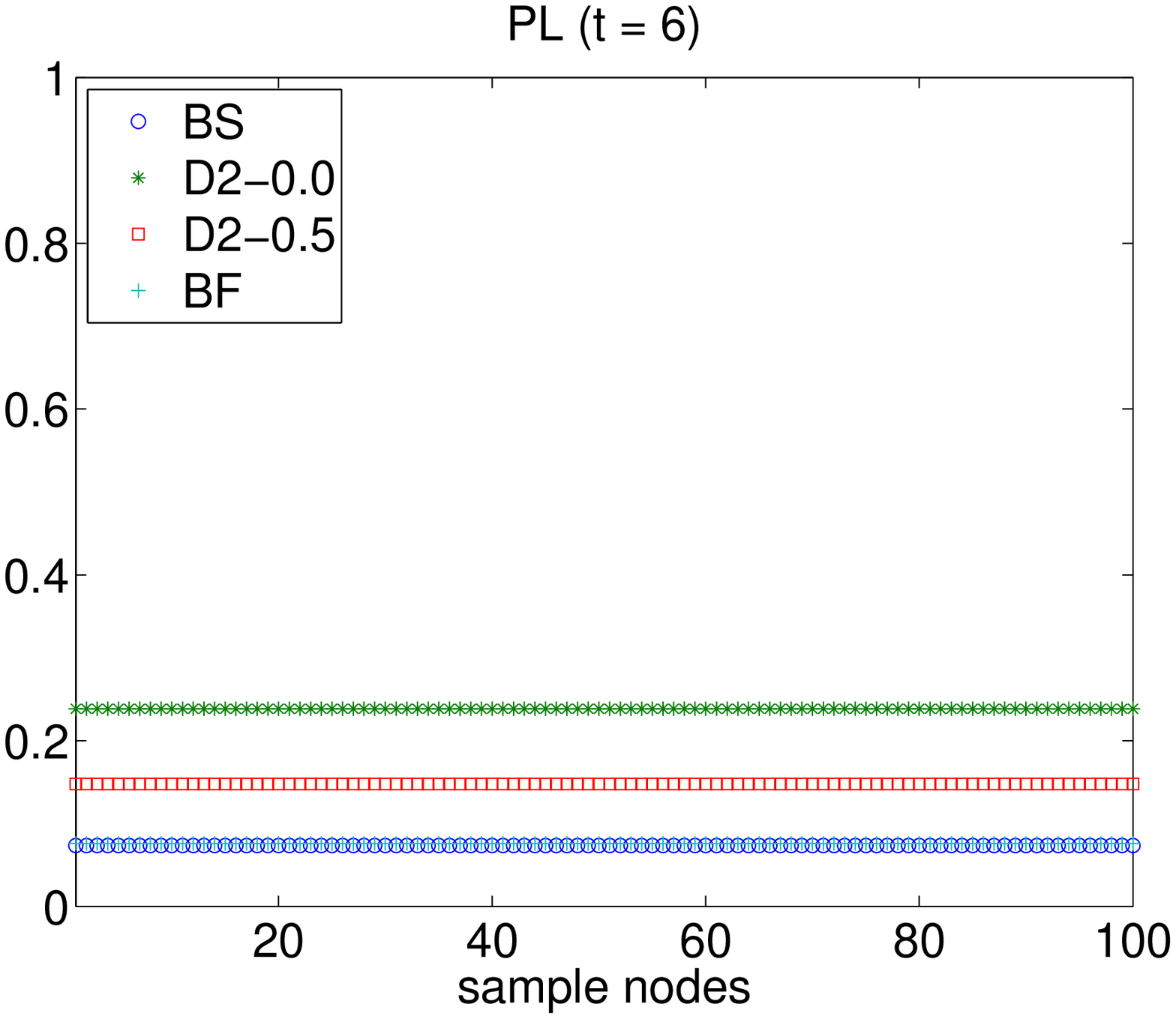, height=1.2in}
                 \setlength{\abovecaptionskip}{0pt}
                 \caption{}	
                 \label{fig:util-PL-6}
         \end{subfigure}
         \hfill
         \begin{subfigure}[b]{0.22\textwidth}
                 \centering
                 \epsfig{file=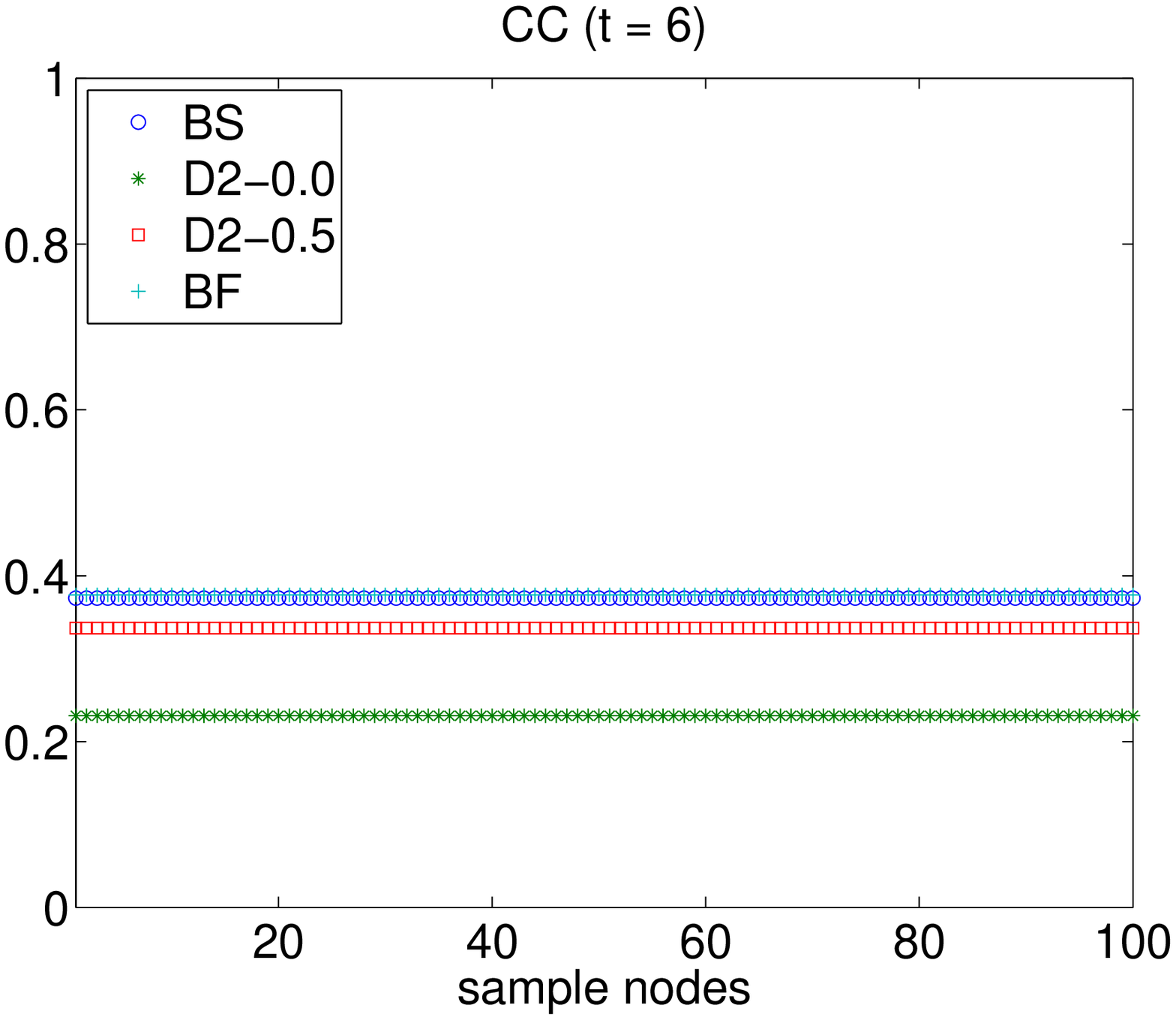, height=1.2in}
                 \setlength{\abovecaptionskip}{0pt}
                 \caption{}
                 \label{fig:util-CC-6}
         \end{subfigure}
         \hfill         
         \begin{subfigure}[b]{0.22\textwidth}
                 \centering
                 \epsfig{file=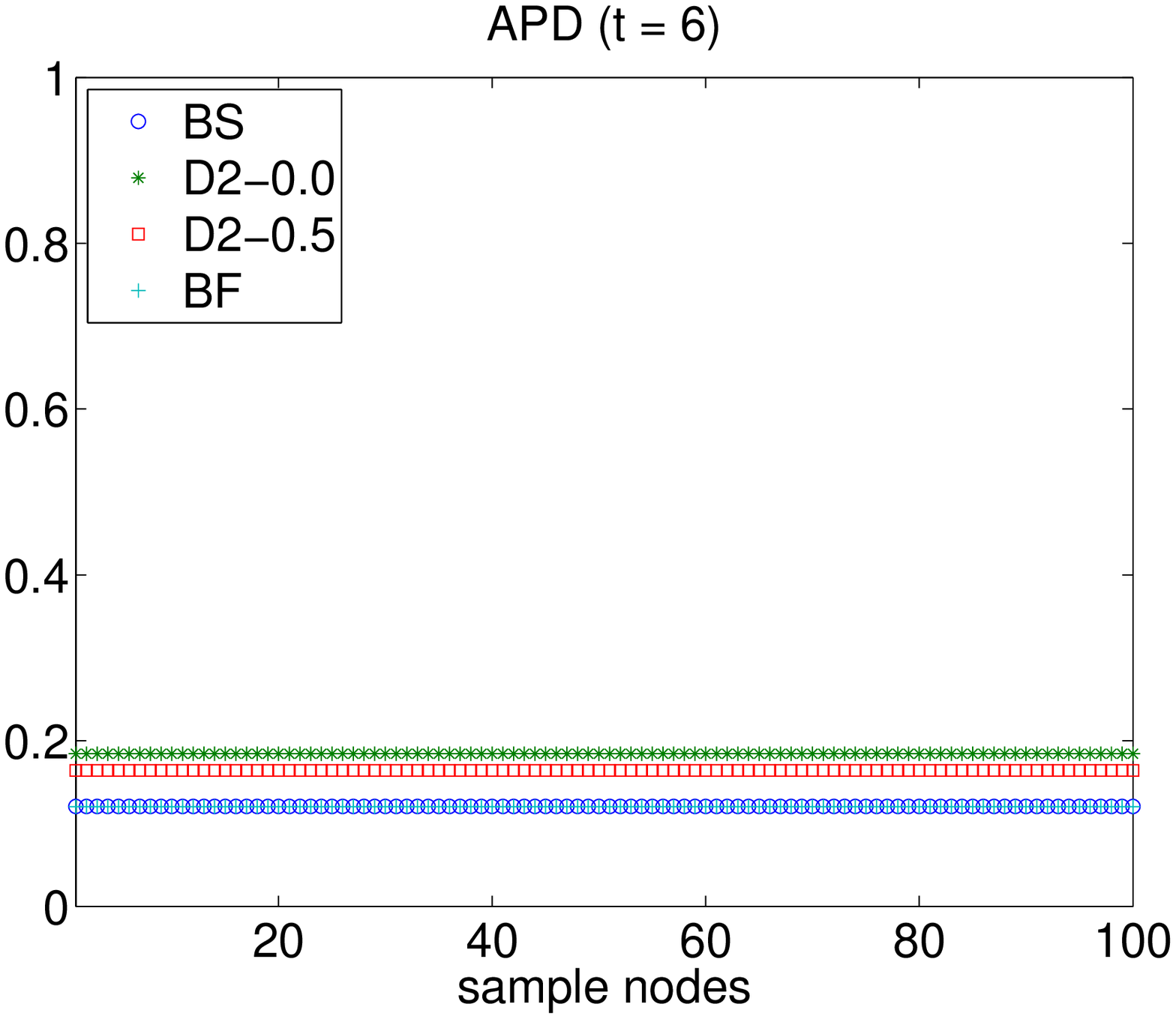, height=1.2in}
                 \setlength{\abovecaptionskip}{0pt}
                 \caption{}
                 \label{fig:util-APD-6}
         \end{subfigure}
         \hfill         
         \begin{subfigure}[b]{0.22\textwidth}
                  \centering
                  \epsfig{file=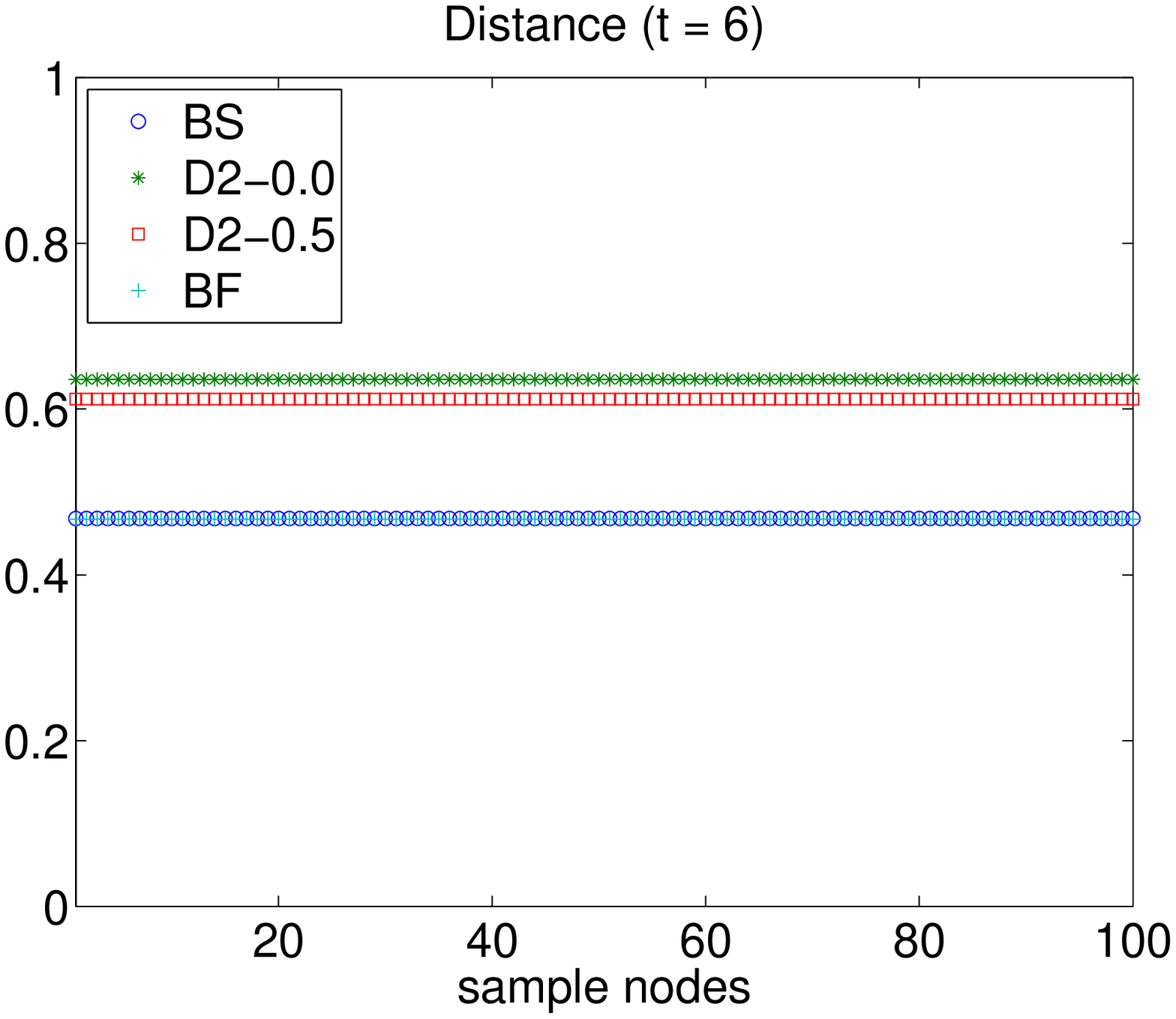, height=1.2in}
                  \setlength{\abovecaptionskip}{0pt}
                 \caption{}
                  \label{fig:util-Dist-6}
         \end{subfigure}           
     \caption{Utility relative errors on PL2 ($\alpha = 1.0, \beta = 0.5$)}
     \label{fig:utility}
\end{figure*}

\subsection{Utility-Oriented Initialization}
In this section, we illustrate the benefit of two-round initialization (Algorithm \ref{algo-init-util}). We set $\gamma = 0.0, 0.5$ and denote the enhanced scheme as \textit{D2}. Several utility metrics are chosen as follows.

- Power-law exponent of degree sequence: $PL$ is the estimate of $\eta$ assuming the degree sequence follows a power-law $n_d \sim d^{-\eta}$ where $n_d$ is the number of $d$-degree nodes.

- Clustering coefficient: $CC = \frac{3N_{\Delta}}{N_3}$ where $N_{\Delta}$ is the number of triangles and $N_3$ is the number of connected triples.

- Average distance: $APD$ is the average distance among all pairs of vertices that are path-connected.

- Distance distribution: $Distance$ is the normalized node-pair shortest-path histogram.

We take 100 sample nodes by degree and compare local aggregated graphs to the ground truth. The ground truth is computed by setting $\beta = 0$ in Baseline scheme. Fig. \ref{fig:utility} shows the benefit of two-round initialization (D2-0.0 and D2-0.5) on PL2 graph in early rounds. D2-0.0 and D2-0.5 schemes result in lower relative errors than Baseline and Bloom Filter in the first and second rounds, especially by $CC$ and $PL$ metrics. All schemes are comparable at $t=3$, except on $CC$ metric. Finally, Baseline and Bloom Filter are almost equivalent in terms of utility and they perform better D2 schemes at $t = Diam(G)$ on $PL$, $APD$ and $Distance$ metrics.

\section{Conclusion}
\label{sec:conclusion}
We motivate the private link exchange problem as an alternative to social graph crawling and centralized anonymization of data. The problem is distributed and provides a privacy/utility trade-off for all nodes. Our proposed problem is unique in the sense that the disseminated data over the links are the links themselves. We present two schemes for $(\alpha,\beta)$-exchange protocol: Baseline and Bloom filter based. 
Experiments on synthetic graphs clarify advantages and drawbacks of both schemes. Baseline scheme keeps link lists in clear form, so its communication cost increases fast. Bloom Filter scheme incurs lower communication complexity but needs an extra recovery step in the final round. Both schemes guarantee link privacy in the range $[\frac{1}{2\beta}, \frac{1}{\beta}]$. In Baseline, the inference attack based on link counting is not much better than the random attack.
For future work, we plan to investigate asynchronous models and node/links failures. We also consider community-based link exchange models in which nodes are gathered in super nodes and the link exchange takes place among super nodes only.

% %
\bibliographystyle{abbrv}
\bibliography{link-exchange-short}

\begin{thebibliography}{10}

\bibitem{fb-dir}
{Facebook Directory}.
\newblock \url{https://www.facebook.com/directory}.

\bibitem{bailey1975mathematical}
N.~T. Bailey et~al.
\newblock {\em The mathematical theory of infectious diseases and its
  applications}.
\newblock 1975.

\bibitem{bloom1970space}
B.~H. Bloom.
\newblock Space/time trade-offs in hash coding with allowable errors.
\newblock {\em Communications of the ACM}, 13(7):422--426, 1970.

\bibitem{broder2004network}
A.~Broder and M.~Mitzenmacher.
\newblock Network applications of bloom filters: A survey.
\newblock {\em Internet mathematics}, 1(4):485--509, 2004.

\bibitem{campan2008clustering}
A.~Campan and T.~M. Truta.
\newblock A clustering approach for data and structural anonymity in social
  networks.
\newblock In {\em PinKDD}, 2008.

\bibitem{dwork2014algorithmic}
C.~Dwork and A.~Roth.
\newblock The algorithmic foundations of differential privacy.
\newblock {\em Foundations and Trends in Theoretical Computer Science},
  9(3-4):211--407, 2014.

\bibitem{fawcett2006introduction}
T.~Fawcett.
\newblock An introduction to roc analysis.
\newblock {\em Pattern recognition letters}, 27(8):861--874, 2006.

\bibitem{ganesh2003peer}
A.~J. Ganesh, A.-M. Kermarrec, and L.~Massouli{\'e}.
\newblock Peer-to-peer membership management for gossip-based protocols.
\newblock {\em Computers, IEEE Transactions on}, 52(2):139--149, 2003.

\bibitem{giakkoupis2015privacy}
G.~Giakkoupis, R.~Guerraoui, A.~J{\'e}gou, A.-M. Kermarrec, and N.~Mittal.
\newblock Privacy-conscious information diffusion in social networks.
\newblock In {\em Distributed Computing}, pages 480--496. Springer, 2015.

\bibitem{huffman1952method}
D.~A. Huffman et~al.
\newblock A method for the construction of minimum-redundancy codes.
\newblock {\em Proceedings of the IRE}, 40(9):1098--1101, 1952.

\bibitem{moffat1998arithmetic}
A.~Moffat, R.~M. Neal, and I.~H. Witten.
\newblock Arithmetic coding revisited.
\newblock {\em ACM Transactions on Information Systems (TOIS)}, 16(3):256--294,
  1998.

\bibitem{moreno2002epidemic}
Y.~Moreno, R.~Pastor-Satorras, and A.~Vespignani.
\newblock Epidemic outbreaks in complex heterogeneous networks.
\newblock {\em The European Physical Journal B-Condensed Matter and Complex
  Systems}, 26(4):521--529, 2002.

\bibitem{newman2003structure}
M.~E. Newman.
\newblock The structure and function of complex networks.
\newblock {\em SIAM review}, 45(2):167--256, 2003.

\bibitem{nguyen2015anonymizing}
H.~H. Nguyen, A.~Imine, and M.~Rusinowitch.
\newblock Anonymizing social graphs via uncertainty semantics.
\newblock In {\em ASIACCS}, pages 495--506. ACM, 2015.

\bibitem{pastor2001epidemic}
R.~Pastor-Satorras and A.~Vespignani.
\newblock Epidemic spreading in scale-free networks.
\newblock {\em Physical review letters}, 86(14):3200, 2001.

\bibitem{shokri2009preserving}
R.~Shokri, P.~Pedarsani, G.~Theodorakopoulos, and J.-P. Hubaux.
\newblock Preserving privacy in collaborative filtering through distributed
  aggregation of offline profiles.
\newblock In {\em RecSys}, pages 157--164. ACM, 2009.

\bibitem{tassa2013anonymization}
T.~Tassa and D.~J. Cohen.
\newblock Anonymization of centralized and distributed social networks by
  sequential clustering.
\newblock {\em TKDE}, 25(2):311--324, 2013.

\bibitem{voulgaris2005cyclon}
S.~Voulgaris, D.~Gavidia, and M.~Van~Steen.
\newblock Cyclon: Inexpensive membership management for unstructured p2p
  overlays.
\newblock {\em Journal of Network and Systems Management}, 13(2):197--217,
  2005.

\end{thebibliography}

\end{document}